\documentclass{article}

\newcommand*{\Scale}[2][4]{\scalebox{#1}{$#2$}}%
\newtheorem{theorem}{Theorem}
\newtheorem{corollary}{Corollary}[theorem]
\newtheorem{lemma}{Lemma}

\usepackage{arxiv}

\usepackage{authblk}

\usepackage{amsmath}
\usepackage{amssymb}
\usepackage[utf8]{inputenc} 
\usepackage[T1]{fontenc}    
\usepackage{hyperref}       
\usepackage{url}            
\usepackage{booktabs}       
\usepackage{amsfonts}       
\usepackage{nicefrac}       
\usepackage{microtype}      
\usepackage{lipsum}		
\usepackage{graphicx}
\usepackage{natbib}
\usepackage{doi}

\title{Quantile Graph Discovery through QuACC: Quantile Association via Conditional Concordance}

\author[1]{Zain Khan}
\author[2]{Daniel Malinsky}
\author[3, 4]{Martin Picard}
\author[4, 5]{Alan A. Cohen}
\author[4]{Columbia SOH Group}
\author[2]{Ying Wei}
\affil[1]{Department of Biomedical Engineering, Columbia University}
\affil[2]{Department of Biostatistics, Columbia University}
\affil[3]{Department of Psychiatry \&\ Neurology, Columbia University}
\affil[4]{Robert N. Butler Columbia Aging Center, Columbia University}
\affil[5]{Department of Environmental Health Sciences, Columbia University}

\renewcommand{\headeright}{}
\renewcommand{\undertitle}{}
\renewcommand{\shorttitle}{}

\hypersetup{
pdftitle={Quantile Graph Discovery through QuACC: Quantile Association via Conditional Concordance},
pdfauthor={Zain Khan, Daniel Malinsky, Martin Picard, Alan A. Cohen, and Ying Wei},
pdfkeywords={Graphical models, conditional independence, tail dependence, quantile regression},
}

\begin{document}
\maketitle

\begin{abstract}
Graphical structure learning is an effective way to assess and visualize cross-biomarker dependencies in biomedical settings. Standard approaches to estimating graphs rely on conditional independence tests that may not be sensitive to associations that manifest at the tails of joint distributions, i.e., they may miss connections among variables that exhibit associations mainly at lower or upper quantiles. In this work, we propose a novel measure of quantile-specific conditional association called QuACC: Quantile Association via Conditional Concordance. For a pair of variables and a conditioning set, QuACC quantifies agreement between the residuals from two quantile regression models, which may be linear or more complex, e.g., quantile forests. Using this measure as the basis for a test of null (quantile) association, we introduce a new class of quantile-specific graphical models. Through simulation we show our method is powerful for detecting dependencies under dependencies that manifest at the tails of distributions. We apply our method to biobank data from All of Us and identify quantile-specific patterns of conditional association in a multivariate setting.
\end{abstract}


\section{Introduction}
\label{s:intro}

Structure learning and graphical models have seen growing use in discovering relationships in a variety of disciplines \citep{dobra2004sparse,iyer2013inferring,vowels2022d,petersen2021data,runge2019inferring,borsboom2013network}. The popularity of these methods is driven by their potential to uncover important relationships in complex systems and their interpretability, as multivariate dependence and conditional independence are easily visualized. In a graphical model, a missing edge between two variables represents some conditional independence between them. Specific measures of association and tests of independence vary depending on the applications.  Among parametric tests for continuous variables, Pearson's (partial) correlation coefficient is a popular association measure. Many nonparametric tests of conditional independence also exist, though no test will have power to detect all possible forms of dependence \citep{shah2020hardness}.

In many complex biological systems, the dependence and dynamics of their components often depend on their relative positions within their distribution, which can be understood as their quantile levels. For example, it has been reported that high environmental stress could alter gene expressions \citep{heim2012current}. This impact, whether up-regulating or down-regulating, is more evident among highly expressed genes (at upper quantile levels). Similarly, our immune response (antibody production) also depends on pathogen load. When the pathogen load is low, the correlation between pathogen and antibody loads is expected to be weak. When the pathogen load is moderately high (near median), there is a robust positive correlation between the pathogen load and antibody production. When the pathogen load is very high (at upper quantiles), the associations become more volatile and sometimes trigger the overproduction of antibodies (extreme upper quantiles) \citep{leishman2012periodontal,gates2021levels}. Hence, a single association measure would be inadequate to capture such complexity and  miss the opportunities to characterize the boundary behaviors, which is of great scientific interest.

\begin{figure}[hh]
\begin{center}
    
\includegraphics[width=.8\textwidth]{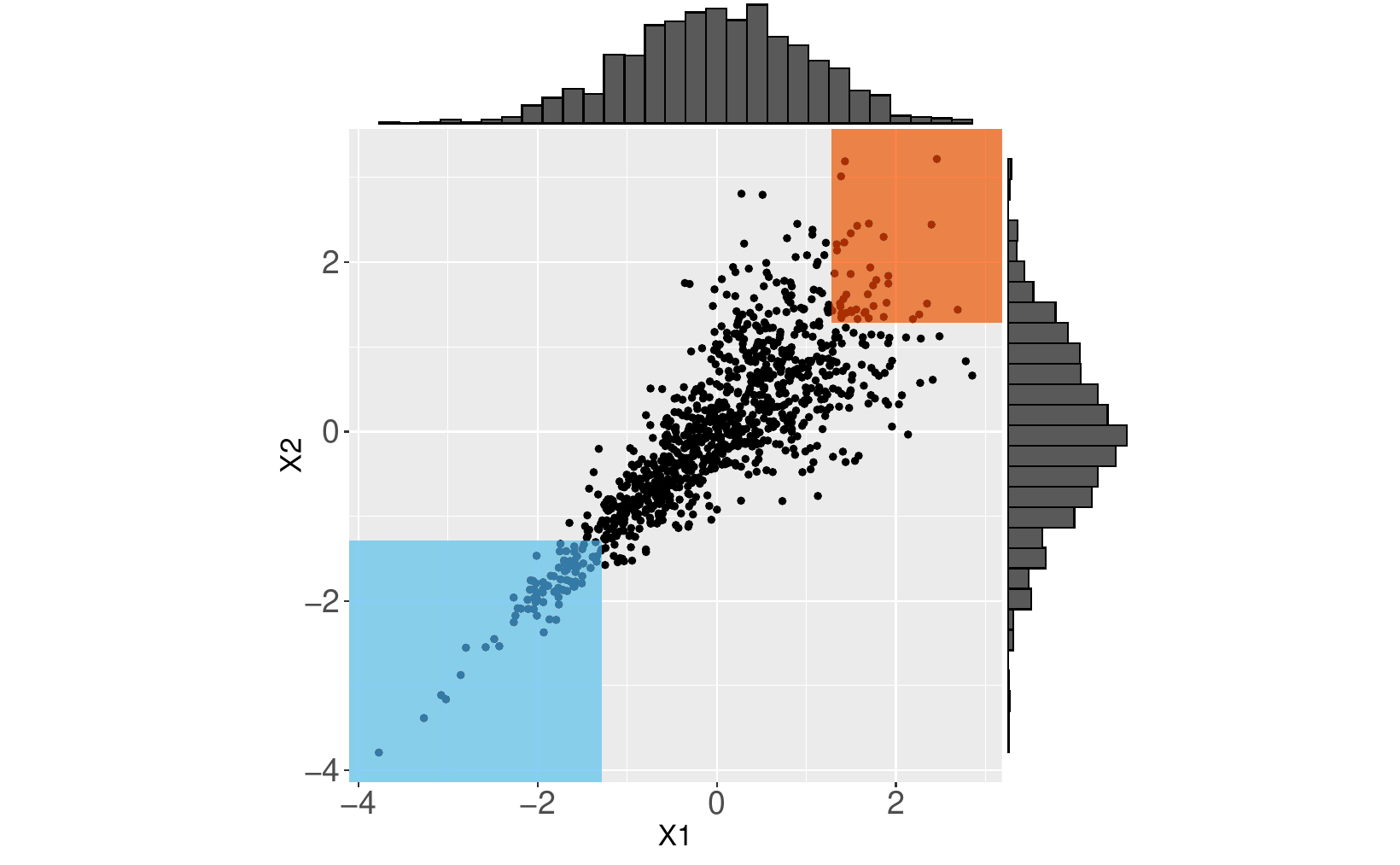}
\caption{Example estimates of the marginal QuACC for two standard normal variables related via a Clayton copula with parameter $\theta=4$. For $\tau = 0.1$, $\hat{\rho}_{0.1}= 0.083$ and for $\tau = 0.9$, $\hat{\rho}_{0.9} = 0.042$. The expected value under the null of independence is $0.01$. Despite the dispersion at the upper tails, the Clayton copula relates the two variables enough to produce a marginal QuACC at the upper tail greater than what would be expected under independence. } 
\label{f:exquacc}

\end{center}
\end{figure}

In this paper, we propose the Quantile Association via Conditional Concordance (QuACC) statistic: 
$$
\rho_{\tau}(Y,X|Z) = \begin{cases}      
P(Y > Q_{Y}(\tau | Z), X > Q_{X}(\tau | Z)) & \text{if } \tau \geq 0.5 \\
P(Y < Q_{Y}(\tau | Z), X < Q_{X}(\tau | Z)) & \text{if } \tau < 0.5  \\
\end{cases}   
$$
where $Y$ and $X$ are two continuous biomarkers of interest, 
and $Q_{Y}(\tau | Z)$ and $Q_{X}(\tau | Z)$ are their conditional quantile functions given a set of covariates $Z$ at the quantile level $\tau$. When the relevant variables are clear, we will sometimes write $\rho_{\tau} \equiv \rho_{\tau}(Y, X|Z)$ for brevity. The QuACC statistic $\rho_{\tau}$ measures the joint tendency of $X$ and $Y$ to approach a boundary (i.e. the limits in their conditional distributions), either at the upper quantile or lower quantile. A positive $\rho_{\tau}$ indicates that $X$ and $Y$ tend to approach their boundaries together, while a negative $\rho_{\tau}$ suggests a discordance between $X$ and $Y$ at their boundaries. The largest possible $\rho_{\tau}$ is $1$, while the smallest possible $\rho_{\tau}$ is $0$. We display an example calculation of $\rho_{\tau}$ for $\tau = 0.1$ and $\tau = 0.9$ in Figure \ref{f:exquacc} and illustrate how the statistic varies with tail dependence constructed from a Clayton copula. In the supplementary materials we describe a normalization procedure to scale $\rho_{\tau}$ to the $[-1, 1]$ interval to more directly resemble a correlation measure, denoted as $\rho_{\tau}^*$.

QuACC forms the basis of a distribution-free approach to measuring dynamic associations across different distributional locations.
In this study, we develop a statistical framework to formally test the null hypothesis $\rho_{\tau}(Y,X|Z)= \theta$, for some $\theta$, and further 
combine this with the PC algorithm \citep{spirtes2000causation} to estimate a graphical structure representing multivariate conditional quantile concordance relationships. The resulting graphical models learned will be referred to as QuACC graphical models (QGMs), which help depict the boundary dynamics in a complex system.

This work is related to copula estimation algorithms which aim to learn boundary behavior. However, copula estimation may require parametric assumptions, namely on the family of copulas being estimated. \citet{gijbels2011conditional} estimate a conditional copula nonparametrically, where the association between two variables is modeled conditional on a univariate conditioning variable and does not extend to multivariate conditioning sets. \citet{veraverbeke2011estimation} explore asymptotic properties of such conditional copula estimators. Quantile-specific association metrics have been developed based on copula constructions, e.g., the quantile odds ratios \citep{li2014quantile}, which have seen application in medical problems such as in survival analysis \citep{chen2021quantile}. The quantile odds-ratio association metric uses quantile regression in a similar manner to QuACC, however it is defined in terms of odds ratios conditional on some $Z$ which differs from our proposed construction. 


The intersection of quantile association and graphical modeling is a nascent field. Conditional Independence Quantile Graphical Models (CIQGM) and Prediction Quantile Graphical Models (PQGM) test conditional independencies among vertices in a graph for a predefined quantile level using penalized quantile regressions \citep{belloni2016quantile}. The Multiple Quantile Graphical Model (MQGM) method generalizes neighborhood selection methods introduced by \citet{meinshausen2006high} and models the conditional quantiles, as opposed to the conditional mean, of a given variable as a sparse function of other variables \citep{ali2016multiple}. \citet{guha2020quantile} study a Bayesian approach to structure learning with respect to quantile graphical models. The operative notion of independence in these works is whether one variable is (conditionally) informative for predicting the quantile level of another variable in the context of a quantile regression; this contrasts with our approach that is based on a joint tendency to for two variables to be on one side of a predicted extreme. Moreover, existing procedures are all focused on graphical structures that are Markov Random Fields (MRFs), in which the only conditional independence hypotheses considered are ones that condition on all variables in the analysis set. Our proposed statistic can be used for testing hypotheses with arbitrary and varying conditioning sets, so can be used in conjunction with procedures that estimated (partially) directed graphs as well as MRFs. We primarily focus on the PC algorithm for structure learning here, though the QuACC test may be used with many other conditional independence based structure learning algorithms as well. 

Existing methods adapt the PC algorithm for structure learning in settings with complex dependence, however these impose semiparametric assumptions on variable behavior, such as the nonparanormal model \citep{harris2013pc}, or study fully nonparametric conditional independence constraints with tests that may not have power to detect quantile-specific associations \citep{chakraborty2022nonparametric}.


In this work, we propose a parametric estimator of the QuACC statistic when conditional quantile functions are modeled via linear quantile regression. An estimator for QuACC when a non-parametric quantile regression models (i.e., random forest) are chosen is discussed in the supplementary materials. We show the asymptotic normality of our proposed estimators, which enables the construction of hypothesis tests. These tests may be straightforwardly incorporated into the PC algorithm for structure learning.

The rest of the paper is organized as follows. In Section \ref{s:methods} we introduce estimators for the QuACC statistic and study their properties. QuACC may be estimated using linear or nonlinear (quantile forest) estimators for the component quantile functions. In Section \ref{s:sim}, the QuACC statistic is tested across different synthetic copulas and QuACC graphical models are applied to a synthetic data generating process. The QuACC graphical models are compared to graphical models constructed via the PC algorithm using a partial correlation conditional independence test. In Section \ref{s:app} we apply QuACC graphical models to large scale health record data from the All of Us Research Program on a population of individuals with mitochondrial disorders  \citep{all2019all}. 

\section{Methods}
\label{s:methods}

\subsection{Estimation of Quantile Association via Conditional Concordance (QuACC)}
\label{s:methods:quacc}

Consider continuous scalar random variables $Y$ and $X$ as well as random vector $Z$, and $(Y_i,X_i, Z_i)$ are $n$ i.i.d.\ observations of the tuple $(Y, X, Z)$. 
We propose a  plug-in estimator for $\rho_{\tau}(Y,X | Z)$ with cross-fitting. We split the sample $(Y_i,X_i, Z_i)$ into $k$ folds where samples outside the $k$-th fold, $S_{-k}$ is used to estimate the quantile functions $Q_Y(\tau | Z)$ and $Q_X(\tau | Z)$, and the $k$-th fold, $S_{k}$, is used for evaluation of QuACC. A plug in estimator for the QuACC statistic is then
\begin{equation}
    \widehat{\rho}_{\tau}^{k}(Y,X|Z) = \begin{cases}      
\frac{1}{n_k} \sum_{j\in S_k} I \{ Y_{j} > \widehat{Q}_{Y}^{-k}(\tau | Z_j), X_{j} > \widehat{Q}_{X}^{-k}(\tau | Z_j) \} & \text{if } \tau \geq 0.5 \\
\frac{1}{n_k} \sum_{j\in S_k} I \{ Y_{j} < \widehat{Q}_{Y}^{-k}(\tau | Z_j), X_{j} < \widehat{Q}^{-k}_{X}(\tau | Z_j) \} & \text{if } \tau < 0.5  \\
\end{cases} 
\label{eq:plugin}
\end{equation}
where $I\{.\}$ denotes the indicator function and $n_k$ denotes the size of fold $S_k$. In practice, we split the data into $K=5$ folds of equal size. To make use of the entire sample, we can then reverse the roles of the folds and define an overall 
$$
\widehat{\rho}_{\tau} = \frac{\sum_{k=1}^K \widehat{\rho}_{\tau}^k }{\sqrt{ \sum_{k=1}^K Var(\widehat{\rho}_{\tau}^k) / n_k  }}
$$

 There is flexibility in the choice of estimator for the conditional quantile models $\widehat{Q}_{Y}^{-k}$ and $\widehat{Q}_{X}^{-k}$ ranging from simple parametric models to complex nonparametric approaches \citep{koenker2001quantile,meinshausen2006high,athey2019generalized}. We select linear quantile regression for this purpose because of its interpretability and linear properties \citep{koenker2001quantile}.

In evaluating QuACC in practice, we take into consideration the problem of non-unique quantiles. In the case where $Y, X$ are not fully continuous, as is common with many clinical lab tests (as instruments have limited precision), tied quantiles will pose a problem for QuACC. To address this, we introduce jittering to the data. 
For each variable $V$ considered, we add a small amount of uniform noise $\tilde{V} = V + \mbox{Unif}(-d/5, d/5)$, where $d$ is the smallest difference between adjacent values in $V$. The added noise is scaled such that quantile crossings will not occur and that the noise does not dominate the original variance. 

We illustrate some key features with a simple example. Consider the marginal QuACC, where $Z = \emptyset$, $Y, X \sim N(0, 1)$ and $Y,X$ are related by a Clayton copula with parameter $\theta = 2$. For this specific copula at $\tau = 0.1$, $\rho_{0.1} = C(u, v) = max\left( u^{-\theta} + v^{-\theta} - 1, 0 \right)^{\frac{-1}{\theta}} = 0.0709$, where $u = v = 0.1$. In estimating $\widehat{\rho_{\tau}}(X,Y)$, the marginal regressions will be equal $Q_{Y}(\tau) = Q_{X}(\tau) = \Phi^{-1} (\tau = 0.1) = -1.28$. Thus $\widehat{\rho_{\tau}}(X,Y)$, assuming one fold for simplicity, will be $\frac{1}{n} \sum_{i} I\{ Y_i < -1.28, X_i < -1.28 \}$ which will converge to $0.0709$.


\subsection{ Asymptotic behavior and inferential properties of $\widehat{\rho}_{\tau} (Y,X|Z)$}
Evaluating the asymptotic distribution of $\widehat{\rho}_{\tau} (Y,X|Z)$ depends both on the rate of convergence of the nuisance estimators, $\widehat{Q}_{Y}$ and $\widehat{Q}_{X}$, to the true conditional quantile functions and on the variance of concordance of the test set of $Y, X$ with its corresponding estimated conditional quantiles.  
Cross-fitting strategies allow for the separation of two sources of uncertainties and consequently 
ensure valid downstream inference. It also helps reduce over-fitting bias and improve the efficiency of the estimates,  particularly when dealing with high-dimensional covariates and using machine learning to learn the conditional quantile function \citep{zivich2021machine,chernozhukov2018double}.

We select $K=5$ folds for cross validation as this number balances robust estimates with reasonable computational cost. Each QuACC estimate requires $2K$ trained conditional quantile models, and training each conditional quantile model varies nonlinearly with sample size. Having $K$ greatly exceed 5 would have diminishing estimation returns while increasing time spent calculating QuACC. Higher values of $K$ lead to less variance in the models for each fold, as each fold is trained on more data and validated on a smaller set. Towards the other extreme, little or no sample splitting encounters the opposite problem where each fold has models that greatly differ due to drastically different training and validation sets \citep{hastie2009elements}.

We denote by $Q_{L}(\tau | Z)$ the true quantile regression for variable $L \in \{Y,X\}$ and the estimated quantile function as $\widehat{Q}_{L}(\tau | Z)$. For an arbitrary partition of size $n_k$ where $n_{-k}$ is the size of data outside the partition, we have $n_k + n_{-k} = n$.  In studying the variance of the plug-in estimator, we first decompose the estimator by
$$ \widehat{\rho}_{\tau}(Y, X|Z) = \rho_{\tau}(Y,X|Z)_0 + (\widehat{\rho}_{\tau}(Y, X|Z) - \rho_{\tau}(Y,X|Z)_0), $$
where $\rho_{\tau}(Y,X|Z)_0$ is the QuACC evaluated under the true conditional quantile $Q_{L}(\tau | Z)), \in \{Y, X\}$. When quantile models are correctly specified, $\rho_{\tau}(Y,X|Z)_0$ is the  the limit of $\widehat{\rho}_{\tau}(Y, X|Z)$.  The variance of QuACC is then composed of two elements: the variance of the approximation error between the estimated versus true quantile regressions, and the variance of concordance. The asymptotic variances and convergence rates of the estimated conditional quantiles vary depending on the choice of regression estimator, so the expression for the variance will need to be adjusted to match the estimators being used. We can state a general theorem for the plug-in estimator under the following assumptions.  

\begin{enumerate}
    \item[C1] $\text{max}_{i=1,..., n} ||Z_i|| / \sqrt{n} \rightarrow 0$
    
    \item[C2] The conditional distributions of $X$ and $Y$ given $Z$ are absolutely continuous, whose densities $f_L(Q_{L}(\tau | Z)), L=Y, X$ and joint density $f_{YX}(Q_{Y}(\tau | Z), Q_{X}(\tau | Z))$ are uniformly bounded away from $0$ and $\infty$ at the respective points $Q_{L}(\tau | Z_i), L \in \{Y,X\}, i=1, ..., n$. 
    
    \item[C3] The conditional quantile estimators, $\widehat{Q}_{L}(\tau | Z), L \in \{Y,X\}$, are consistent. i.e. $|\widehat{Q}_{L}(\tau | Z)- {Q}_{L}(\tau | Z)|=o_p(\sqrt{s/ n})$ for any $Z\in \mathbb{Z}$ where $s$ is the dimension of the training model for $\widehat{Q}_{L}(\tau | Z), L \in \{Y,X\}$ and $lim_{s \rightarrow \infty, n \rightarrow \infty} \frac{s}{n} \rightarrow 0$.

    \item[C4] The joint CDF, or copula, function, $F(Y, X | Z) =  P(Y < Q_{Y}(\tau | Z), X < Q_{X}(\tau | Z))$ is continuously differentiable and has unbounded partial derivatives with respect to the quantiles $Q_{L}(\tau | Z), L \in \{Y,X\}$ for all $Z$.
\end{enumerate}

Condition C1 bounds the moments of the covariate space and is a weaker assumption than assuming the covariate space is uniformly bounded. The combination of C1 and C2 ensure the estimated quantiles have valid asymptotic properties. C2 also guarantees a valid asymptotic covariance matrix for the parameters in the conditional quantile models. C3 makes the estimator for the conditional quantiles converge to the true quantiles. C4 is necessary for evaluating the concordant behavior between $X$ and $Y$, since if these partial derivatives are incalculable then QuACC is not estimable.

Recall that $\widehat{\rho}^k_{\tau}(Y, X|Z)$ is the estimated QuACC on the $k$-th partition, and $\rho_{\tau}(Y,X|Z)_0 $ is the true QuACC. We now introduce Theorem 1 on the asymptotic normality of $\widehat{\rho}^k_{\tau}(Y, X|Z)$.
\begin{theorem}
Suppose conditions C1 - C4 hold for an arbitrary partition of size $n_k$, then the asymptotic distribution of $\left\{ \widehat{\rho}^k_{\tau}(Y, X|Z) - \rho_{\tau}(Y,X|Z)_0 \right\}$ is determined by the model selection and convergence rate of the conditional quantile estimators, $b_n = n_k / s$.
\begin{equation}
(b_n)^{1/2} \left\{ \widehat{\rho}^k_{\tau}(Y, X|Z) - \rho_{\tau}(Y,X|Z)_0 \right\}  \overset{p}{\to}
N(0, \Sigma_\tau)
\end{equation}
When estimating the conditional quantiles via parametric models, if root-n consistency is achieved, the limiting distribution of $b_n$ is $n_k$. In more flexible models where the dimension of the model depends on sample size, the limiting distribution of $b_n$ is $n_k / s$.

The variance $\Sigma_\tau$ holds the same form across parametric and flexible models: 
$$ \Sigma_\tau =  \kappa_{Y} {\sigma}^2_{{Q}_{Y}} \kappa_{Y} + \kappa_{X} {\sigma}^2_{{Q}_{X}} \kappa_{X} 
+ 2 \kappa_{Y} \kappa_{X} V_{XY} (\tau) + V(\tau)$$
However, in the case of flexible models, $V(\tau)$ contributes negligibly to the overall variance and can be ignored.

$\kappa_{X}$ is defined as
\begin{equation}
\kappa_{X} = 
\begin{cases}
     lim_{n_k \rightarrow \infty} \frac{1}{n_k} \sum_{j = 1}^{n_k} f_{X} (Q_{X}(\tau | Z_j) | Y > Q_{Y}(\tau | Z_j)) (1 - F_{Y} (Q_{Y}(\tau | Z_j))) & \text{if } \tau \geq 0.5 \\
     lim_{n_k \rightarrow \infty} \frac{1}{n_k} \sum_{j = 1}^{n_k} f_{X} (Q_{X}(\tau | Z_j) | Y < Q_{Y}(\tau | Z_j)) F_{Y} (Q_{Y}(\tau | Z_j)) & \text{if } \tau < 0.5
\end{cases}
\end{equation}
and $\kappa_{Y}$ is defined similarly by exchanging $X$ and $Y$ in the definition of $\kappa_{X}$.
The covariance of the conditional quantile estimators are captured by $V_{XY}(\tau)$
$$V_{XY} (\tau) = \lim_{n_k\rightarrow \infty} \frac{1}{n_k} \sum_{i=1}^{n_k} (\widehat{Q}^{-k}_{Y} (\tau | Z_i) - \overline{\widehat{Q}^{-k}_{Y}} (\tau | Z_i)) ( \widehat{Q}^{-k}_{X} (\tau | Z_i) - \overline{\widehat{Q}^{-k}_{X}} (\tau | Z_i)),$$

where $\overline{\widehat{Q}^{-k}_{L}} (\tau | Z)$ is the mean of the $\tau$th conditional quantile estimates for $L$ and the variance of concordance of the true conditional quantile functions are given by $V(\tau)$ as follows:
$$
V(\tau) = \begin{cases}      
\Scale[0.75]{ (1 - \tau)^2 (1 - (1 - \tau))^2 + 
(1 - 4 ( 1 -\tau) + 4 (1 - \tau)^2) [p(Y > {Q}_{Y}(\tau | Z), X > {Q}_{X}(\tau | Z)) - (1 - \tau)^2]} & \text{if } \tau \geq 0.5 \\
\Scale[0.75]{ \tau^2 (1 - \tau)^2 + (1 - 4 \tau + 4 \tau^2) [ p(Y < {Q}_{Y}(\tau | Z), X < {Q}_{X}(\tau | Z))  - \tau^2 ]} & \text{if } \tau < 0.5.  \\
\end{cases}  
$$

\label{theorem:conv}
\end{theorem}

Theorem \ref{theorem:conv} shows that the estimated QuACC converges to the true QuACC and allows us to construct Wald confidence intervals. We note that the convergence rate is dependent on the limiting distribution of the conditional quantile estimator convergence rates.  

The weights $\kappa_{Y}, \kappa_{X}$ capture the partial derivative of concordance w.r.t the true quantile function (e.g. $\kappa_{X}$ corresponds to the partial derivative of quantile convergence, $(\widehat{Q}^{-k}_{X}(\tau | Z) - Q_{X}(\tau | Z))$, w.r.t. the true quantile function $Q_X(\tau | Z_j)$).

Regarding the variance of concordance of the true conditional quantile functions, $V(\tau)$, when $\tau < 0.5$, any value in $[0, \tau]$ can be substituted in for $p(Y < {Q}_{Y}(\tau | Z), X < {Q}_{X}(\tau | Z))$ to test that specific level of concordance. Likewise, when $\tau \geq 0.5$, any value in $[0, 1 - \tau]$ can be substituted in for $p(Y > {Q}_{Y}(\tau | Z), X > {Q}_{X}(\tau | Z))$. 

There are many choices in selecting an estimator for the conditional quantile functions, $\widehat{Q}^{-k}_{L}(\tau | Z)$. If linear quantile regression is selected as the estimator, we have a special case of Theorem \ref{theorem:conv} under a replacement assumption.



\begin{description}
    \item[C5] For $L \in \{Y,X\}$, there exist positive definite matrices $D_{0}$ and $D_{1, L}(\tau)$ such that 
        \begin{itemize}
            \item[1.] $\text{lim}_{n \to \infty} n^{-1} \sum Z_i Z_i^T = D_{0}$
            \item[2.] $\text{lim}_{n \rightarrow \infty} n^{-1} \sum f_{L}(Q_{L}(\tau| Z_i)) Z_i Z_i^T = D_{1, L}(\tau)$
        \end{itemize}
\end{description}

C5(i) and C5(ii) replace C2 and are typical assumptions that are necessary to ensure the asymptotic behavior of linear quantile regression. C3 is not necessary as linear quantile regression is already a consistent estimator by C5. 
\begin{corollary}
Suppose conditions C1 and C4-C5 hold for an arbitrary partition of size $n_k$ and the conditional quantiles are estimated via linear quantile regression, then
\begin{multline}
n_k^{1/2} \left\{ \widehat{\rho}^k_{\tau}(Y, X|Z) - \rho_{\tau}(Y,X|Z)_0 \right\}   
\overset{p}{\to} N(0, \kappa_{Y} {\sigma}^2_{{Q}_{Y}} \kappa_{Y} + \kappa_{X} {\sigma}^2_{{Q}_{X}} \kappa_{X} + 2 \kappa_{Y} \kappa_{X} V_{XY} (\tau) + V(\tau))
\end{multline}
Where ${\sigma}^2_{{Q}_{Y}}, {\sigma}^2_{{Q}_{X}}$ are the asymptotic variances of quantile convergence for linear quantile regression estimators \citep{koenker2001quantile}: 
\begin{align*}
\begin{split}
{\sigma}^2_{{Q}_{Y}} =  \tau(1 - \tau) Z D_{1, Y}^{-1} D_{0} D^{-1}_{1, Y} Z^T \\
{\sigma}^2_{{Q}_{X}} =  \tau(1 - \tau) Z D_{1, X}^{-1} D_{0} D^{-1}_{1, X} Z^T  
\end{split}
\end{align*}

and $V_{XY} (\tau)$ and $V(\tau)$ as defined as seen in Theorem \ref{theorem:conv}. 

\textbf{Proof:} Under C1 and C4-C5, ${\sigma}^2_{{Q}_{Y}}, {\sigma}^2_{{Q}_{X}}$ assume the above functional form. Substituting in these terms in the result for Theorem \ref{theorem:conv} completes this proof.  

\label{corollary:linearconv}
\end{corollary}

For theoretical results when generalized random forests \citep{athey2019generalized} are used in place of linear models for the conditional quantile estimators, see the supplementary materials.

\paragraph{Hypothesis test}

For fixed quantile level $\tau$, we focus on the null hypothesis
$$
H_0: \begin{cases}      
P(X>Q_{X}(\tau|Z), Y>Q_{Y}(\tau|Z)) = P(X>Q_{X}(\tau|Z)) P( Y>Q_{Y}(\tau|Z)) & \text{if } \tau \geq 0.5 \\
P(X < Q_{X}(\tau|Z), Y < Q_{Y}(\tau|Z)) = P(X < Q_{X}(\tau|Z)) P( Y < Q_{Y}(\tau|Z)) & \text{if } \tau < 0.5.  \\
\end{cases}  
$$

Under the independence null, when $\tau \geq 0.5$, the probability that realizations of $Y$ and $X$ are both jointly above their conditional quantiles is $(1 - \tau)^2$, since the probability that some observation lies above a quantile plane may be modeled as a Bernoulli random variable with parameter $(1 - \tau)$. By symmetry, when $\tau < 0.5$, the probability that realizations of $Y$ and $X$ are jointly below their conditional quantiles is $\tau^2$. Thus, under independence, the null hypothesis amounts to a test of 
$$
H_0: \begin{cases}      
\rho_{\tau}(Y,X|Z) = (1 - \tau)^2 & \text{if } \tau \geq 0.5 \\
\rho_{\tau}(Y,X|Z) = \tau^2 & \text{if } \tau < 0.5.  \\
\end{cases}  
$$

For each fold, under the null hypothesis we have 
$$
n_k^{1/2} \widehat{\rho}^k_{\tau}(Y, X|Z) \sim \begin{cases}      
N( (1 - \tau)^2, \kappa_{Y} {\sigma}^2_{{Q}_{Y}} \kappa_{Y} + \kappa_{X} {\sigma}^2_{{Q}_{X}} \kappa_{X} + (1 - \tau)^2(1- (1 - \tau))^2) & \text{if } \tau \geq 0.5 \\
N( \tau^2 , \kappa_{Y} {\sigma}^2_{{Q}_{Y}} \kappa_{Y} + \kappa_{X} {\sigma}^2_{{Q}_{X}} \kappa_{X} + \tau^2(1-\tau)^2) & \text{if } \tau < 0.5.  \\
\end{cases}  
$$
This results in the following test statistic under the null hypothesis  
$$\widehat{\rho}_{\tau} = \frac{\sum_{k=1}^K \widehat{\rho}_{\tau}^k - \rho_{\tau} }{\sqrt{ \sum_{k=1}^K Var(\widehat{\rho}_{\tau}^k) / n_k  }} \sim \: N(0, 1) $$.

Regarding inference in practice, requisite densities are estimated using the Hendricks-Koenker sandwich \citep{koenker2017handbook}. We denote an arbitrary density function $f_L$ and its corresponding quantile as $Q_{L}(\tau|.)$, without loss of generality. The Hendricks-Koenker sandwich uses the following representation:  
\begin{equation*}
f_L(Q_{L}(\tau| .)) = \frac{2 h_n}{ Q_{L}(\tau + h_n| .) - Q_{L}(\tau - h_n| .)}
\end{equation*}
    
Where $h_n$ is a bandwidth parameter that corresponds to either the \citet{hall1988distribution} bandwidth ($O^{-1/3}$) or the \citet{bofinger1975optimal} bandwidth ($O^{-1/5}$).

\subsection{QuACC Graphical Models (QGMs)}

We chose graphical models due to their ability to visualize complex conditional independence relationships, which has proved useful in various biomedical applications. Two commonly employed model classes are Markov Random Fields (MRFs), based on undirected graphs, and Bayesian network models, based on directed acyclic graphs (DAGs). In an MRF, a missing edge between some pair of variables corresponds to conditional independence given a specific conditioning set: the set of all remaining variables represented in the graph. In contrast, a missing edge in a DAG (Bayesian network model) represents a conditional independence where the conditioning set is some subset of the remaining variables; this could be the empty set (marginal independence), some small number of variables, or many. The correspondence between pathways in a graph and conditional independence relationships encoded by that graph is captured by the criterion of d-separation \citep{pearl2009causality}. We focus on structure learning for DAGs here, rather than MRFs for two reasons: 1) algorithms that learn MRFs can produce output that is overly dense in some biomedical applications, due to ``spurious’’ dependence induced by conditioning on colliders \citep{sanchez2021combining}, and 2) the tests involved in learning MRFs require conditioning on potentially high-dimensional covariate sets (all remaining variables) which leads to tests that may have low power. In any case, the QuACC statistic can be used as the basis of procedures that learn MRFs or Bayesian network structures, as long as these procedures are based on hypothesis tests of independence. For the difference between models based on MRFs and models based on DAGs, see \citet{lauritzen1996graphical}.

The PC algorithm is widely studied graphical structure learning algorithm that estimates an equivalence class of DAGs \citep{spirtes2000causation, kalisch2007estimating}. The procedure begins with a complete graph, where each pair of vertices is connected by an undirected edge. A sequence of conditional independence hypotheses are tested, beginning with marginal independence tests, and then with conditioning sets of increasing cardinality. For every pair of variables, if any independence hypothesis is accepted (not rejected at the chosen significance level), the edge between the corresponding pair of vertices is removed. PC then performs additional steps to orient some edges as directed based on a series of graphical rules which we do not discuss here; we limit ourselves to the ``adjacency'' phase of PC that is concerned only with the presence or absence of edges. Thus, we use PC only to estimate what is called the undirected skeleton of a DAG. Though the resulting graph is undirected, it is important to keep in mind that this is not a Markov Random Field. In the output of PC, missing edges correspond to conditional independence given some subset of the remaining variables.

\section{Simulation Studies}
\label{s:sim}

\subsection{QuACC Simulations}

We conduct simulations to show that QuACC correctly identifies quantile dependence when it exists under different conditions. Specifically, we assume a modified version of the simulated data generating processes from \citet{li2014quantile}. Suppose that $Y, X$ share the same marginal behavior that depends on $Z_1, Z_2$, where $Z_1$ follows a standard normal variable truncated at -2 and 2, and $Z_2 \sim Bernoulli(0.5)$.
\begin{align*}
\begin{split}
Q_Y(u | Z) = & \: Qnorm(u; 0, 0.2^2) + \alpha_1 Z_1  + \alpha_2 \{ Qnorm(u; 0, 0.4^2) - Qnorm(u; 0, 0.2^2) \} Z_2 \\
Q_X(u | Z) = & \: Qnorm(u; 0, 0.3^2) - \alpha_3 Z_1 + \alpha_4 Z_2
\end{split}
\end{align*}

Here $Qnorm(u; \mu, \sigma^2)$ denotes the $u$th quantile of a standard normal distribution with mean $\mu$ and variance $\sigma^2$. 

We consider similar associations between $Y, X$ as seen in \citet{li2014quantile} through three data generating processes labelled $S_1, S_2,$ and $S_3$. The simulation parameters have been adjusted such that quantile dependence at particular $\tau$ levels are magnified or attenuated.

Under $S_1$, $Y, X$ are related through a Plackett copula with parameter $\theta = e^{1.42}$ and $\alpha_1 = \alpha_2 = 0.25; \alpha_3 = \alpha_4 = 0$. Such a DGP exhibits quantile association throughout the distribution but notably symmetrically at the quantile extremes.

The setting $S_2$ exhibits heterogeneity in dependence across values of $Z_2$. When $Z_2 = 1$, $Y, X$ are associated through a flipped Clayton copula with parameter $\theta = 1$. When $Z_2 = 0$, $Y, X$'s marginals are independent and not related through a copula. Here $\alpha_1 = \alpha_2 = 0.25; \alpha_3 = 1; \alpha_4 = 0.5$. $Z_2$ influences association strength, through the presence or absence of the flipped Clayton copula, and structure, through its effect on the marginal distributions of $Y, X$. We expect associational strength at the extreme upper quantiles via the flipped Clayton and not elsewhere due to the independence when $Z_2=0$. 

$S_3$ is identical to $S_2$ except for the case when $Z_2 = 0$. Instead of allowing conditional independence between $Y, X$ in this case, we let $Y, X$ be associated through a (regular) Clayton copula with parameter $\theta=1$. Since the strength of association is identical when $Z_2 =1$ and $Z_2 = 0$, this case signifies an example where a covariate influences the structure of association without influencing strength of association. In this setting, we would expect to detect significant dependence with QuACC across all quantiles except the nonzero effect from $Z_2$ since $\alpha_3=1; \alpha_4=0.5$ confounds lower quantile association, and instead we see a linearly increasing association as the quantile level rises.

\begin{figure}[htbp]
\centering
\begin{tabular}{cc}
      \includegraphics[width=0.45\textwidth]{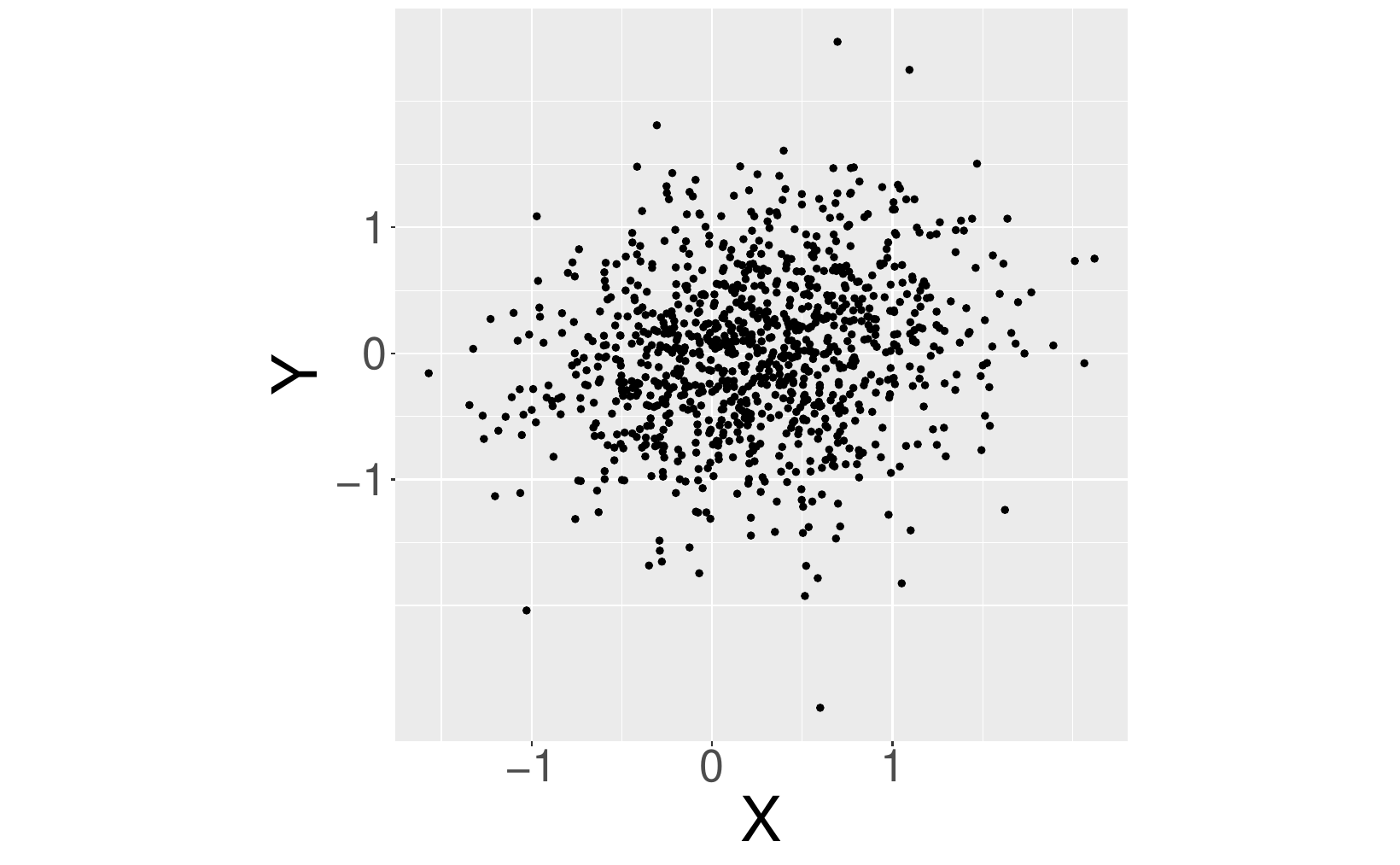} &  
      \includegraphics[width=0.45\textwidth]{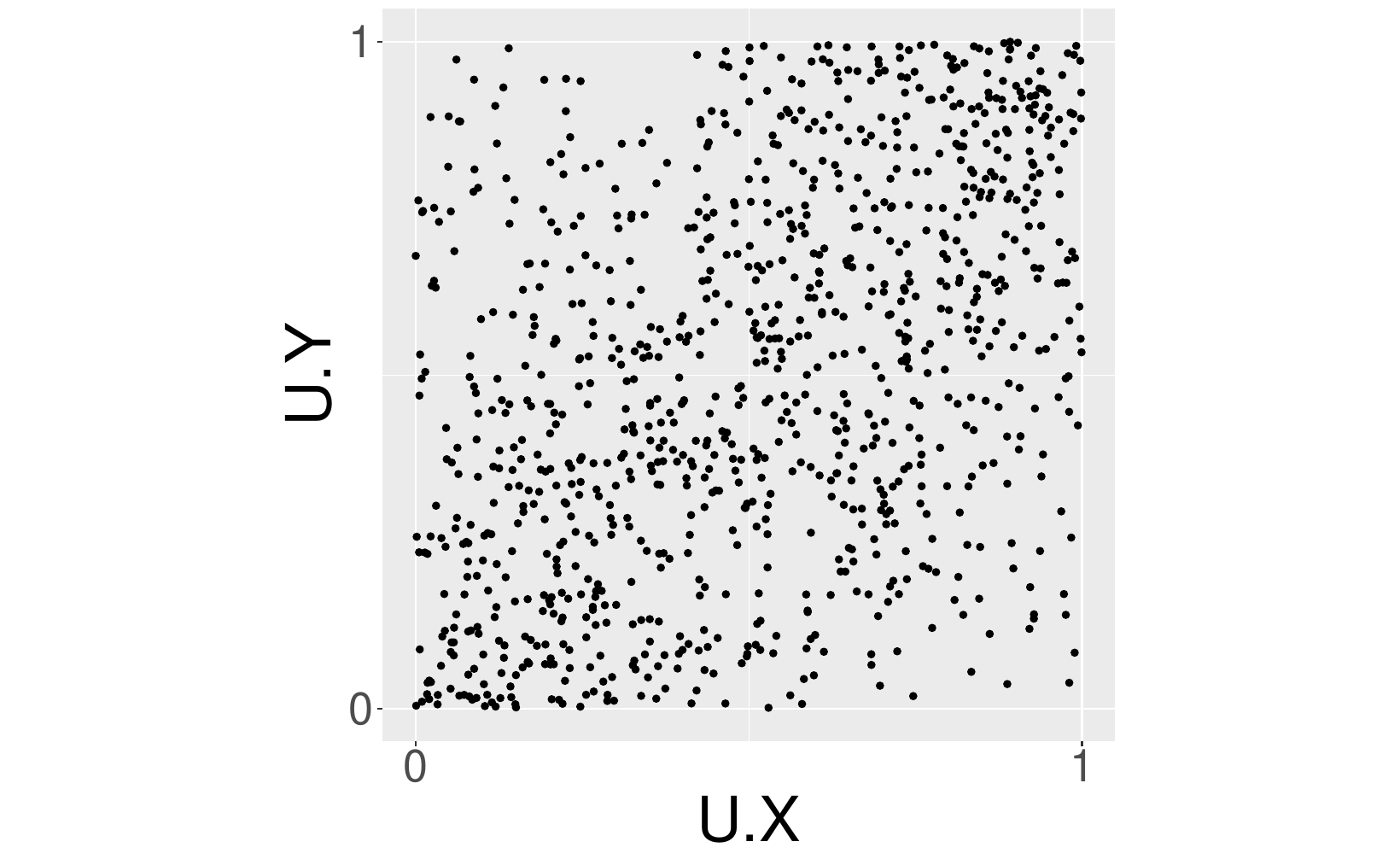} \\
    (a) $S_1$ & (b) Plackett in $S_1$ \\[6pt]
    
      \includegraphics[width=0.45\textwidth]{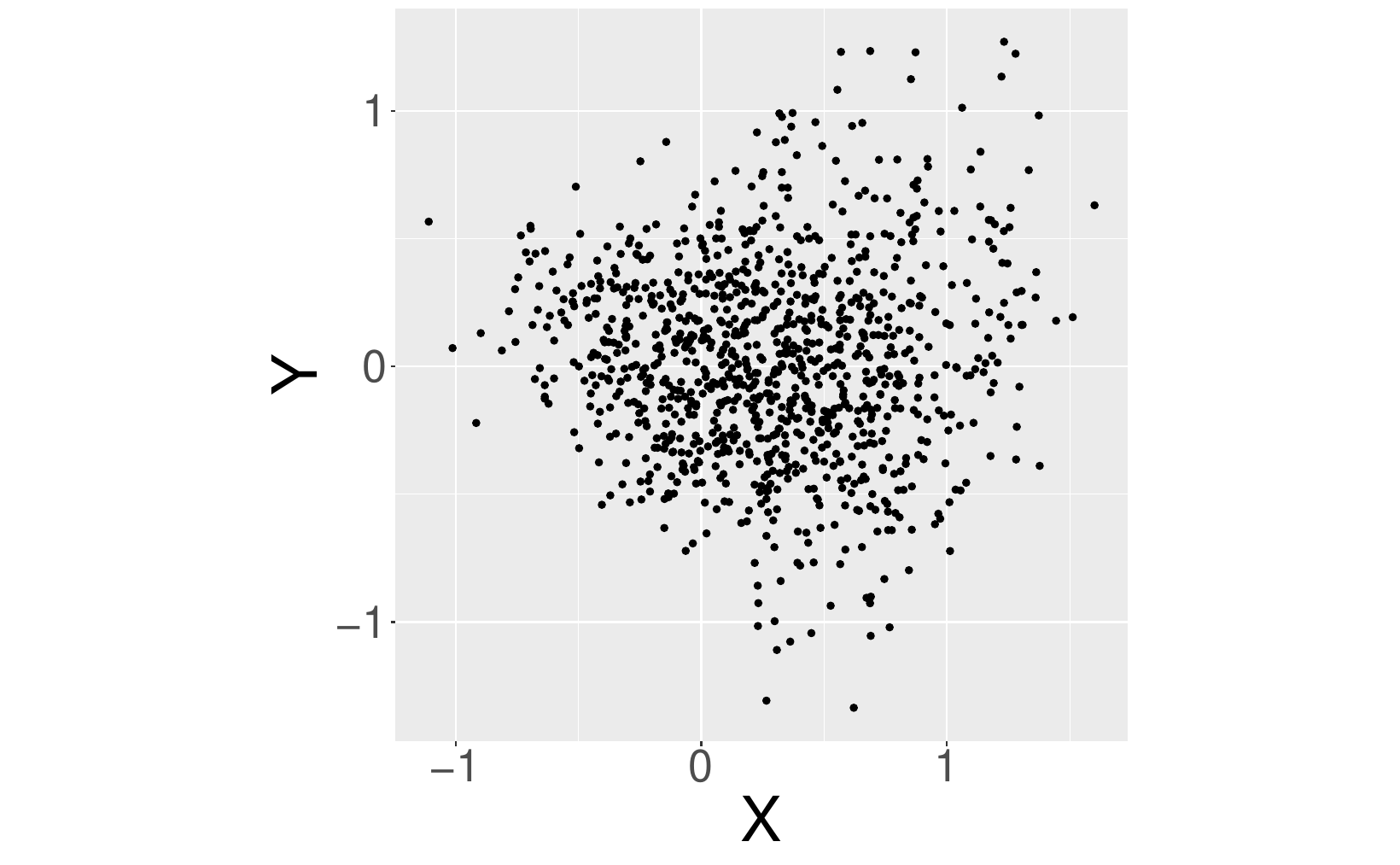} &   
      \includegraphics[width=0.45\textwidth]{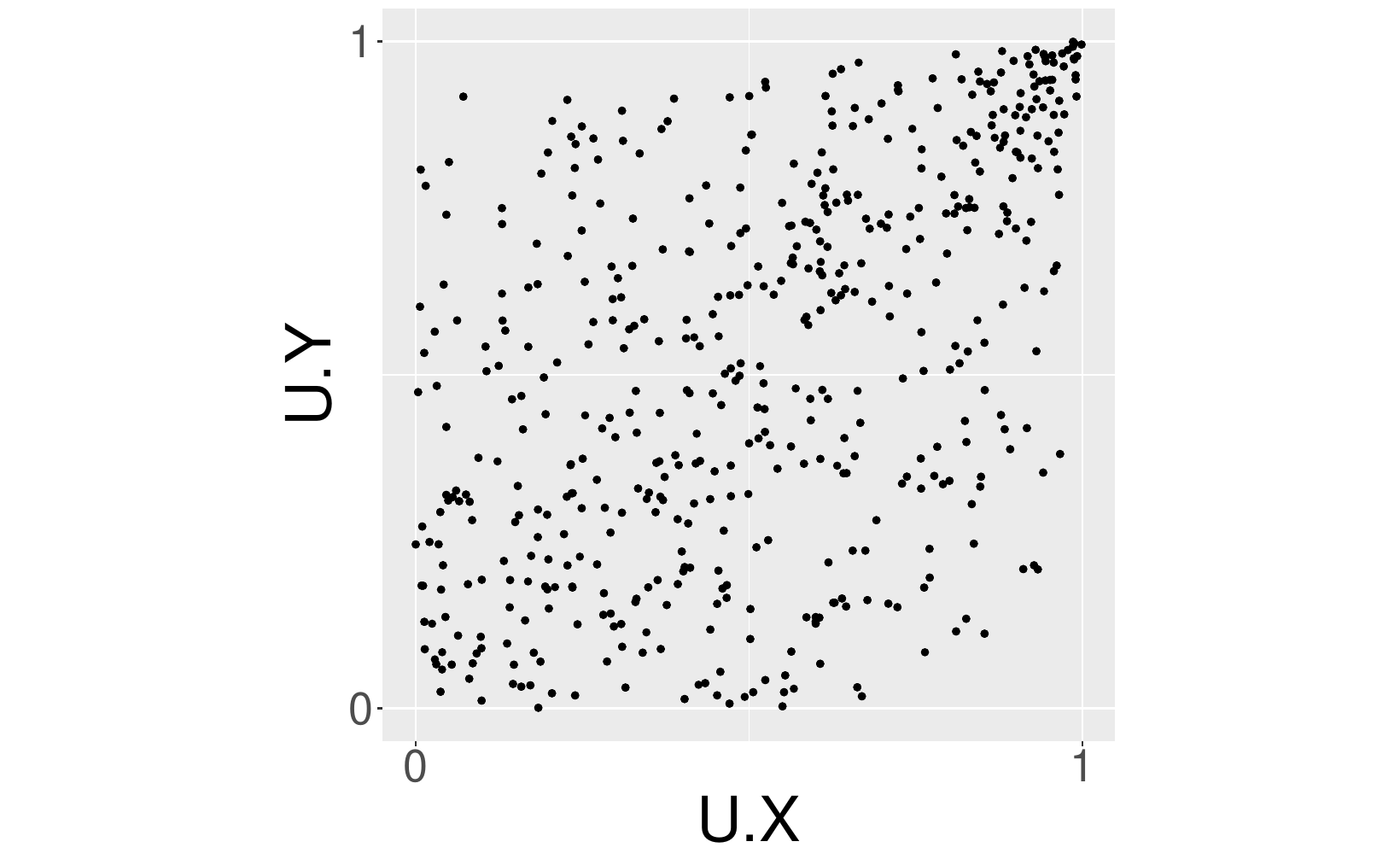} \\   
    (c) $S_2$ & (d) Flipped Clayton + independence in $S_2$  \\[6pt]
          
    \includegraphics[width=0.45\textwidth]{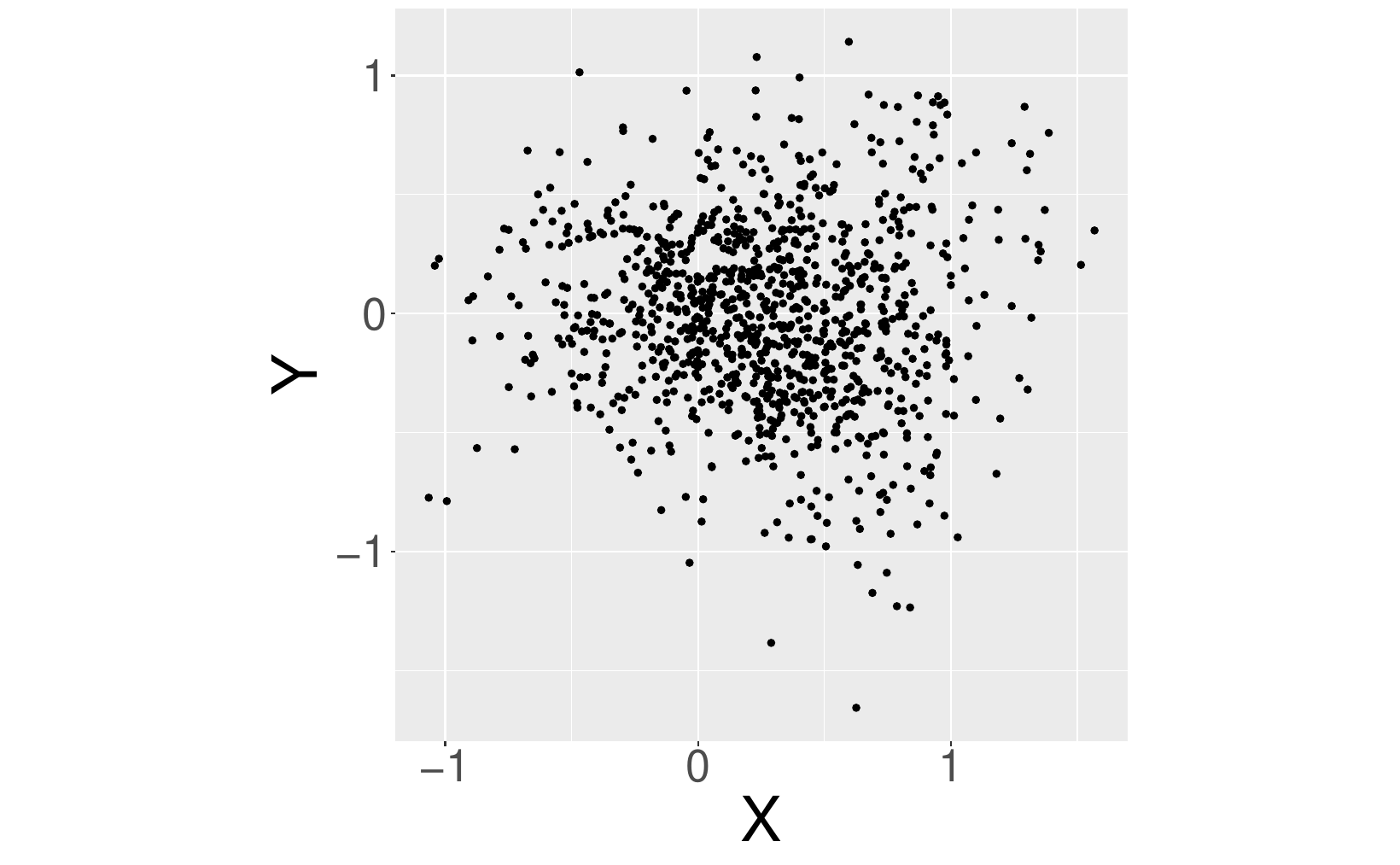} &  \includegraphics[width=0.45\textwidth]{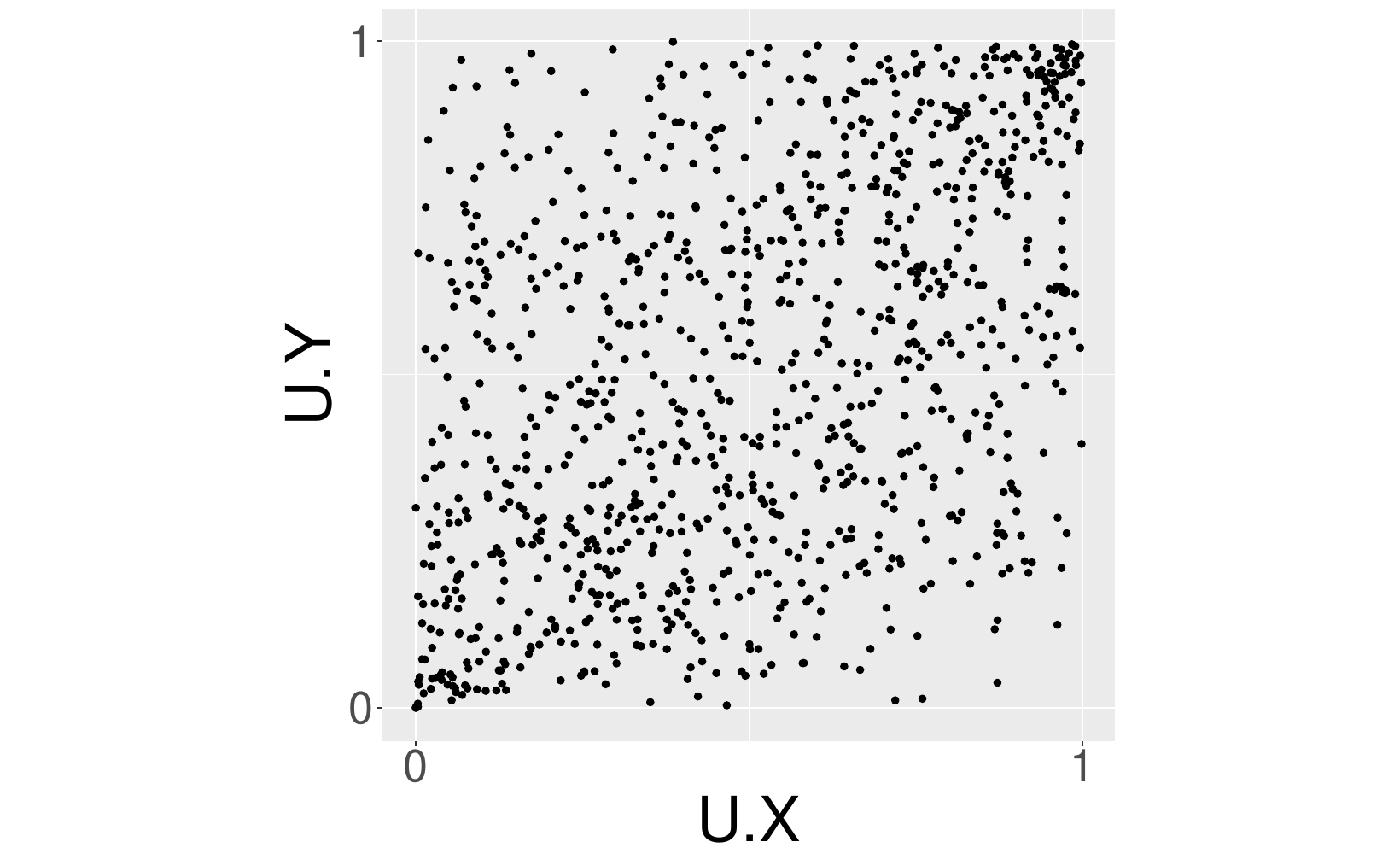} \\
    (e) $S_3$ & (f) Flipped Clayton and Clayton in $S_3$ \\[6pt]
    
    & 
\end{tabular}
\caption{Example draws from simulation settings $S_1, S_2, S_3$ alongside the copula relationship present in each. In $S_1$ and $S_3$ there is strong quantile concordance at each quantile level, although it is heightened at the tails. In $S_2$, the concordance increases with the quantile level.}
\label{fig:examplesims}
\end{figure}

For more details on these data-generating processes, refer to \citet{li2014quantile}. Samples from $S_1, S_2, S_3$ are visualized in Figure \ref{fig:examplesims}.

\begin{figure}[htbp]
    \centering
    
    \begin{minipage}[b]{0.3\textwidth}
        \centering
        \textbf{\quad \quad $\tau = 0.1$}
    \end{minipage}
    \begin{minipage}[b]{0.3\textwidth}
        \centering
        \textbf{\quad \quad $\tau = 0.5$}
    \end{minipage}
    \begin{minipage}[b]{0.3\textwidth}
        \centering
        \textbf{\quad \quad $\tau = 0.9$}
    \end{minipage}
    
    \vspace{0.5cm} 
    
    \begin{minipage}[b]{0.05\textwidth}
        \raggedright
        \textbf{S1}
    \end{minipage}
    \begin{minipage}[b]{0.3\textwidth}
        \centering
        \includegraphics[width=\textwidth]{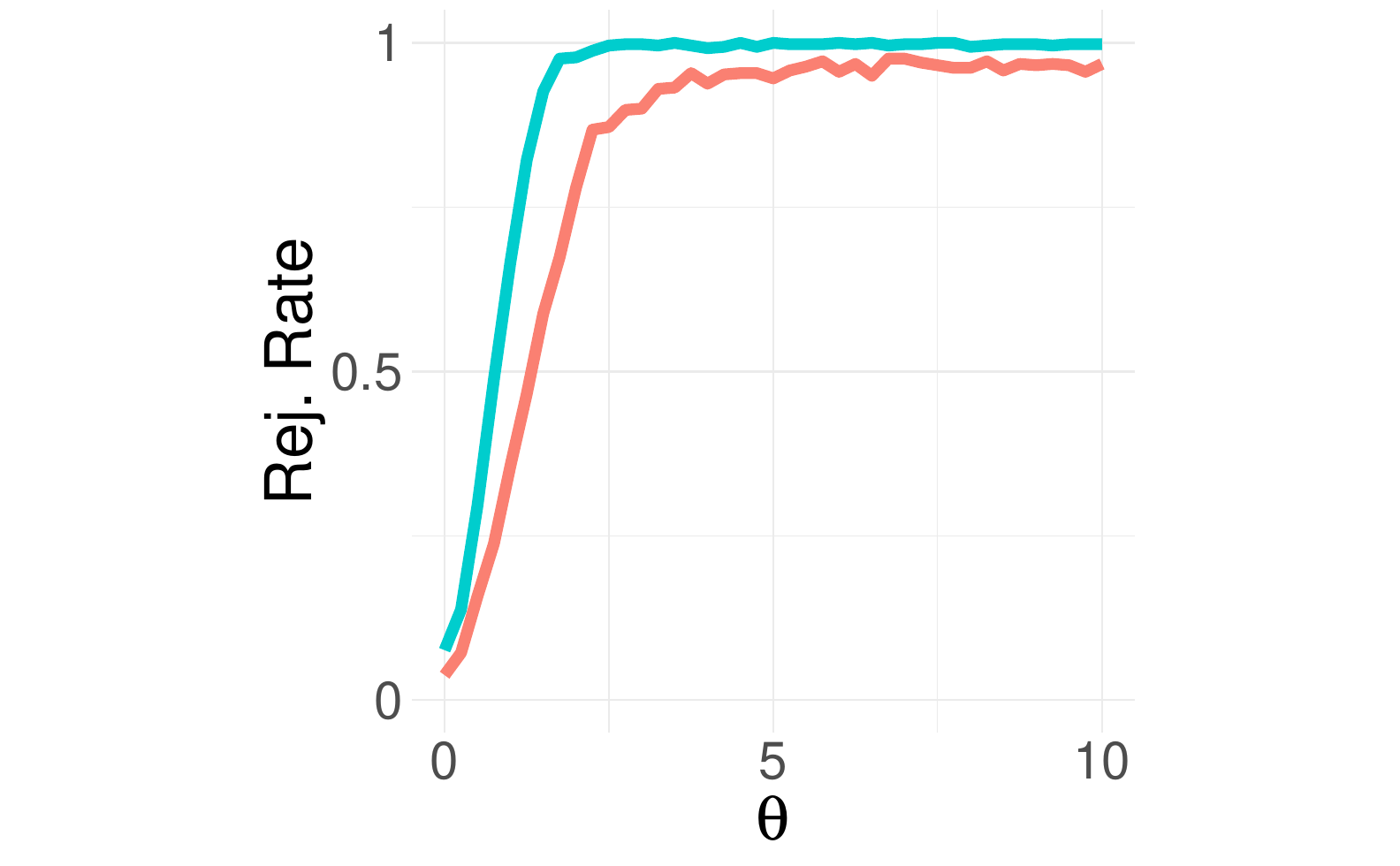}
    \end{minipage}
    \begin{minipage}[b]{0.3\textwidth}
        \centering
        \includegraphics[width=\textwidth]{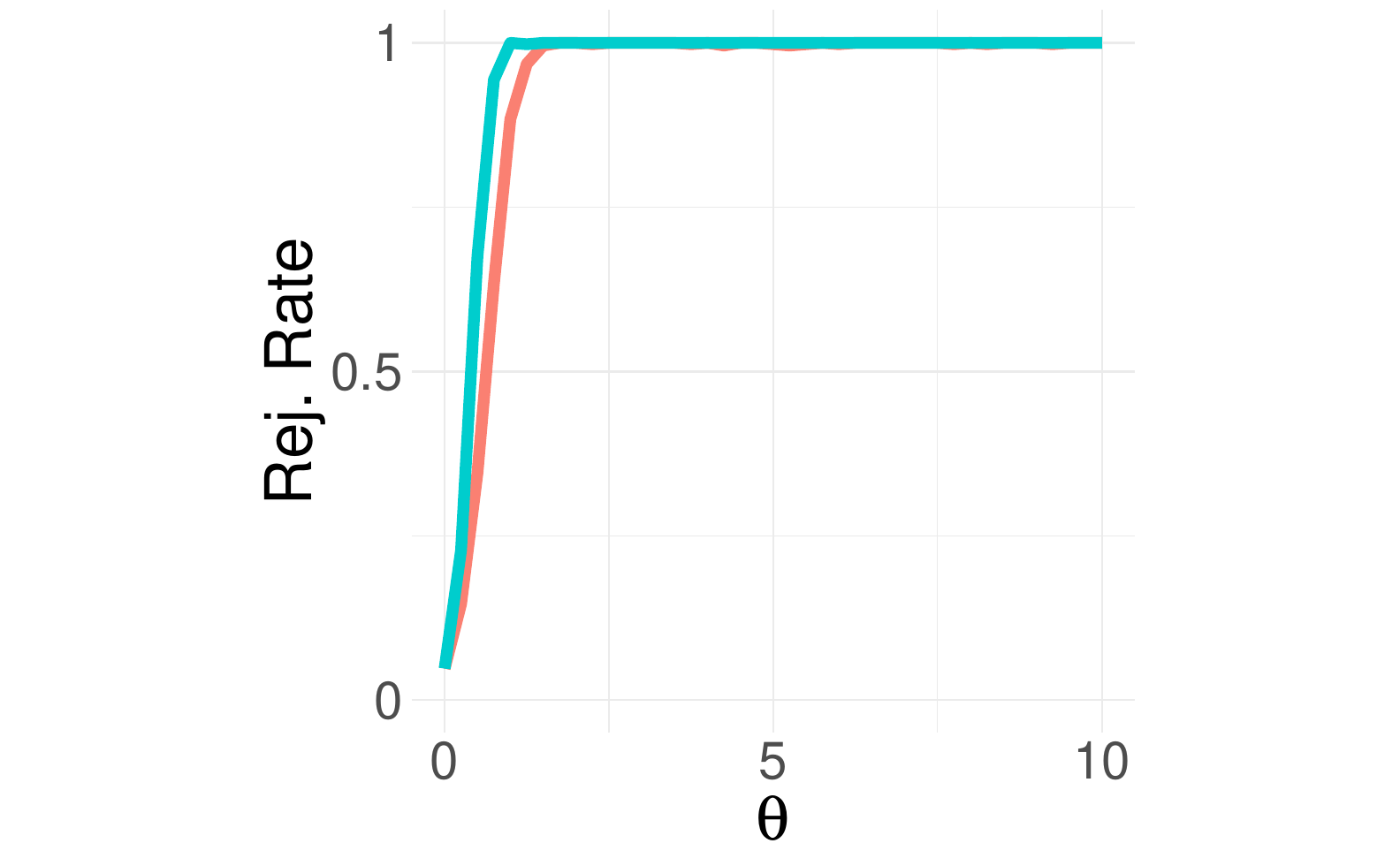}
    \end{minipage}
    \begin{minipage}[b]{0.3\textwidth}
        \centering
        \includegraphics[width=\textwidth]{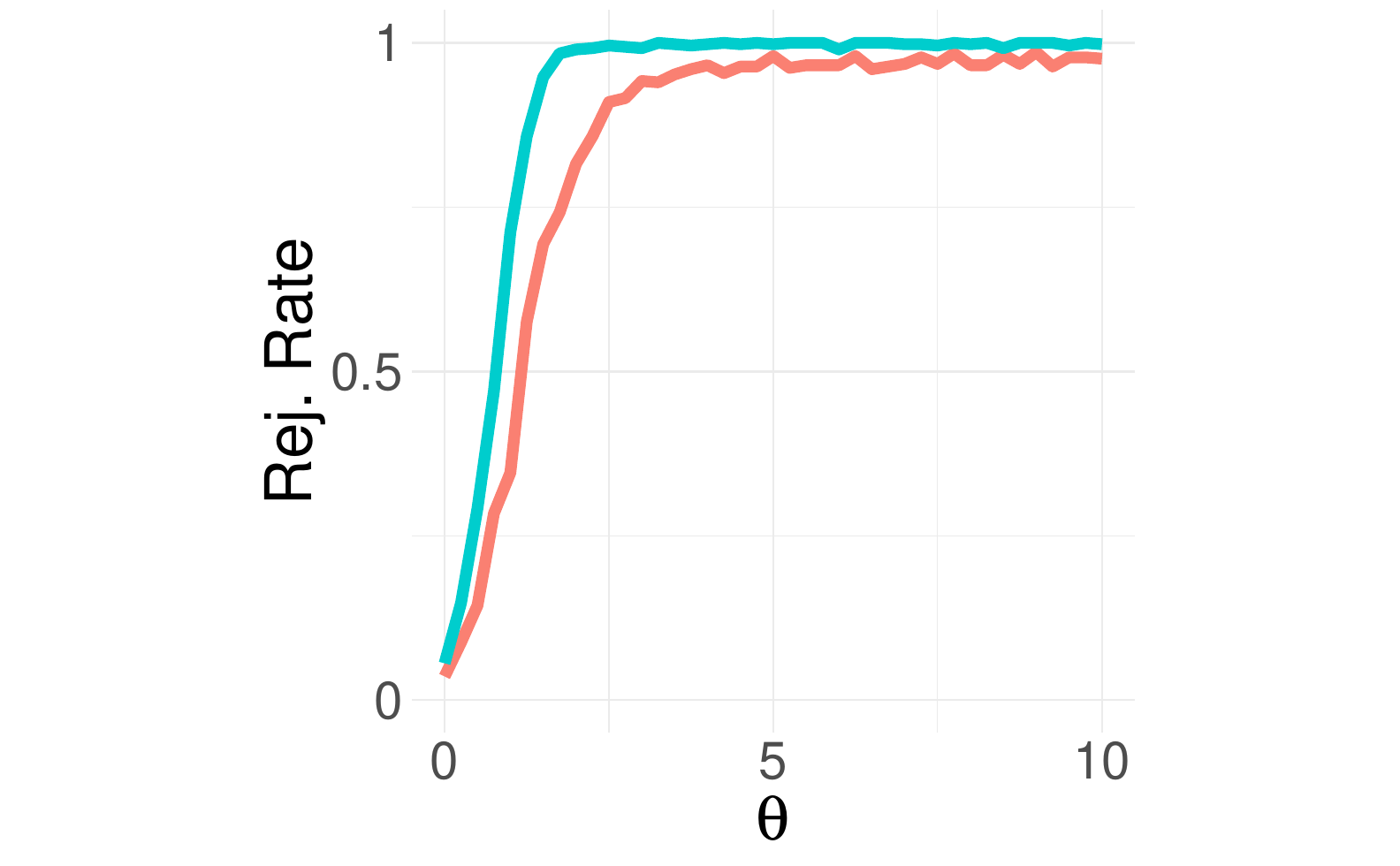}
    \end{minipage}

    \vspace{0.5cm} 

    \begin{minipage}[b]{0.05\textwidth}
        \raggedright
        \textbf{S2}
    \end{minipage}
    \begin{minipage}[b]{0.3\textwidth}
        \centering
        \includegraphics[width=\textwidth]{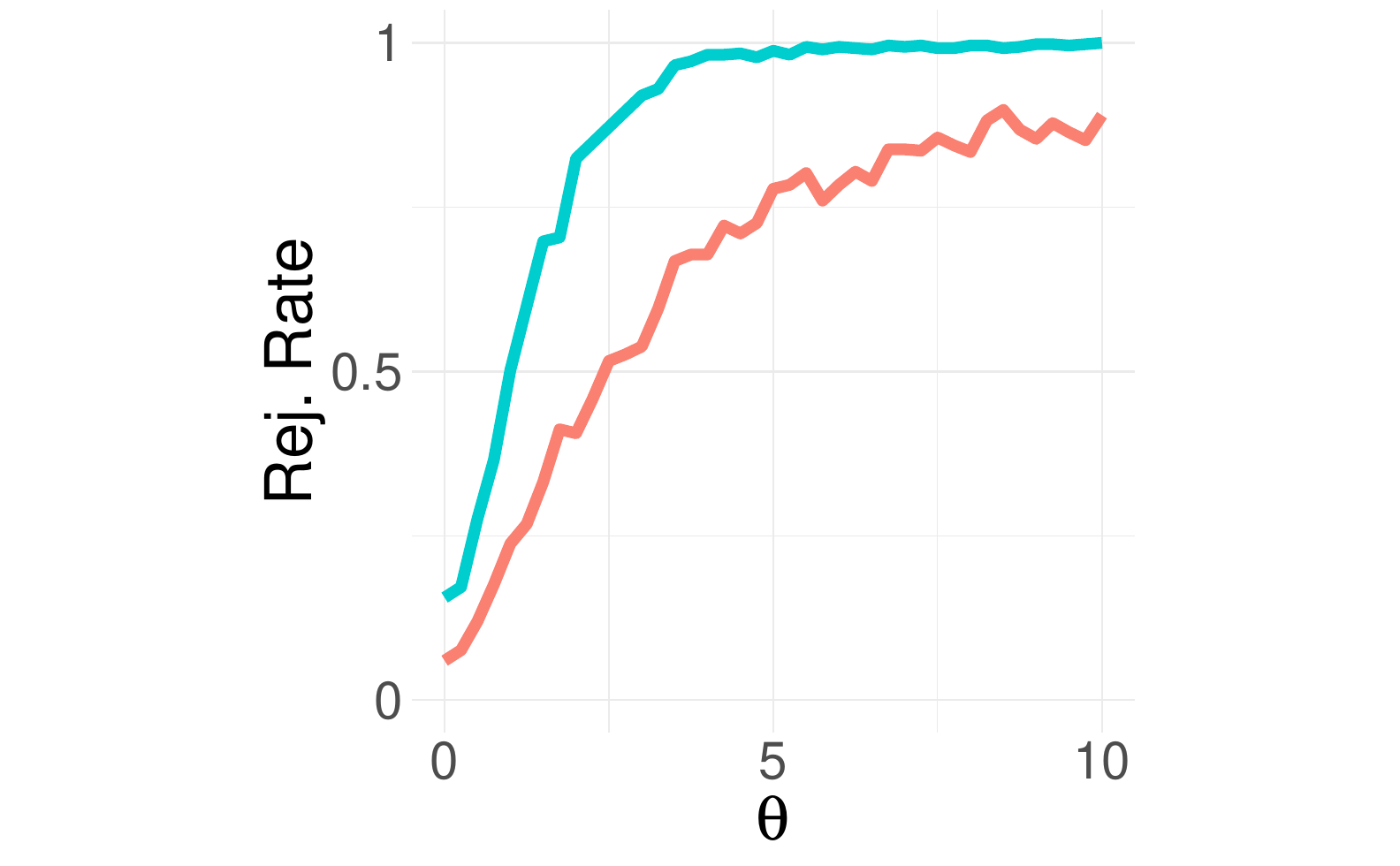}
    \end{minipage}
    \begin{minipage}[b]{0.3\textwidth}
        \centering
        \includegraphics[width=\textwidth]{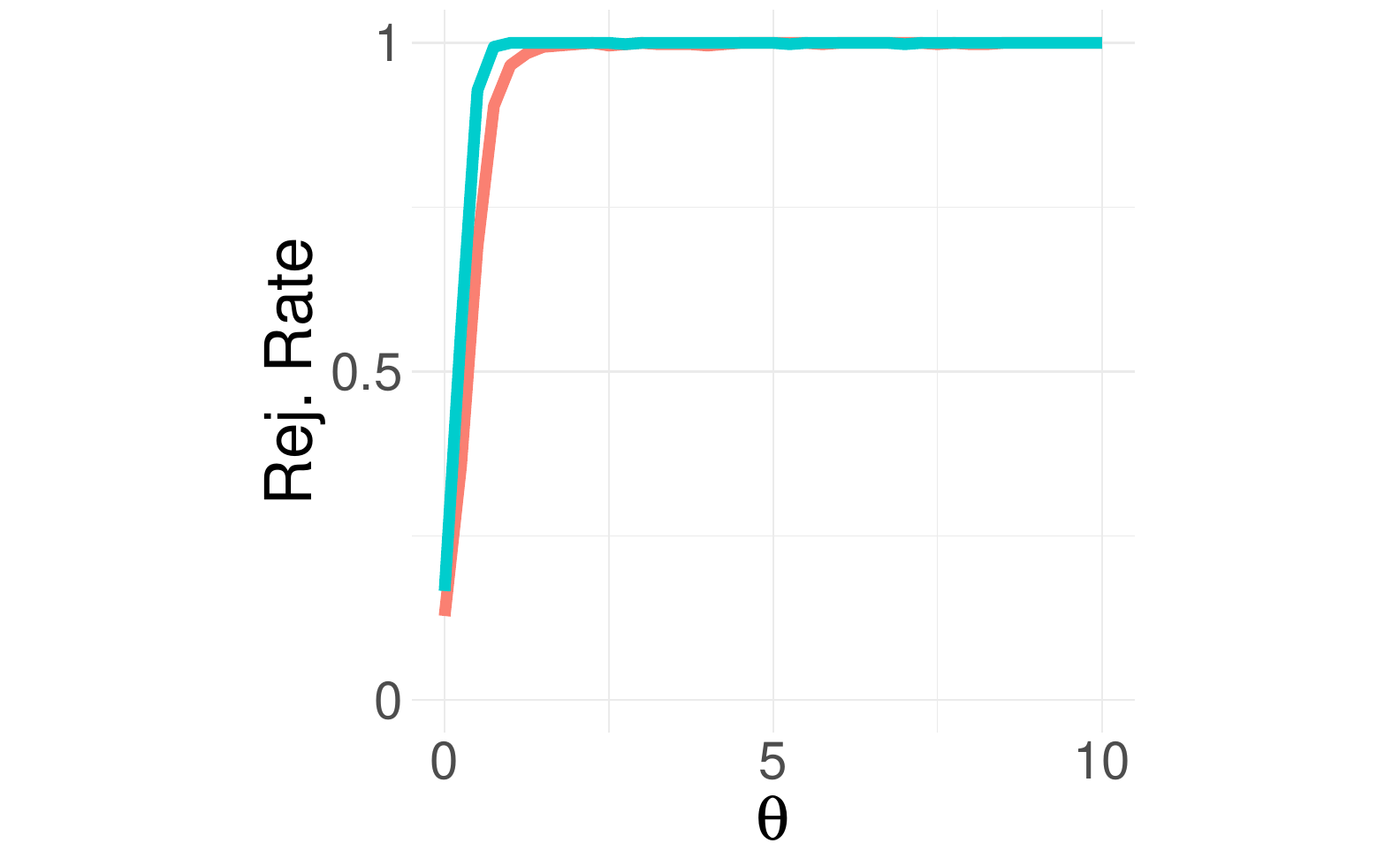}
    \end{minipage}
    \begin{minipage}[b]{0.3\textwidth}
        \centering
        \includegraphics[width=\textwidth]{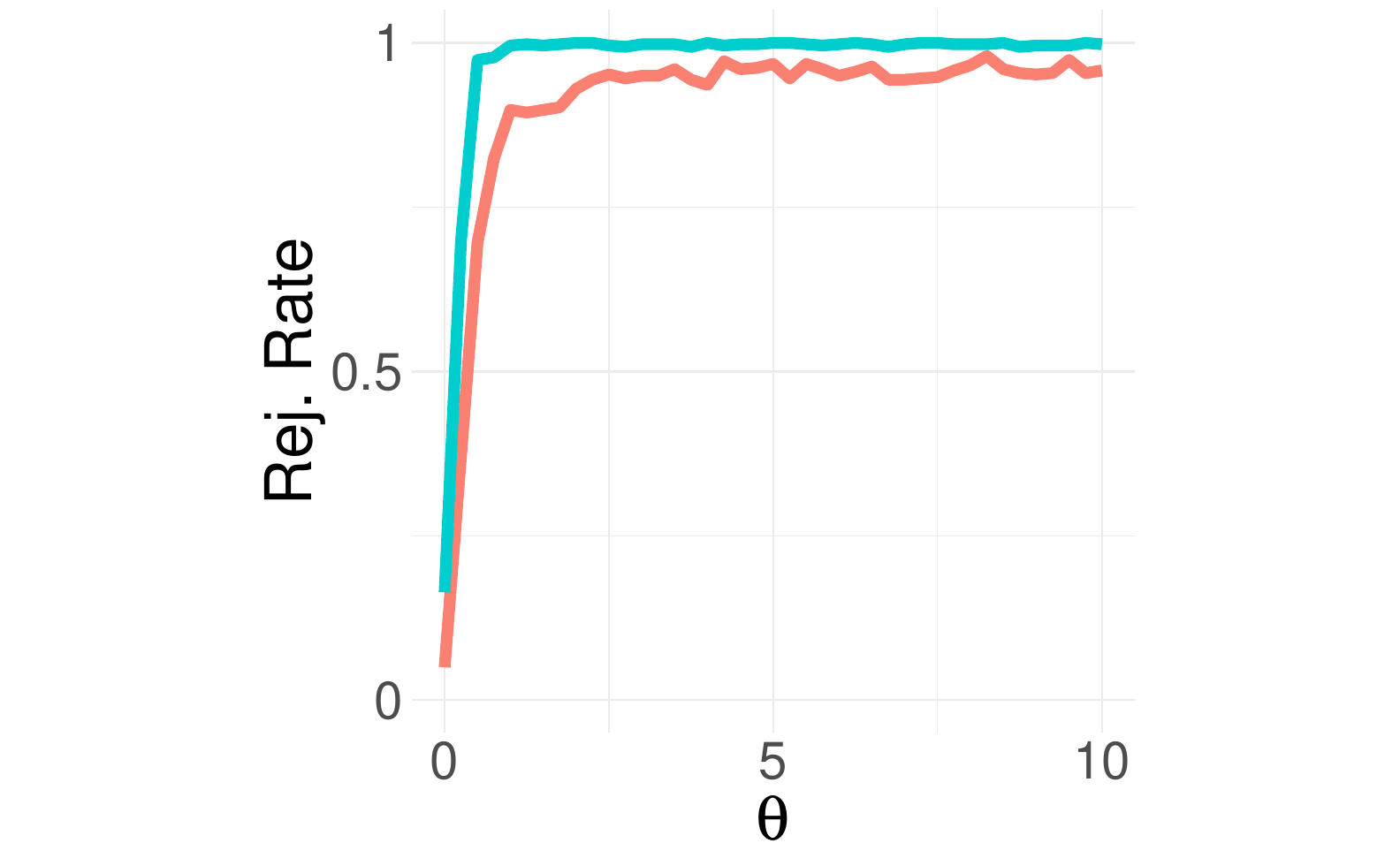}
    \end{minipage}

    \vspace{0.5cm} 

    \begin{minipage}[b]{0.05\textwidth}
        \raggedright
        \textbf{S3}
    \end{minipage}
    \begin{minipage}[b]{0.3\textwidth}
        \centering
        \includegraphics[width=\textwidth]{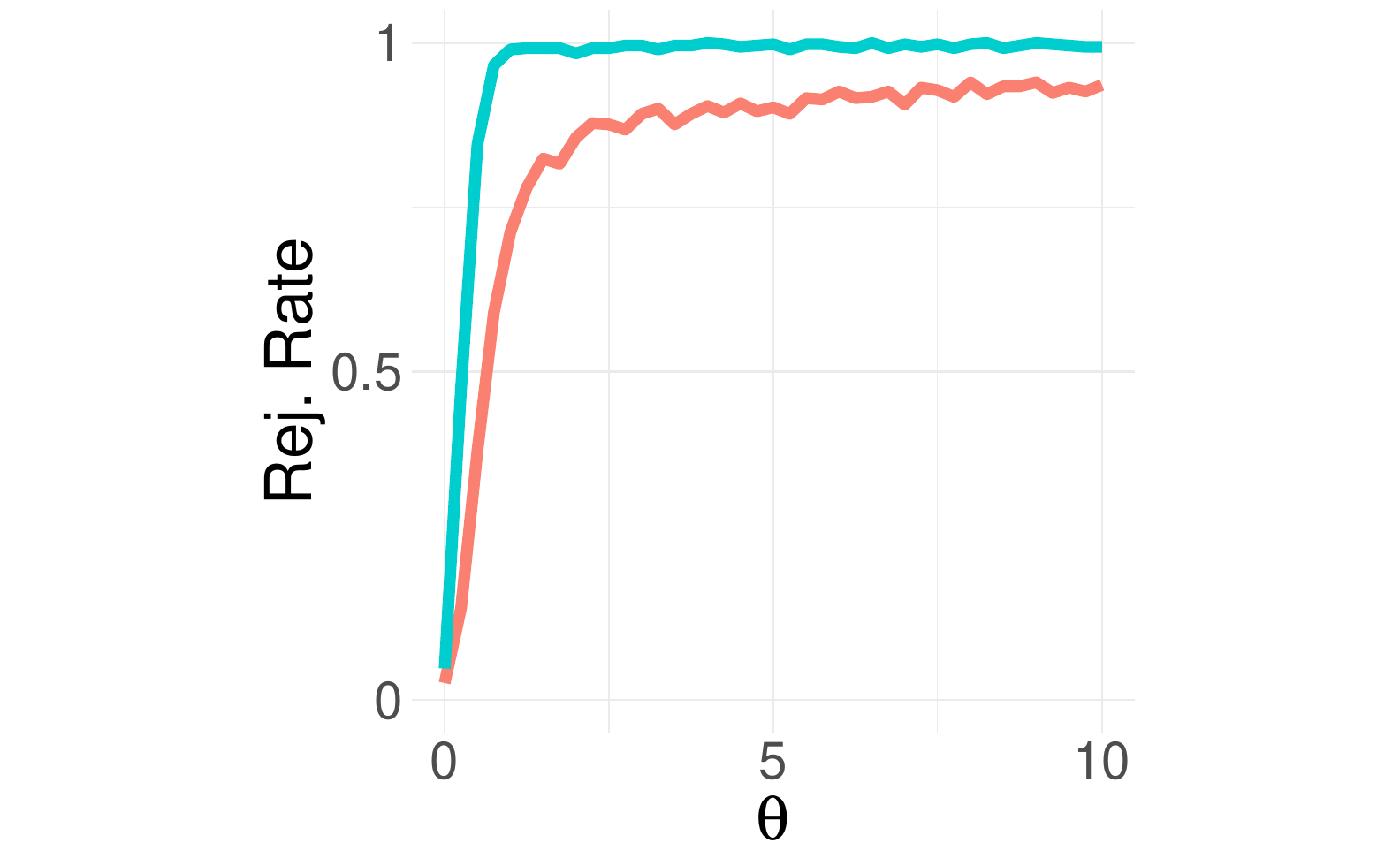}
    \end{minipage}
    \begin{minipage}[b]{0.3\textwidth}
        \centering
        \includegraphics[width=\textwidth]{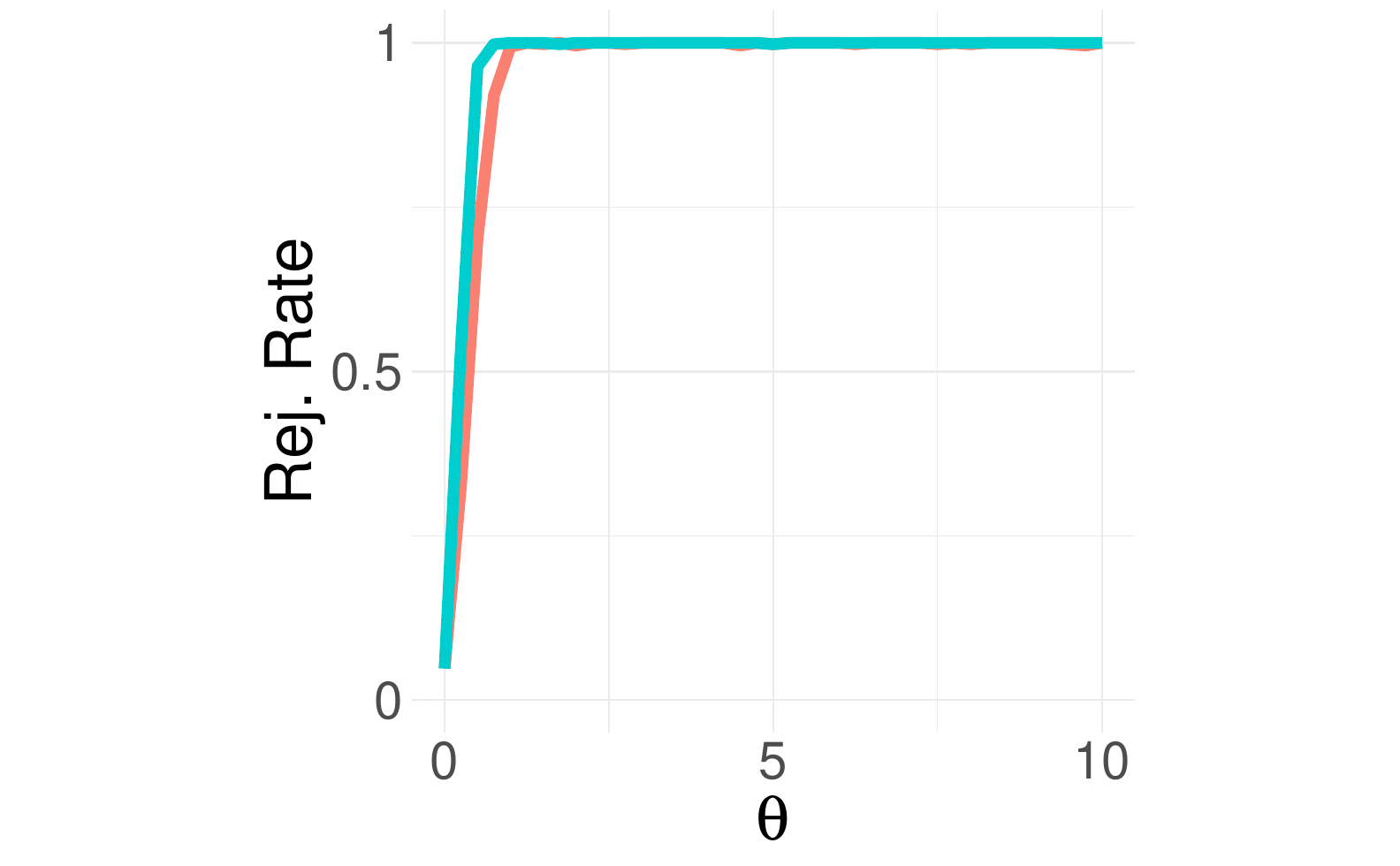}
    \end{minipage}
    \begin{minipage}[b]{0.3\textwidth}
        \centering
        \includegraphics[width=\textwidth]{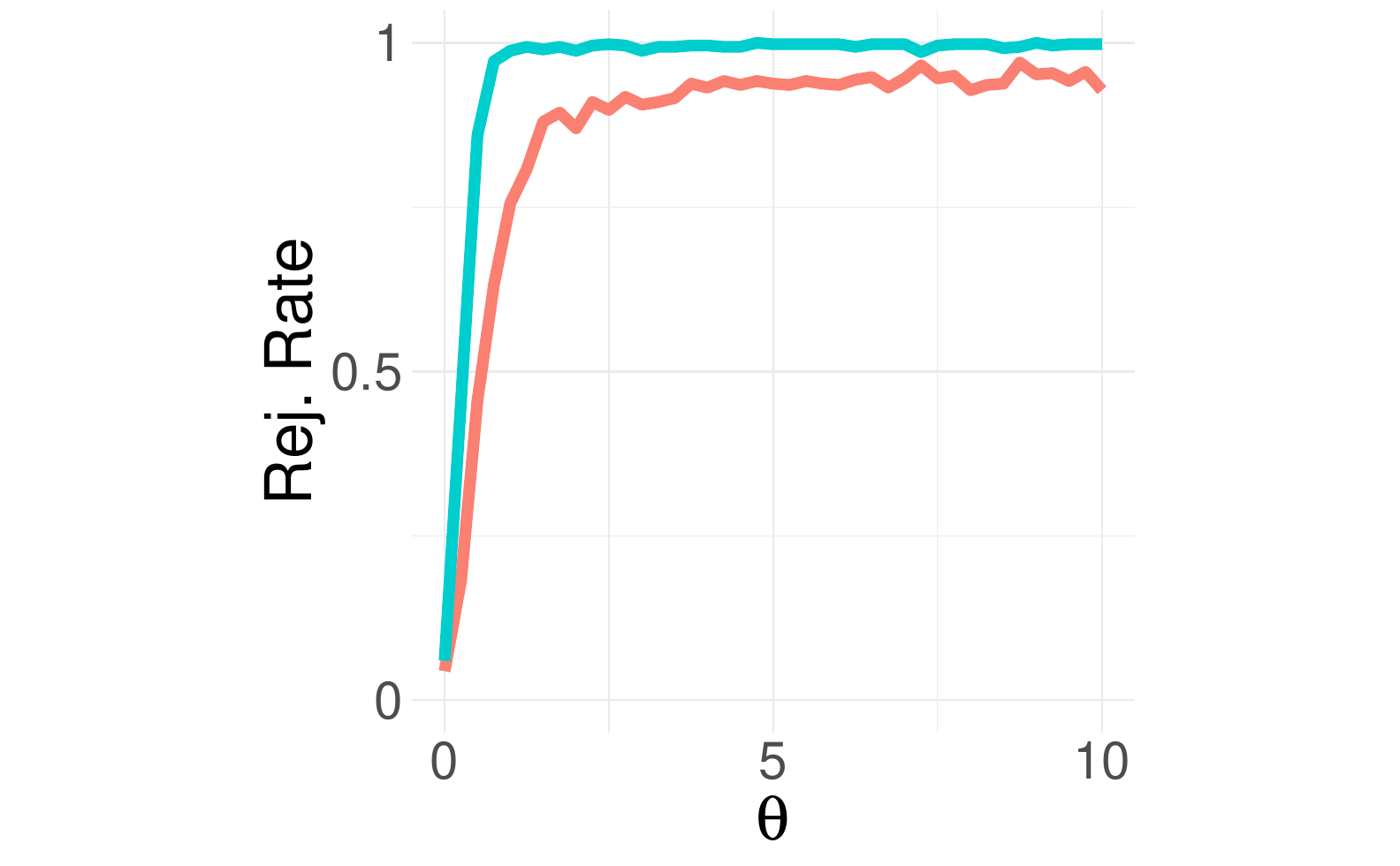}
    \end{minipage}

    \begin{minipage}{0.5\textwidth}
        \centering
        \includegraphics[width=\textwidth,trim={3cm 3cm 3cm 3cm}]{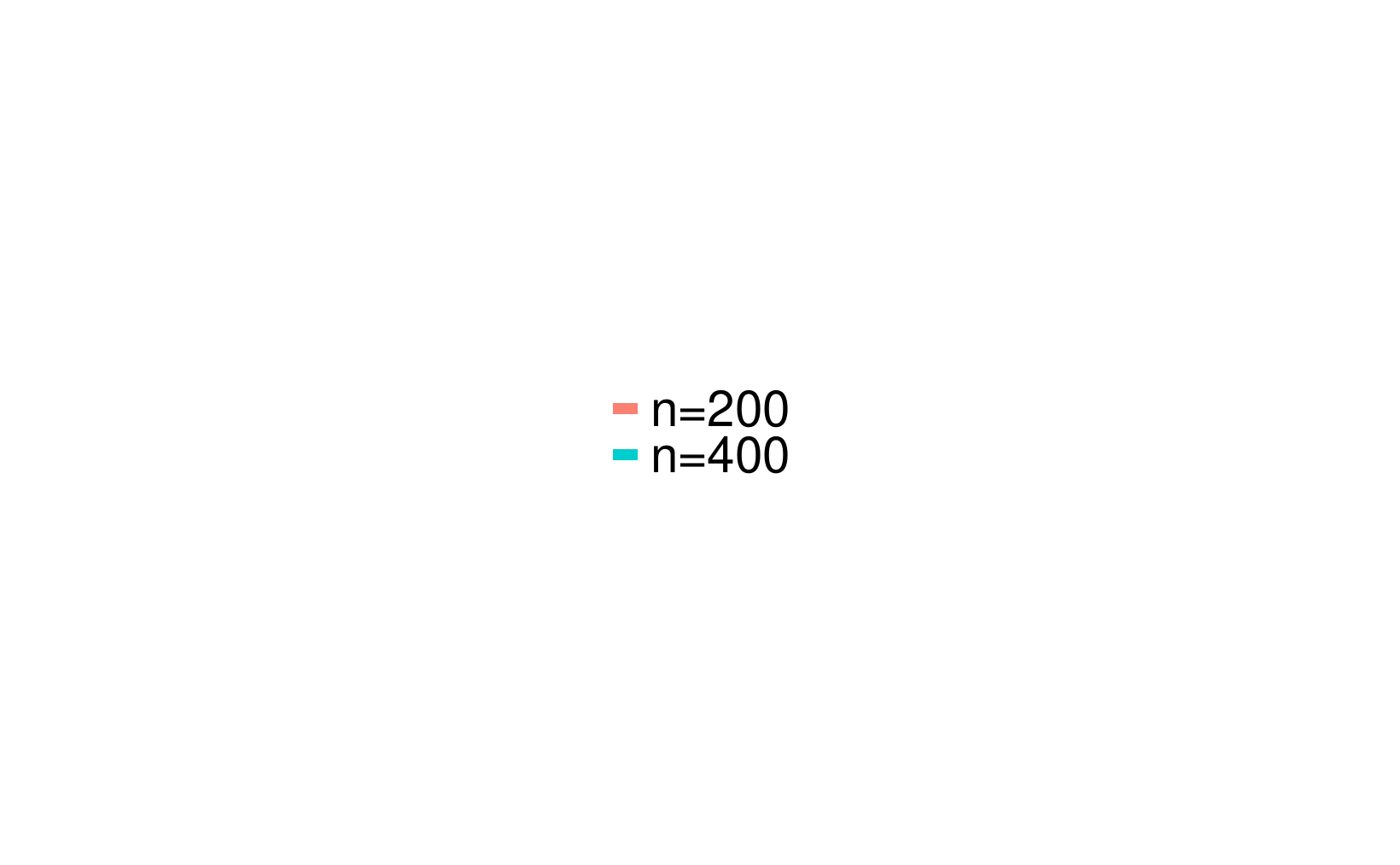}
    \end{minipage}

\caption{Rejection rate of null hypothesis versus copula parameter across different simulation settings $S_1, S_2, S_3$. Note that in $S_1$ and $S_3$ there is strong quantile concordance across all quantile levels while in $S_2$ the concordance increases after the lower quantile level as the copula strengthens the association towards the upper tail. Note that this is not a strict increase in rejection rate as the median ``sees’’ the upper tail concordance in its calculation and has more power in seeing the copula concordance from the 0.5 quantile and above.}
\label{fig:rejrate}

\end{figure}

We evaluate the empirical rejection rate of the null hypothesis, that $\rho_{\tau}(Y,X|Z_1,Z_2)=\tau^2$ if $\tau \leq 0.5$ or $\rho_{\tau}(Y,X|Z_1,Z_2)= (1 -\tau)^2$ otherwise, conditioned on $Z_1, Z_2$. The null hypothesis is tested using the linear QuACC statistic across $500$ iterations with sample sizes of $n=200$ and $n=400$ for a grid of copula parameters from $0$ to $10$ with a spacing of $0.25$. Results are presented in Figure \ref{fig:rejrate}.

\begin{figure}[htbp]
\centering
\begin{tabular}{c}
    \includegraphics[width=0.45\textwidth]{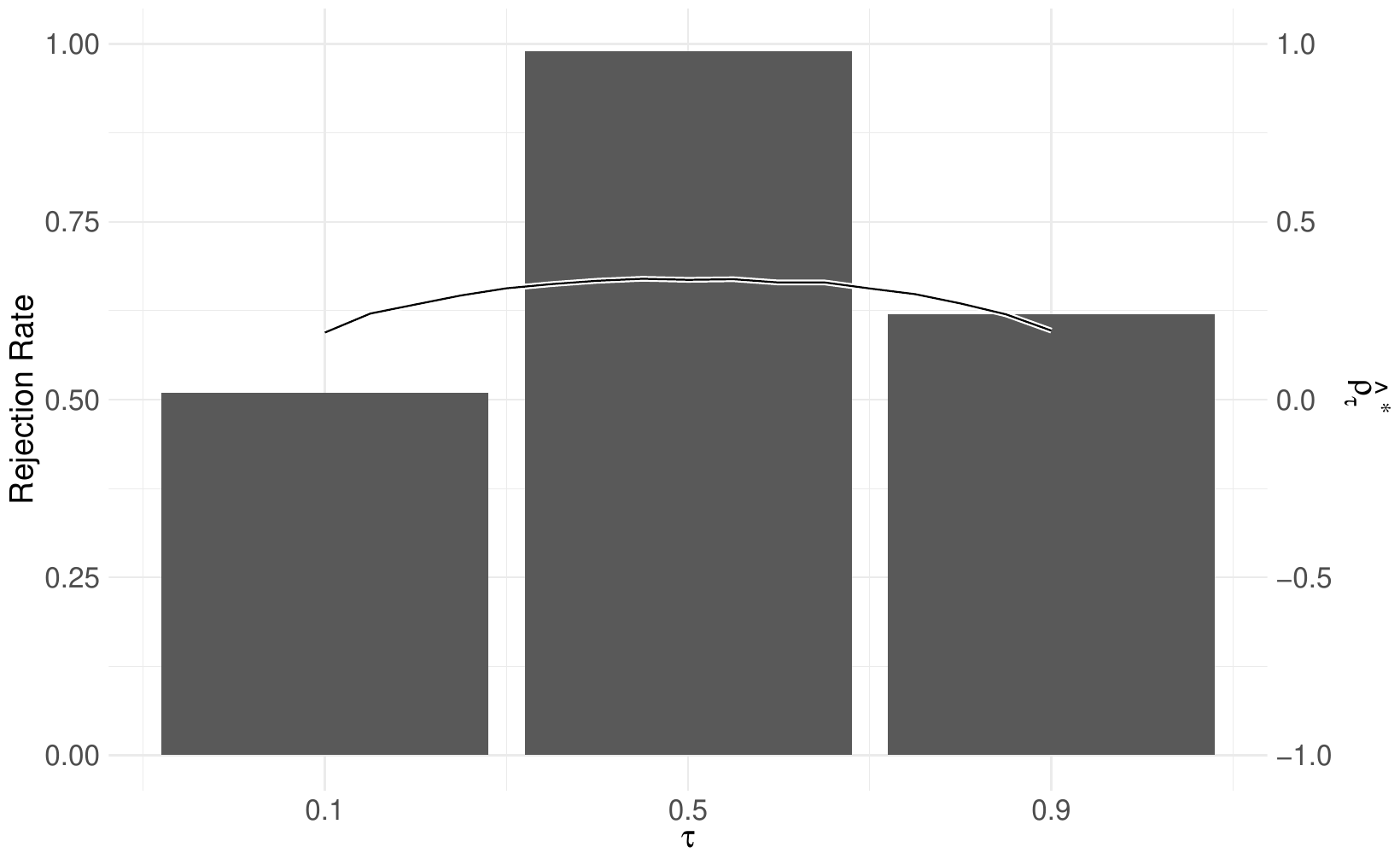} \\
    (a) $S_1$ \\[6pt]
    
    \includegraphics[width=0.45\textwidth]{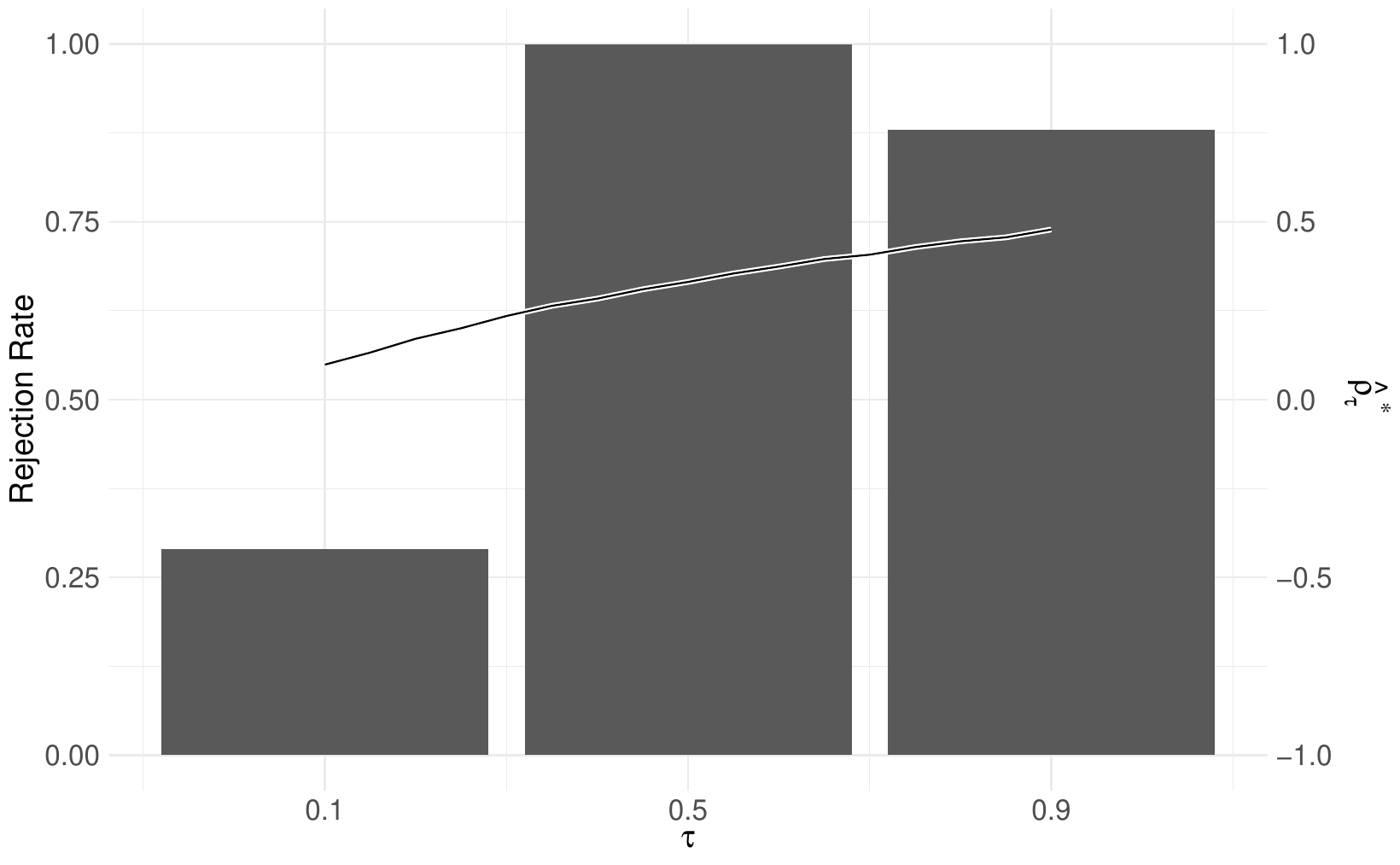} \\ 
    (b) $S_2$ \\[6pt]

     \includegraphics[width=0.45\textwidth]{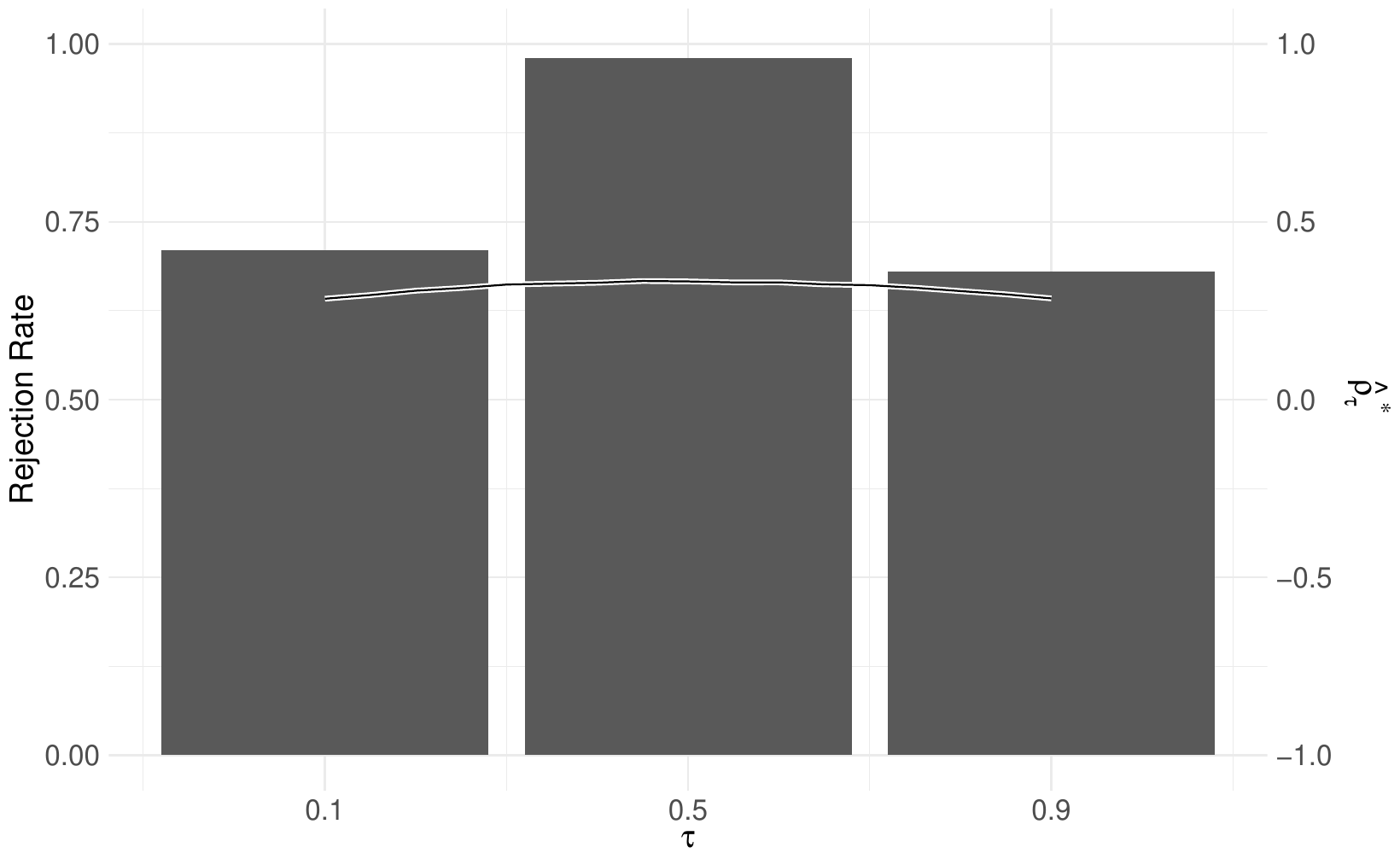} \\
    (c) $S_3$ \\[6pt]
    
    \\
\end{tabular}
\caption{Rejection rates across $\tau$ levels for simulation settings $S_1, S_2, S_3$. Empirical $\widehat{\rho}^*$ visualized alongside rejection rate showcasing the dependence across $\tau$ and how it aligns with rejection rate. For $S_1$ and $S_3$, there is strong concordance at each quantile level and we find the null hypothesis is rejected across all quantile levels - although note that the median has the most power for detecting association and thereby we see greater rejection there. In the case of $S_2$, concordance is strongest at the median, which sees the upper quantile dependence, followed by the upper quantile and finally the lower quantile. Thus we expect to see null hypothesis rejection rate generally increase with quantile level, with a peak at the median, which is illustrated in the figure. }
\label{fig:quantdependence}
\end{figure}

The rejection rate of the null hypothesis, $H_0$, for the copula parameters defined above for $S_1, S_2, S_3$ is also evaluated across $\tau$ from $0.1$ to $0.9$, with a spacing of $0.05$, for the linear QuACC statistic across 100 iterations for a sample size of $n=200$. Alongside the rejection rates we show the normalized $\hat{\rho}$ (scaled to be in [-1,1]), calculated across 100 iterations and $n=5000$ samples to showcase the underlying statistic. These results are seen in Figure \ref{fig:quantdependence}.

\subsection{QuACC Graphical Model Simulations}
\begin{figure}[hh]
\begin{center}
    
\includegraphics[width=.8\textwidth]{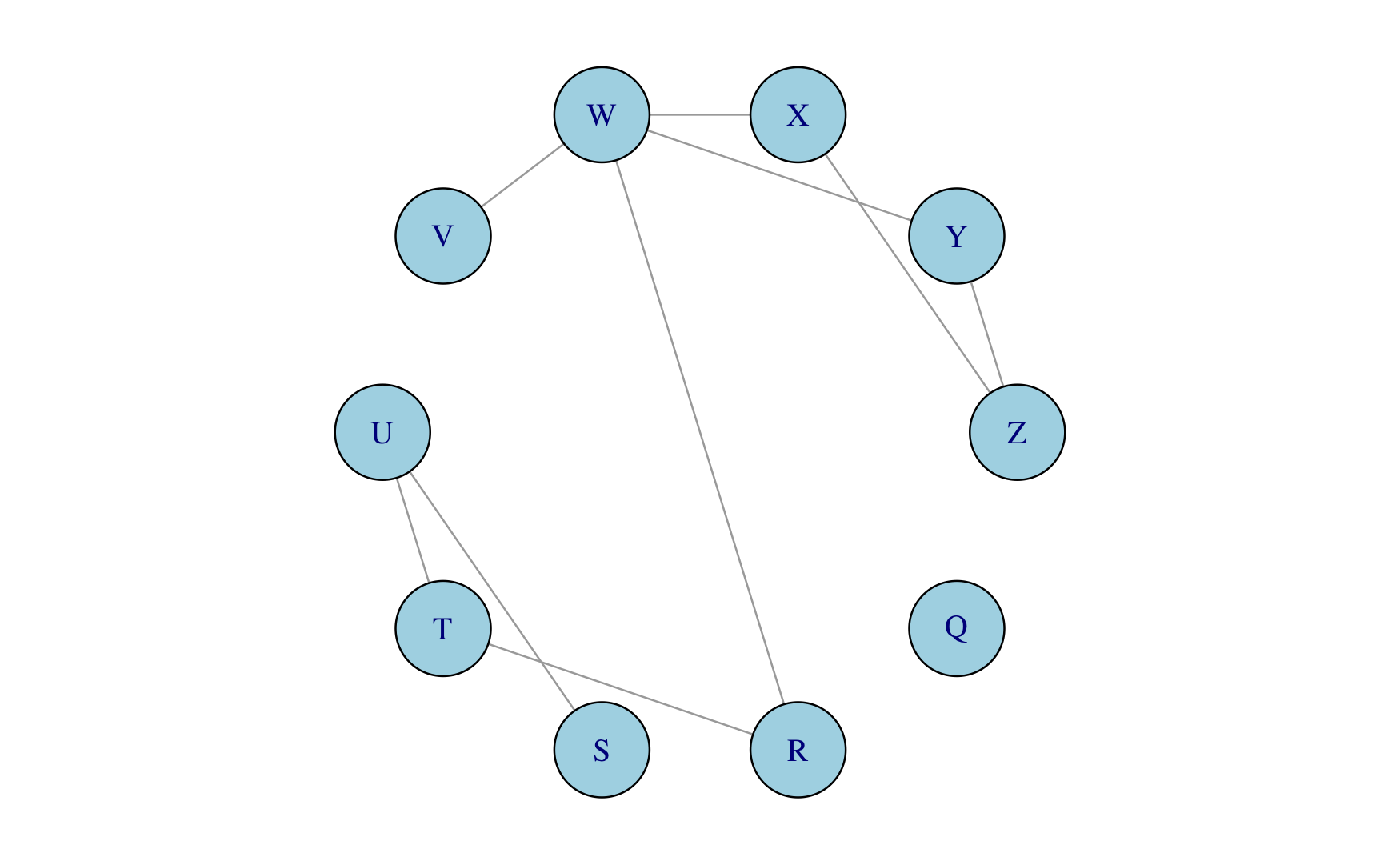}
\caption{True graph underlying the data generating process in simulation studies.}
\label{f:simgraph}

\end{center}
\end{figure}

To test the ability of our statistic to recover graphical relationships we apply a combination of the PC algorithm with QuACC to a known multivariable data-generating process that exhibits quantile dependence. The true (synthetic) graph, $G$, contains 5 vertices: $Z,X,Y,W$, and $R$, with edges as displayed in Figure \ref{f:simgraph}. 

The data is generated according to the following system of equations, which includes linear relationships alongside indicator terms that induce tail dependence. The coefficients that determine the strength of tail relationships are sampled $\beta_i, \gamma_i \sim Uniform(0.3, 0.8)$. In the main text we present results in the setting where the linear coefficients $\alpha_i=0$, and in the supplementary materials we present additional results where linear coefficients are sampled $\alpha_i \sim Uniform(-0.4,0.4)$.

\begin{align*}
\begin{split}
Z, U, Q \sim & \: TruncN(0, 1, min=-2, max=2) \\
Y \sim & \: \alpha_1 Z + \beta_1 Z I\{Z \geq Q_Z(0.9)\} + \gamma_1 Z I\{Z \leq Q_Z(0.1)\} + N(0,1) \\
X \sim & \: \alpha_2 Z + \beta_2 Z I\{Z \geq Q_Z(0.9)\} + \gamma_2 Z I\{Z \leq Q_Z(0.1)\} + N(0,1) \\
W \sim & \: \alpha_3 X + \beta_3 X I\{X \geq Q_X(0.9)\} + \gamma_3 X I\{X \leq Q_X(0.1)\} + \\ 
       & \alpha_4 Y +  \beta_4 Y I\{Y \geq Q_Y(0.9)\} + \gamma_4 Y I\{Y \leq Q_Y(0.1)\} + N(0,1) \\ 
V \sim & \: \alpha_5 W + \beta_5 W I\{W \geq Q_W(0.9)\} + \gamma_5 W I\{W \leq Q_W(0.1)\} + N(0,1) \\
T \sim & \: \alpha_6 U + \beta_6 U I\{U \geq Q_U(0.9)\} + \gamma_6 W I\{U \leq Q_U(0.1)\} + N(0,1) \\
S \sim & \: \alpha_7 U + \beta_7 U I\{U \geq Q_U(0.9)\} + \gamma_7 W I\{U \leq Q_U(0.1)\} + N(0,1) \\
R \sim & \: \alpha_8 T + \beta_8 T I\{T \geq Q_T(0.9)\} + \gamma_8 T I\{X \leq Q_T(0.1)\} + \\ 
       & \: \alpha_9 W +  \beta_9 W I\{W \geq Q_W(0.9)\} + \gamma_9 W I\{W \leq Q_W(0.1)\} + N(0,1) \\ 
\end{split}
\end{align*}

$Q_L(\tau)$ denotes the $\tau$th marginal quantile of vertex $L$. We generate $n \in \{500, 1000, 5000\}$ samples from the above process across 100 replicates. For each of these replicates, we use QuACC in conjunction with the PC algorithm to estimate the underlying edges. The mean and standard deviation of precision, recall, and structural hamming distance of the actual edges being present across these replicates for each $\tau$ is displayed in Table \ref{t:discoverymetricsnewquacc}. Simulation results obtained via partial correlation conditional independence tests in place of QuACC are displayed in Table \ref{t:discoverymetricsnewgaussian}. The results show partial correlation tests are not robust to spurious correlation at higher sample sizes and cannot detect tail dependence at extreme quantiles nearly as well as QuACC. In contrast, our method only improves as sample size increases.

\begin{table}[h!]
\centering
\begin{tabular}{|| c | c c c c||} 
 \hline
 $n$ & $\tau$ & Precision & Recall & Structural Hamming Distance \\ 
 \hline\hline
 500 & 0.1 & 0.897 (0.139) & 0.524 (0.149) & 0.108 (0.035) \\ 
 \hline
 & 0.5 & 0.747 (0.209) & 0.368 (0.145) & 0.152 (0.039)  \\
 \hline
  & 0.9 & 0.881 (0.139) & 0.496 (0.146) & 0.115 (0.034) \\ 
 \hline

 1000 & 0.1 & 0.923 (0.089) & 0.710 (0.130) & 0.070 (0.032) \\ 
 \hline
 & 0.5 & 0.775 (0.182) & 0.472 (0.167) & 0.133 (0.044) \\
 \hline
  & 0.9 & 0.942 (0.086) & 0.704 (0.132) & 0.069 (0.030) \\ 
 \hline
 
 5000 & 0.1 & 0.892 (0.081) & 0.990 (0.039) & 0.028 (0.022) \\ 
 \hline
 & 0.5 & 0.721 (0.161) & 0.581 (0.158) & 0.129 (0.049) \\
 \hline
  & 0.9 & 0.907 (0.083) & 0.987 (0.036) & 0.024 (0.023) \\ 
 \hline
\end{tabular}
\caption{Mean and standard deviation of precision, recall, and Hamming distance for predicted edges obtained via QuACC versus true edges across 100 replicates of a linear quantile dependent dataset with underlying graph depicted in Figure \ref{f:simgraph}. }
\label{t:discoverymetricsnewquacc}
\end{table}

\begin{table}[h!]
\centering
\begin{tabular}{|| c | c c c||} 
 \hline
 $n$ & Precision & Recall & Structural Hamming Distance \\ 
 \hline\hline
 500  & 0.792 (0.140) & 0.593 (0.154) & 0.114 (0.043) \\ 
 \hline

 1000 & 0.765 (0.141) & 0.67 (0.141) & 0.110 (0.044) \\ 
 \hline
 
 5000 & 0.689 (0.098) & 0.797 (0.132) & 0.114 (0.041) \\ 
 \hline
\end{tabular}
\caption{Mean and standard deviation of precision, recall, and Hamming distance for predicted edges obtained via partial correlation conditional independence testing versus true edges across 100 replicates of a linear quantile dependent dataset with underlying graph depicted in Figure \ref{f:simgraph}.}
\label{t:discoverymetricsnewgaussian}
\end{table}

\section{Application}
\label{s:app}

The All of Us research program features a diverse set of participants spanning a myriad of backgrounds \citep{all2019all}. Among the conditions present in the All of Us dataset related to mitochondrial disorders. These disorders are prevalent in a minimum of 1 in 5000 individuals with often no clear treatment path and debilitating symptoms including nervous system and muscular disorders, such as paralysis of key bodily functions \citep{ng2016mitochondrial}. Inappropriate energy transformation by the mitochondria has many expressions; however, we focus on individuals in All of Us with mitochondrial cytopathy, mitochondrial metabolism defect, mitochondrial myopathy, depletion of mitochondrial DNA, Kearns-Sayre syndrome, Ceber's optic atrophy, progressive external ophthalmoplegia, and inborn errors of metabolism, all of which are forms of mitochondrial disorders. Consistent with the mitochondrial research literature, we denote this population as the MitoD group.

The benefit of biobank data such as All of Us is that they enable large scale statistical analyses. Our motivation in applying QuACC graphical models to this afflicted subpopulation is to compare graphical structures estimated from this cohort and a match ``healthy’’ cohort (i.e. without mitochondrial disorders). This comparison can elucidate which biomarker relationships manifest in those with mitochondrial disorders compared to a control sample. This is of special interest at extreme tails of biomarker relationships, as it is well understood that many biomarkers have significantly different values in the exposed population compared to the control, and thus we hypothesize that behavior at these quantile extremes will differ between these groups \citep{sharma2021circulating}. We consider the following biomarkers in our comparison: albumin, calcium, chloride, cholesterol, blood pressure (systolic and diastolic), erythrocytes, leukocytes, hematocrit, hemoglobin, and glucose. It is known that biomarkers from complete blood counts (erythrocytes, hematocrit, and hemoglobin) are significantly different in those with mitochondrial disorders \citep{sharma2021circulating}. These biomarkers were also selected to maximize sample size (since not all biomarkers are measured on the same set of participants in All of Us) while elucidating potentially meaningful relationships.

While biobank datasets have compelling positives, we note that our analysis has limitations with respect to data collection. The lack of standardization of biomarker collection can confound biomarker levels (e.g. the time of day the biomarkers were collected and a subjects eating habits before collection are not typically standardized in biobank data). One additional limitation is that there is clinical bias in when blood tests were ordered for which biomarkers. As a result, intervals between measurements could differ systematically for some groups of patients, for specific biomarker pairs. This could lead to systematic bias across groups (i.e., control vs. exposed) in the detection of associations that might not be constant across temporal scales. However, this limitation is minor given that our objective here is not to draw strong biological conclusions but to establish the potential utility of the QuACC method.

Our preprocessing of the data included a quantile-quantile transform on each column independent of the rest. To do this, the rank of each observation in every column was found and normalized by the total number of observations (with ties being averaged in rank) giving us an estimated quantile for each observation. This estimated quantile is projected onto a standard normal variable using the standard normal quantile function. We did not remove any outliers as our method is aiming to detect boundary behavior.

The MitoD population will represent the ``exposed’’ population ($n = 1968$). A corresponding control group was constructed from first encounter visits across All of Us that had data from at least one of our biomarkers listed above and that did not have any participant overlap with the exposed group ($n = 138043$). Because our hypothesis test is capable of testwise deletion, we do not require that an individual in the control/exposed group has data from all biomarkers simultaneously. 

\subsection{Pairwise Effects}

\begin{figure}[hh!]

\centering
\begin{tabular}{cc}
    \includegraphics[width=0.4\textwidth]{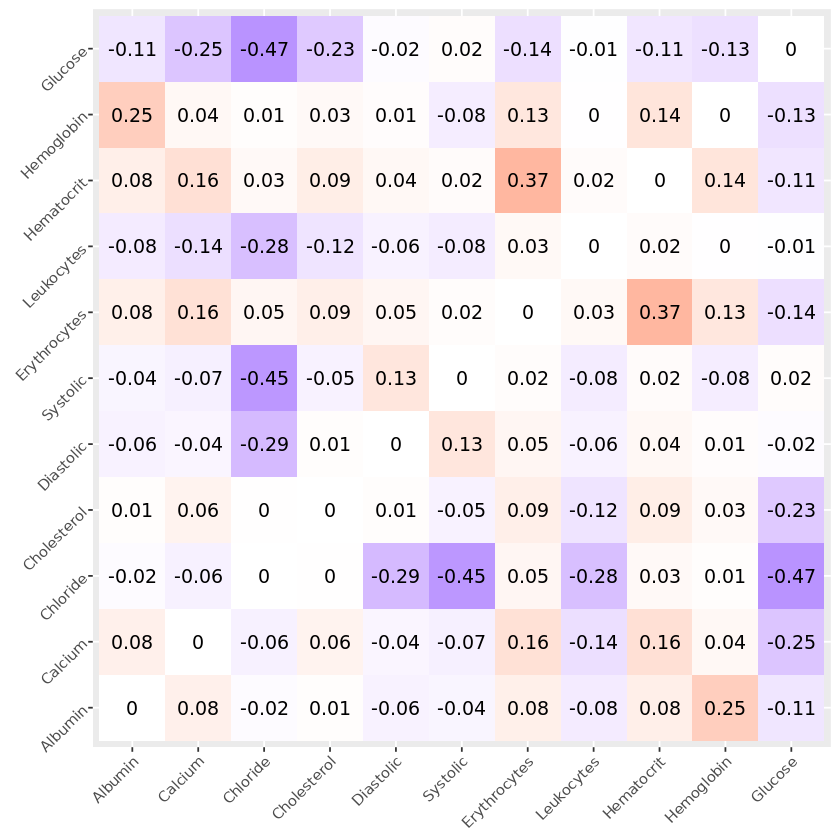} & \includegraphics[width=0.4\textwidth]{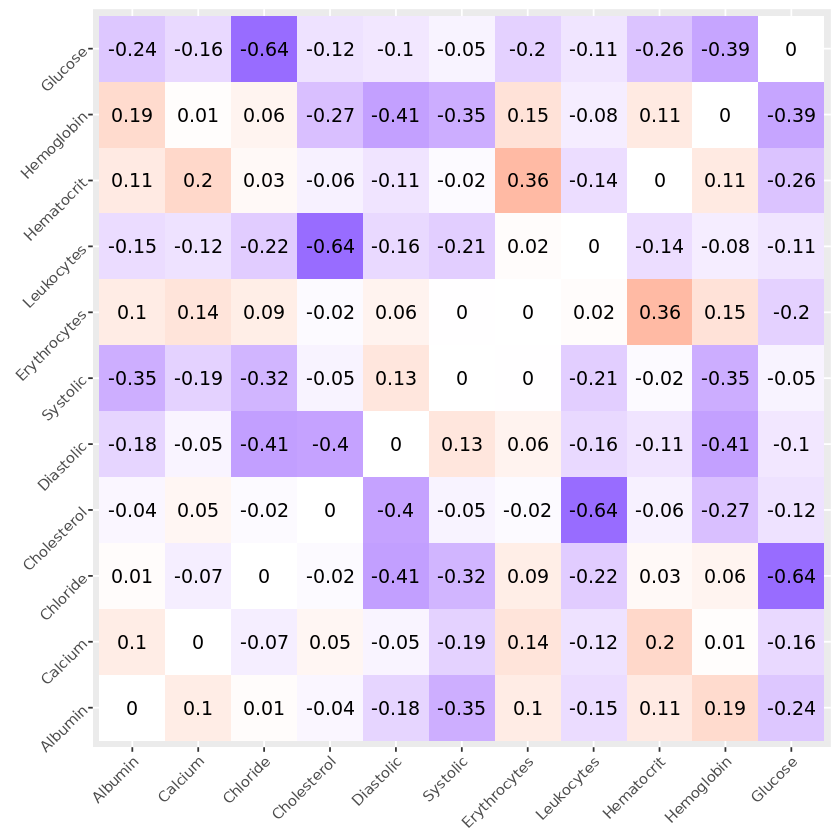}\\
    (a) $\tau = 0.1$, Control & (b) $\tau = 0.1$, Exposed \\[6pt]
    
    \includegraphics[width=0.4\textwidth]{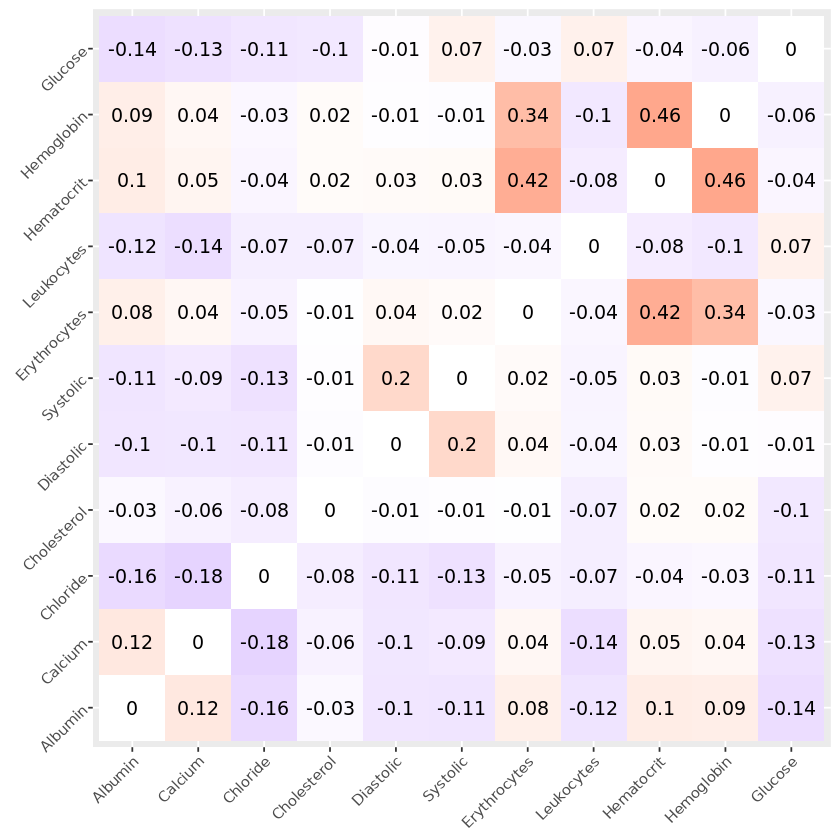} & \includegraphics[width=0.4\textwidth]{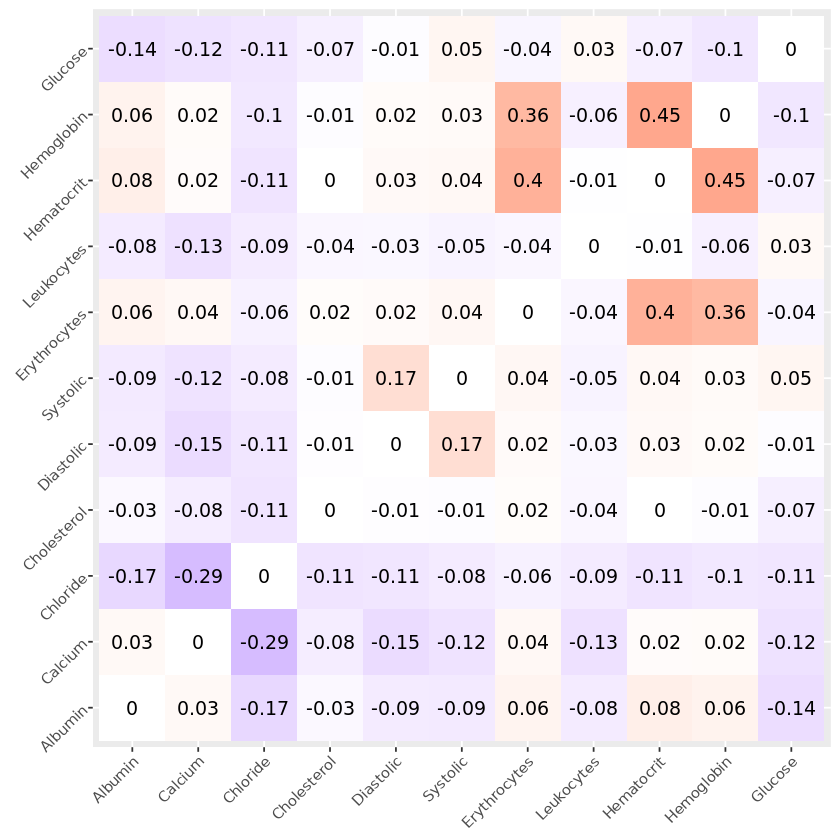}\\
    (c) $\tau = 0.5$, Control & (d) $\tau = 0.5$, Exposed  \\[6pt]
          
    \includegraphics[width=0.4\textwidth]{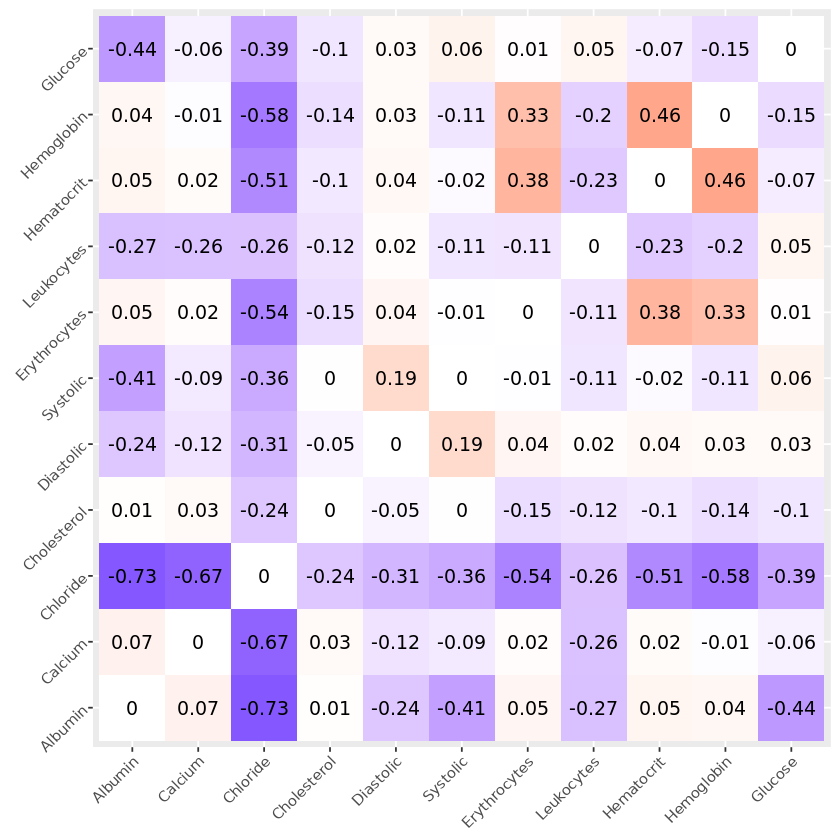} & \includegraphics[width=0.4\textwidth]{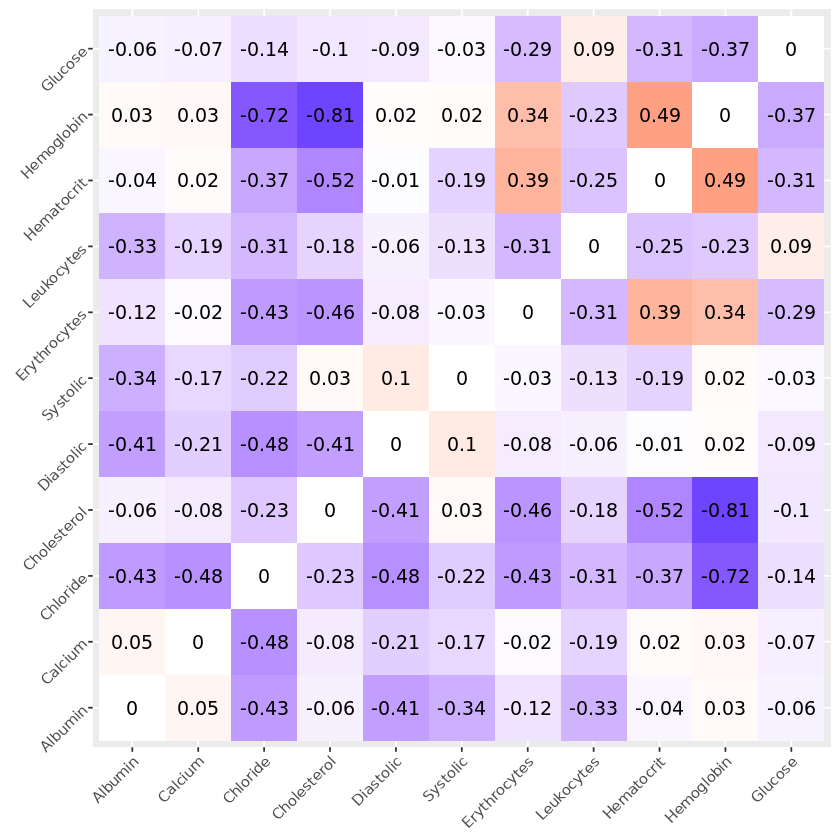} \\
    (e) $\tau = 0.9$, Control & (f) $\tau = 0.9$, Exposed \\[6pt]
    
    \\
\end{tabular}
    
\caption{The graphs on the left are the pairwise marginal ($Z = \{ \}$) QuACC statistics on the All of Us control population for $\tau \in \{ 0.1, 0.5, 0.9\}$. The right side features the pairwise marginal QuACC statistics for the MitoD population for the same $\tau$. Note that testwise deletion was performed in calculating the QuACC for each pair. }
\label{f:pairwise-marginal-control-and-exposed}

\end{figure}

Before we perform structure learning, we calculate the marginal ($Z = \emptyset)$ QuACC pairwise for all covariates. These pairwise associations simply show what proportion of the time the quantiles of two biomarkers are concordantly extreme. We compared the pairwise relationships from the control group ($n = 138043$) to the pairwise concordance proportions on the MitoD population in Figure \ref{f:pairwise-marginal-control-and-exposed}.

The differences in Figure \ref{f:pairwise-marginal-control-and-exposed} (a) and (b) as well as Figure \ref{f:pairwise-marginal-control-and-exposed} (e) and (f) support the hypothesis that individuals with mitochondrial deficiencies experience biomarker dysregulation at extremes relative to their control counterparts. Certain biomarker QuACC values reinforce our method, such as those of systolic and diastolic blood pressure. While the median graphs, Figure \ref{f:pairwise-marginal-control-and-exposed} (c) and (d), appear similar, there are differences at the quantile extremes for most other pairwise biomarker combinations.

We include pairwise QuACC plots where $Z = V \setminus \{ X, Y\}$, similar to MRFs, in the supplementary material.

\subsection{Structure Learning}

Due to technical limitations on the virtual environments needed to access All of Us, running QGM for a sample size above $50000$ was not possible, therefore our control graphs were aggregated over 15 replicates with $n = 1968$ independent draws from the total control group where edges were only kept in the final graph if a majority of the replicates included said edge. Keeping the same sample size as the exposed population was done to preserve the same power of detecting effects for each replicate.

\begin{figure}[hh]

\centering
\begin{tabular}{cc}
    \includegraphics[width=0.4\textwidth]{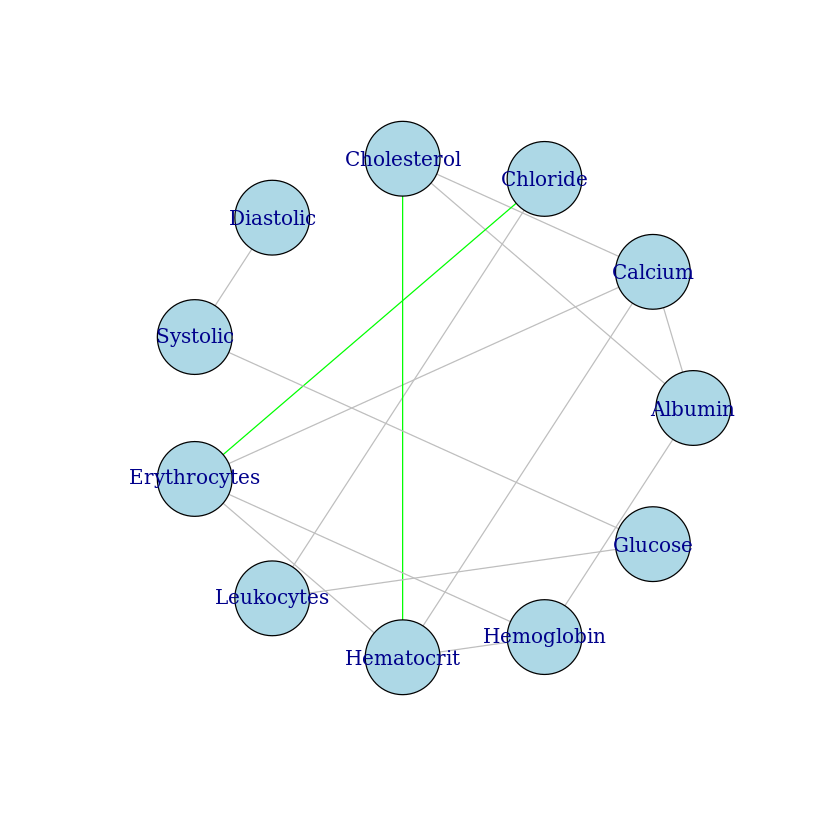} & \includegraphics[width=0.4\textwidth]{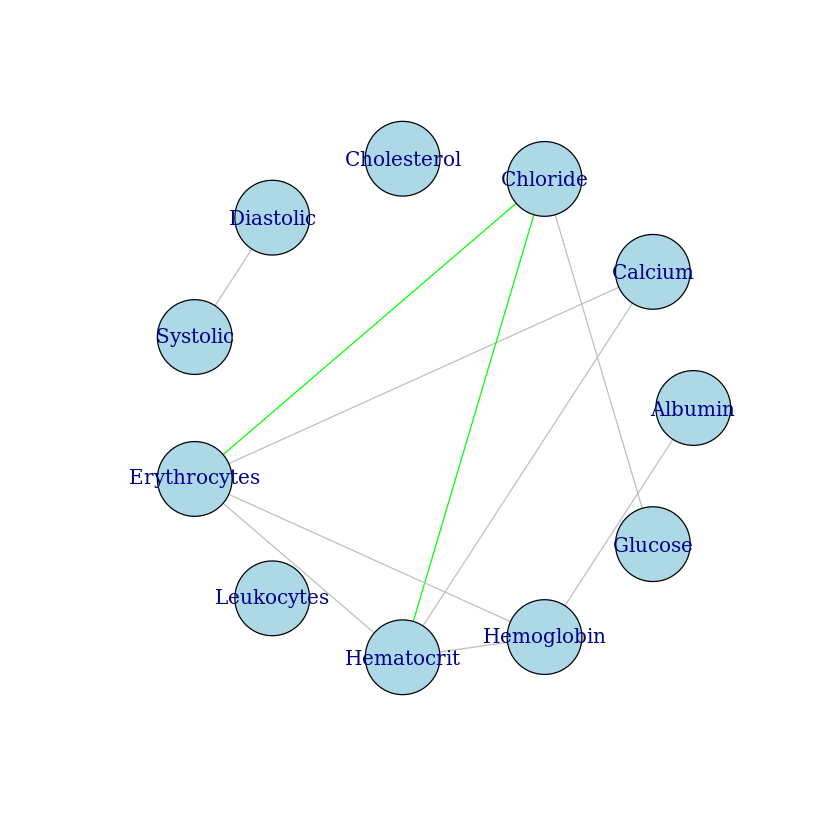}\\
    (a) $\tau = 0.1$, Control & (b) $\tau = 0.1$, Exposed \\[6pt]
    
    \includegraphics[width=0.4\textwidth]{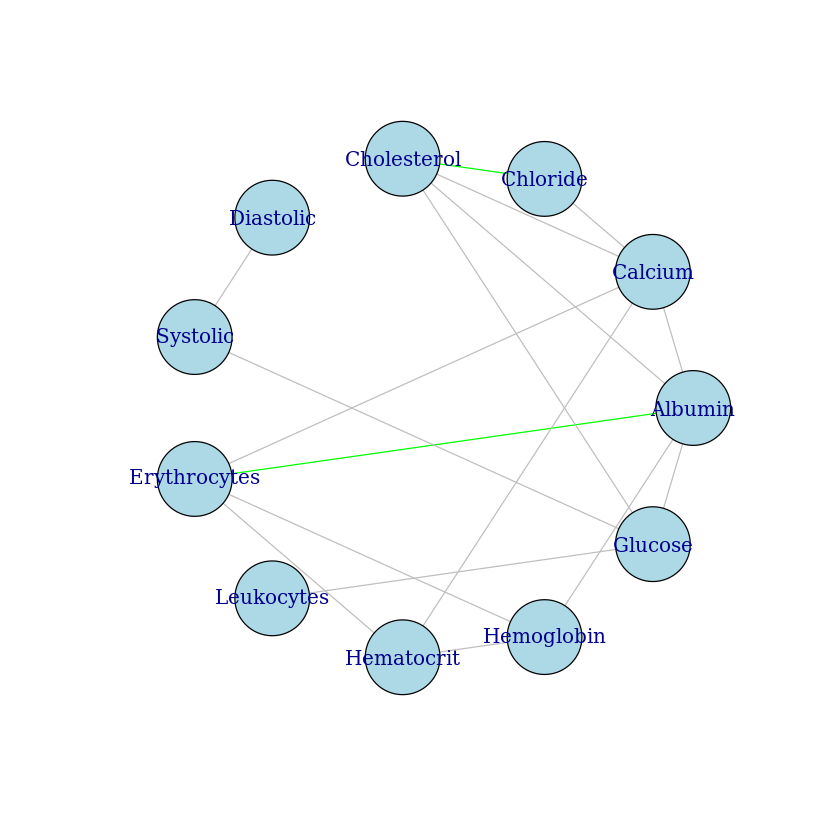} & \includegraphics[width=0.4\textwidth]{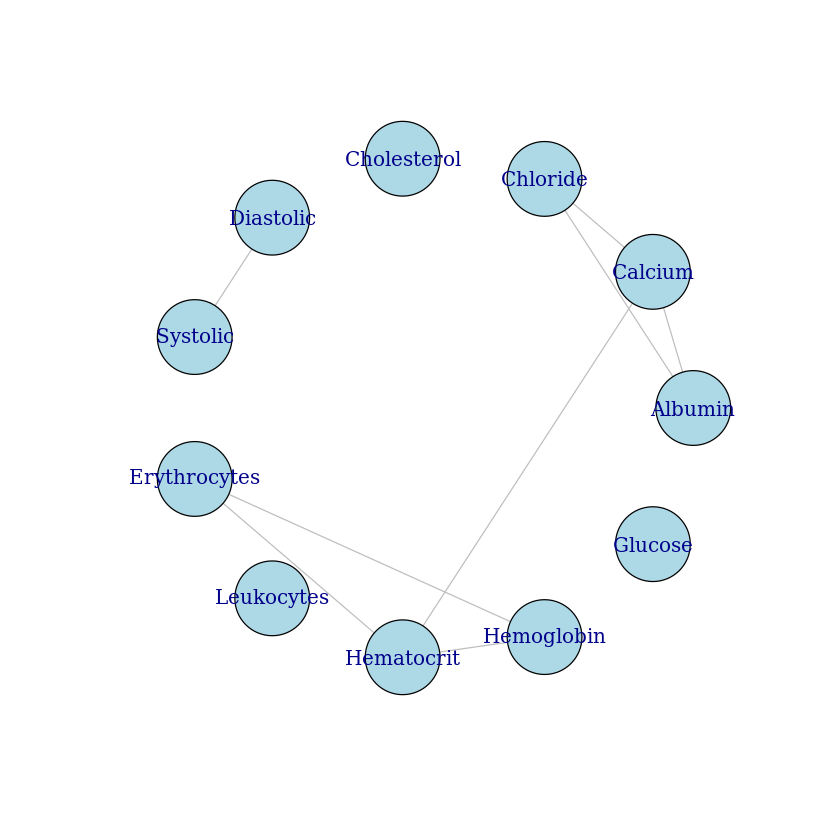}\\
    (c) $\tau = 0.5$, Control & (d) $\tau = 0.5$, Exposed  \\[6pt]
          
    \includegraphics[width=0.4\textwidth]{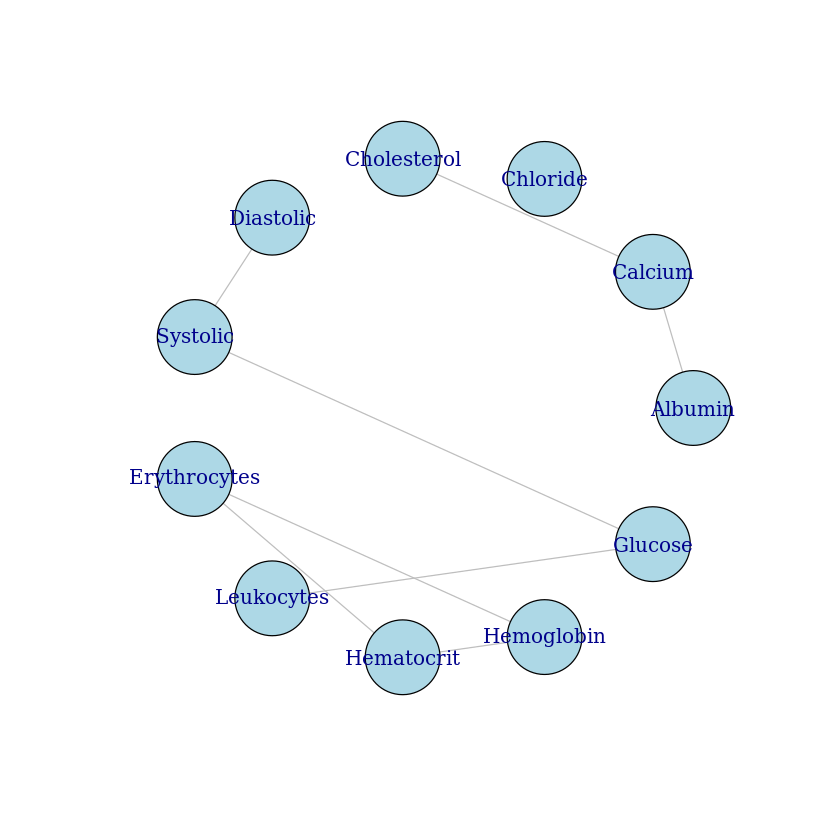} & \includegraphics[width=0.4\textwidth]{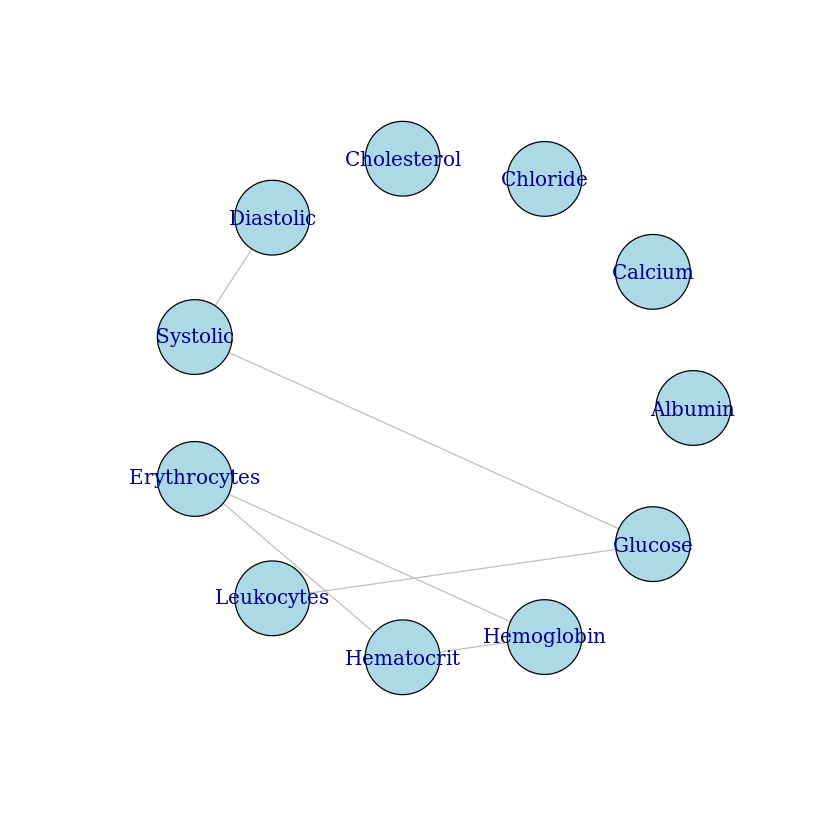} \\
    (e) $\tau = 0.9$, Control & (f) $\tau = 0.9$, Exposed \\[6pt]
    
    \\
\end{tabular}
    
\caption{The graphs on the left are the result of aggregating 15 replicates with majority voting for edges in the All of Us control population for $\tau \in \{ 0.1, 0.5, 0.9\}$. The right side features the QuACC graphical models in the All of Us MitoD population for the same $\tau$. Edges that are colored green do not appear in the graphs obtained using Pearson correlation versus QuACC for the same population. Edges that are gray appear in both QuACC and Pearson correlation graphs. }
\label{f:allofusqgmcontrol-and-exposed}

\end{figure}

The left side of Figure \ref{f:allofusqgmcontrol-and-exposed} displays the QuACC graphical models results across $\tau \in \{0.1, 0.5, 0.9\}$ for the constructed control group while the right side displays the corresponding QuACC graphical models for the group with mitochondrial disease. The edge between systolic and diastolic blood pressure is unsurprising and expected across all quantile levels across both populations, given that blood pressure during the systolic and diastolic phases of the cardiac cycle are largely determined by the same parameters (cardiac output, vascular elasticity and resistance). Likewise, we see a connection between erythrocytes, hematocrit, and hemoglobin, all being red blood cell features that are biologically interdependent (e.g., the more erythrocytes in circulation, the higher the hematocrit [$\%$ of blood volume accounted for by erythrocytes] and the more hemoglobin per volume of blood). Note that the control population is not necessarily ``healthy’’ and therefore the edge between glucose and (systolic) blood pressure may be indicative of the first encounters population having patterns of hyperglycemia and its effects on hypertension, or of a shared regulatory factor (e.g., sympathetic nervous system activity as a driver of both parameters). In the diseased population, we see a subgraph that includes albumin, calcium, and chloride at the median. Excess energy likely needs to be expended to regulate dysregulated biomarkers in mitochondrial disorders. For those with mitochondrial disease where energy production is impaired, this subgraph suggests that the joint regulation of these biomarkers cannot be performed adequately due to energy constraints. These findings display how mitochondrial energy transformation defects induce dysregulation of biomarker networks asymmetrically with respect to biomarker extremes. 

\begin{figure}[hh]

\centering
\begin{tabular}{cc}
    \includegraphics[width=0.5\textwidth]{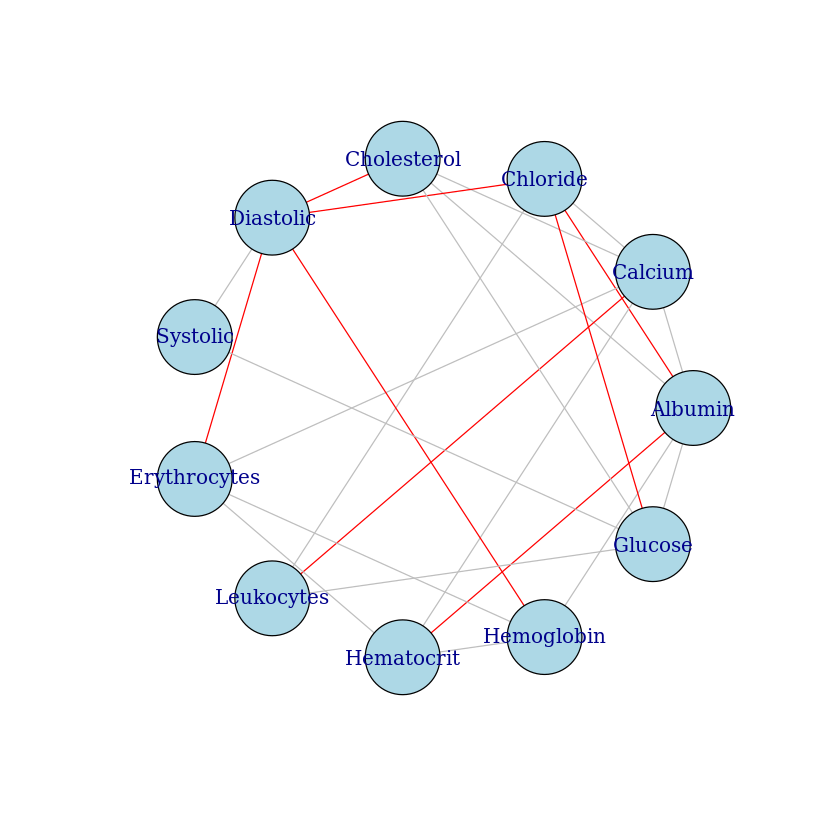} & \includegraphics[width=0.5\textwidth]{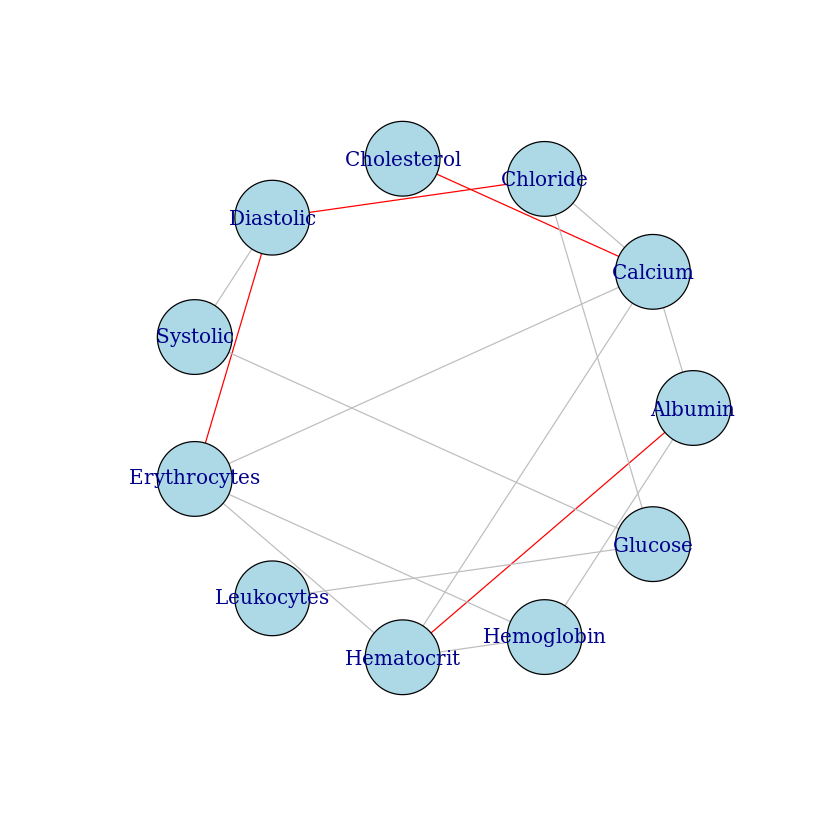}\\
    (a) Control & (b) Exposed \\[6pt]
    \\
\end{tabular}
    
\caption{The graph on the left is the result of aggregating 15 partial correlation PC graph replicates with majority voting for edges in the All of Us control population. The right graph features the same graphical model result in the All of Us MitoD population. Edges that are colored red do not appear in the QuACC graphs for the corresponding population. Edges that are gray appear in both QuACC and Pearson correlation graphs. }
\label{f:allofusgaussian}

\end{figure}

For comparison, Figure \ref{f:allofusgaussian} displays partial correlation graphical model (detecting linear associations) results for both the control and mitochondrial diseases group. While the partial correlation model is much more dense, there are differences in the graph such as the lower tail graph in the exposed population (e.g. Figure \ref{f:allofusqgmcontrol-and-exposed} (b)) : the green edges from chloride to erythrocytes and hematocrit are missing in the partial correlation model for the exposed group. Overall, this models' graphs are more dense as the threshold for significance scales with $\frac{1}{\sqrt{n}}$, although most of the edges in the QuACC graphs appear in the partial correlation graphs. While the partial correlation structure learning method is standard, there are limitations in comparing it to QuACC as the underlying parameters differ in purpose: QuACC is designed to detect tail behavior while partial correlations summarize linear associations. While QuACC may have weaker detection of overall dependence, the trade off is that QuACC can generate graphs at arbitrary $\tau$ values, namely at the tails, and has more power to detect effects there.

To illustrate uncertainty in structure learning, we additionally provide QuACC graphs when the threshold of significance, $\alpha$, is doubled and halved from the default $\alpha=0.05$, in the supplementary materials.

\section{Conclusion}
\label{s:conclusion}

The aim of our method is to provide a novel hypothesis test of quantile association and to integrate this with graph learning algorithms to learn graphical models at specific quantile levels. The asymptotic properties of such a test show that it is flexible and able to be applied to popular quantile regression model types, including linear quantile regression and quantile random forest models. Our robust synthetic results and application on real world medical data reveals that this method has potential for uncovering new quantile relationships in multivariable settings. 

There are many elements of the statistical theory here that can be expanded. Relaxing the assumption that quantile levels need to be the same for $Y$ and $X$, such as is done in the quantile association test by \citet{li2014quantile}, or even across conditioning sets, is one such example. Additionally, theory for broader classes of quantile regression, e.g. neural network quantile regressions \citep{cannon2011quantile}, requires further consideration. Application of this quantile association test outside of graphical learning is another topic that may be worth pursuing, as there are many other potential applications. Future work will be motivated by these considerations.

\section*{Acknowledgments}
We are grateful for the support provided by the National Institutes of Health, specifically through award R01AG087496 from the National Institute on Aging (NIA) and award K25ES034064 from the National Institute of Environmental Health Sciences (NIEHS).

We would like to acknowledge \textit{All of Us} and its participants for making this paper possible. Additionally, we would like to thank the National Institute of Health's \textit{All of Us} Research Program for making the cohort data used in our analyses accessible.

The Columbia Science of Health (SOH) Group provided valuable inspiration and guidance for developing this method. We thank SOH members Linda P. Fried, Nour Makarem, Daniel Belsky, Sen Pei, Molei Liu, Kiros T. Berhane, Julie Herbstman, and John R. Beard.

\bibliographystyle{plainnat} 
\bibliography{references}

\newpage
\appendix
\section*{Appendix A: Uncertainty in Structure Learning}
\label{aou-uncertainty}

\begin{figure}[hh]

\centering
\begin{tabular}{cc}
    \includegraphics[width=0.4\textwidth]{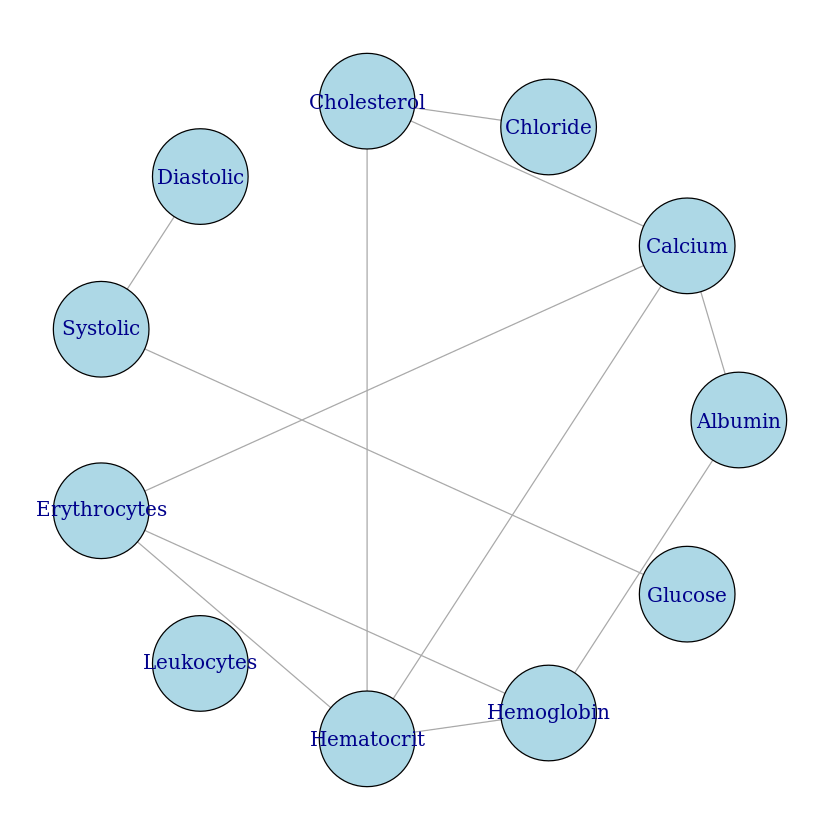} & \includegraphics[width=0.4\textwidth]{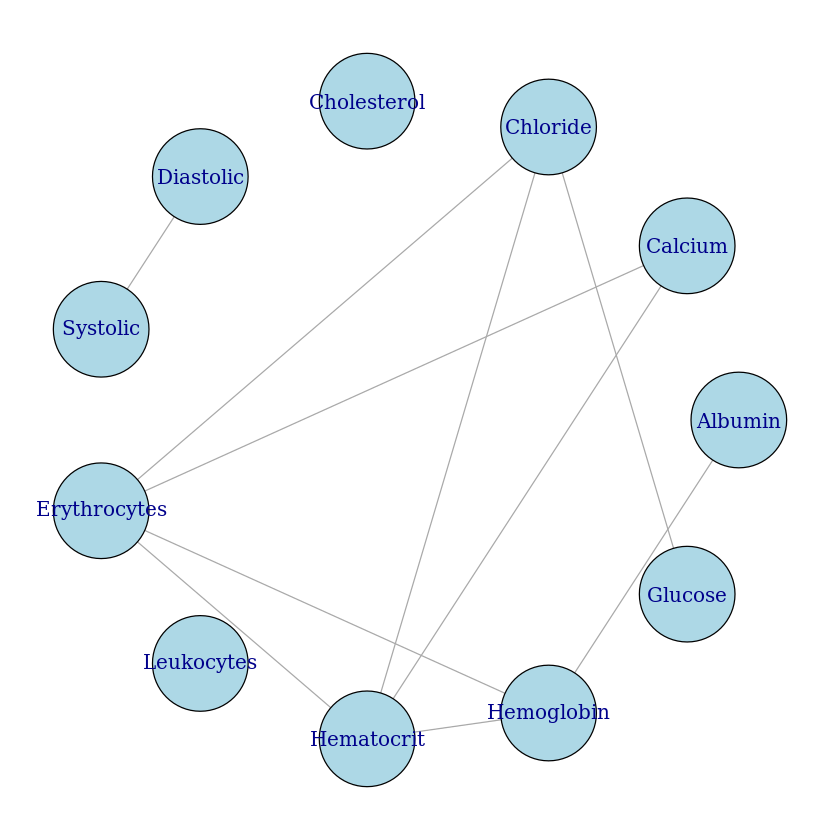}\\
    (a) $\tau = 0.1$, Control & (b) $\tau = 0.1$, Exposed \\[6pt]
    
    \includegraphics[width=0.4\textwidth]{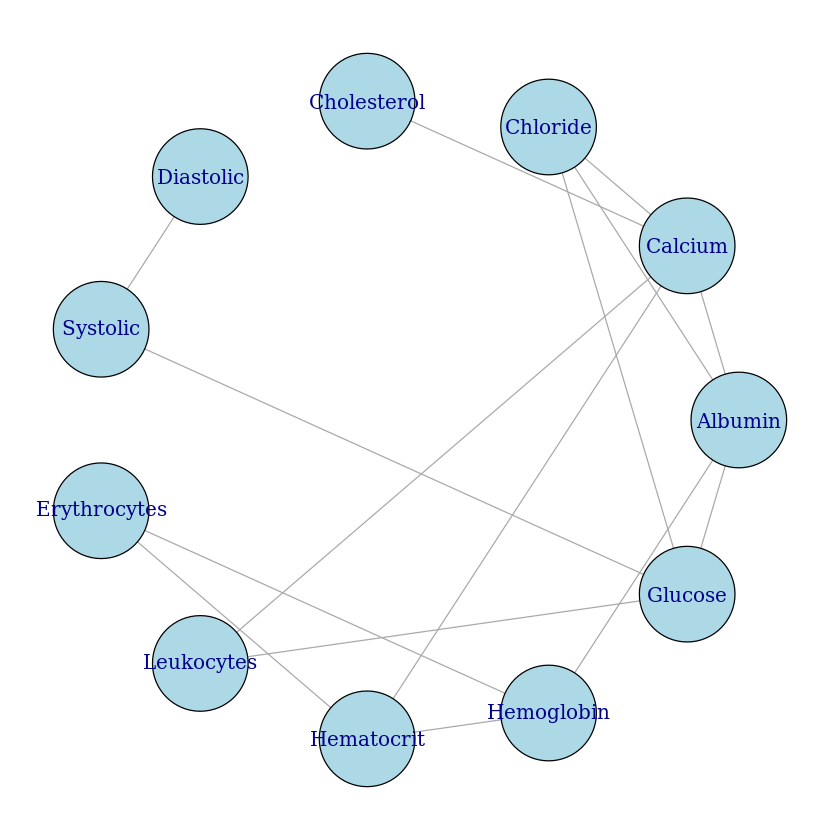} & \includegraphics[width=0.4\textwidth]{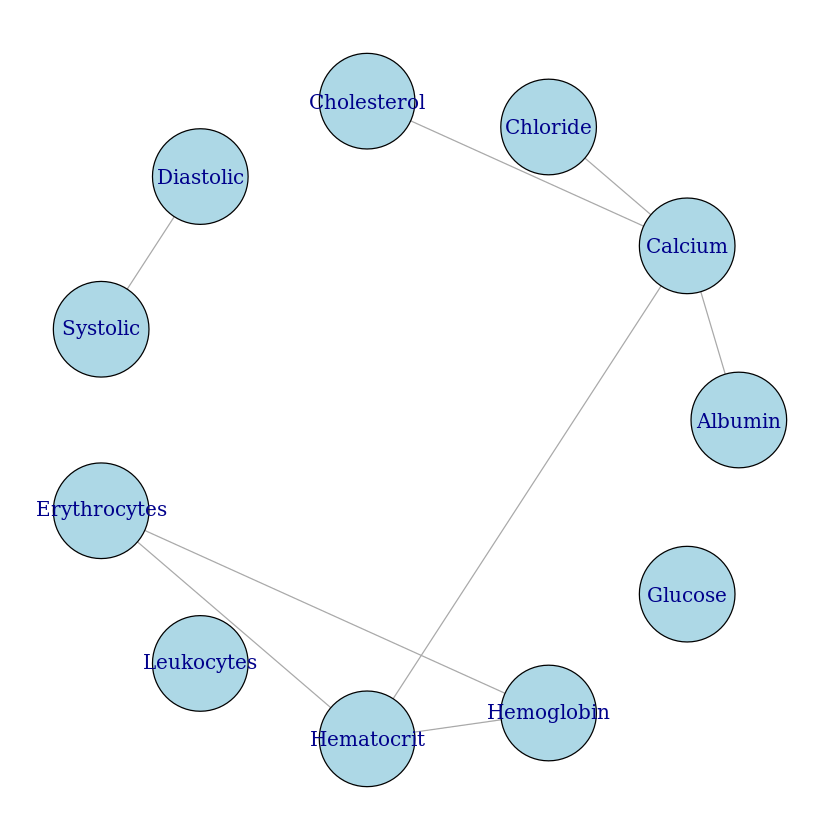}\\
    (c) $\tau = 0.5$, Control & (d) $\tau = 0.5$, Exposed  \\[6pt]
          
    \includegraphics[width=0.4\textwidth]{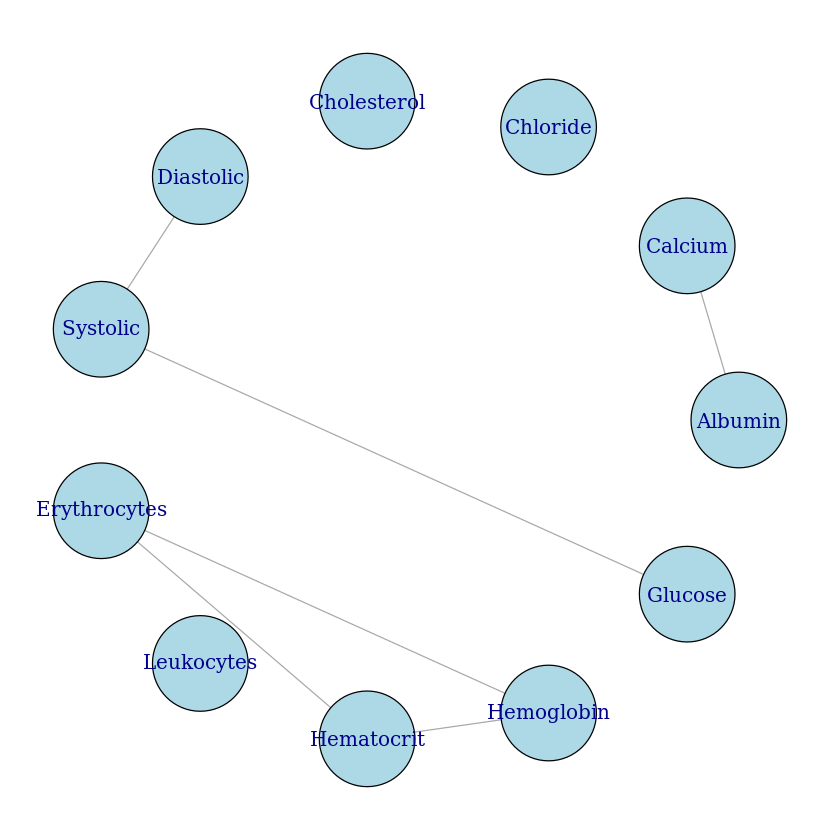} & \includegraphics[width=0.4\textwidth]{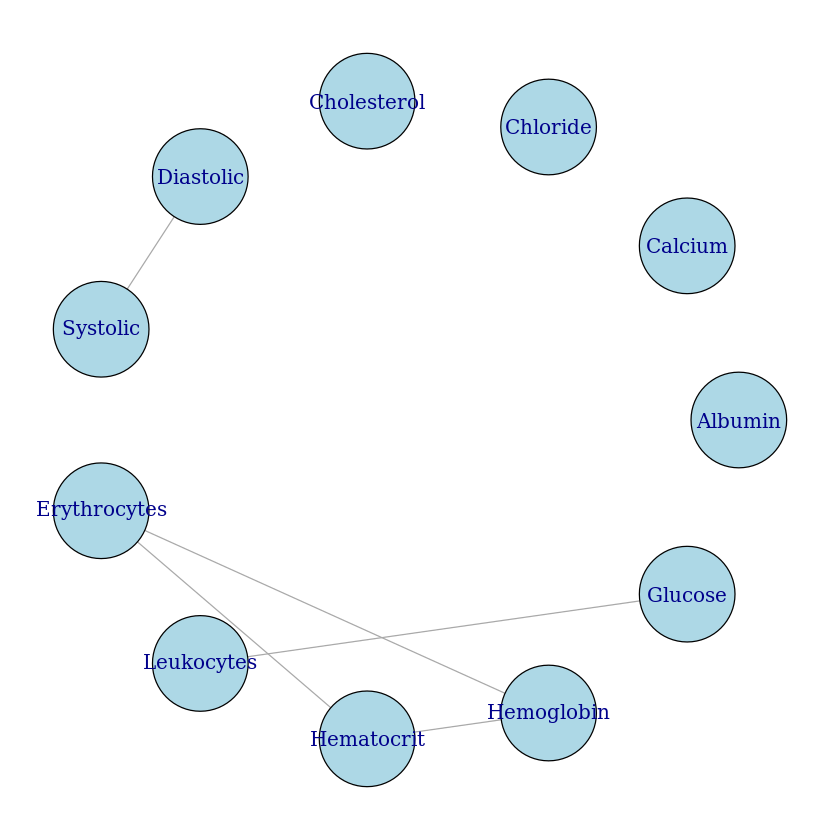} \\
    (e) $\tau = 0.9$, Control & (f) $\tau = 0.9$, Exposed \\[6pt]
    
    \\
\end{tabular}
    
\caption{The graphs on the left are the result of aggregating 15 replicates with majority voting for edges in the All of Us control population for $\tau \in \{ 0.1, 0.5, 0.9\}$. The right side features the QuACC graphical models in the All of Us MitoD population for the same $\tau$. All hypothesis testing in these graphs was performed with $\alpha = 0.025$ as the threshold for significance.  }
\label{f:allofusqgmchalved}

\end{figure}

\begin{figure}[hh]

\centering
\begin{tabular}{cc}
    \includegraphics[width=0.4\textwidth]{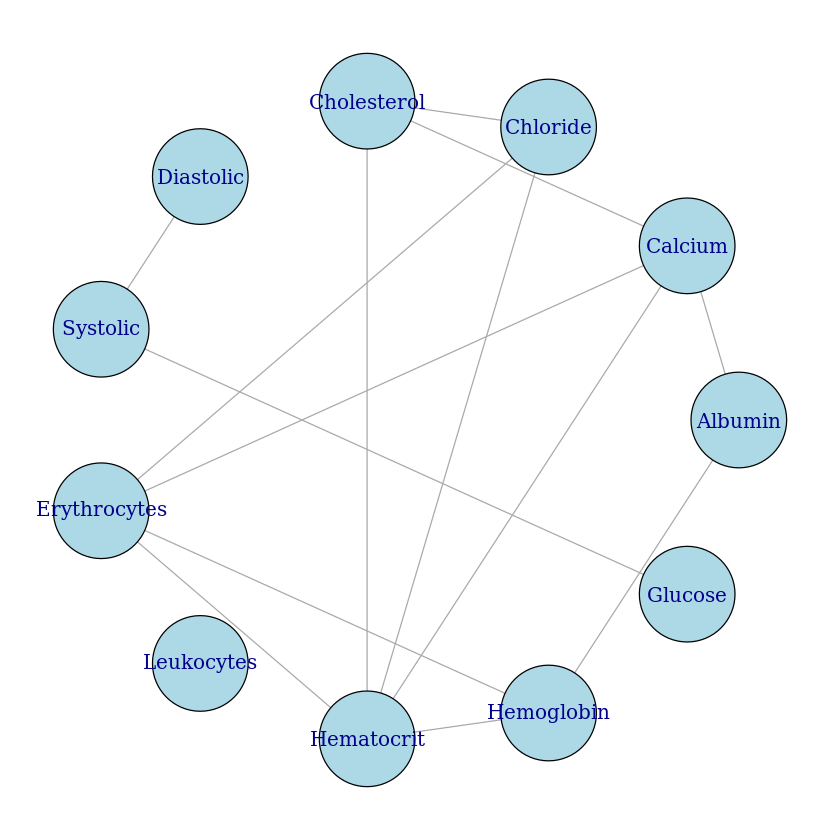} & \includegraphics[width=0.4\textwidth]{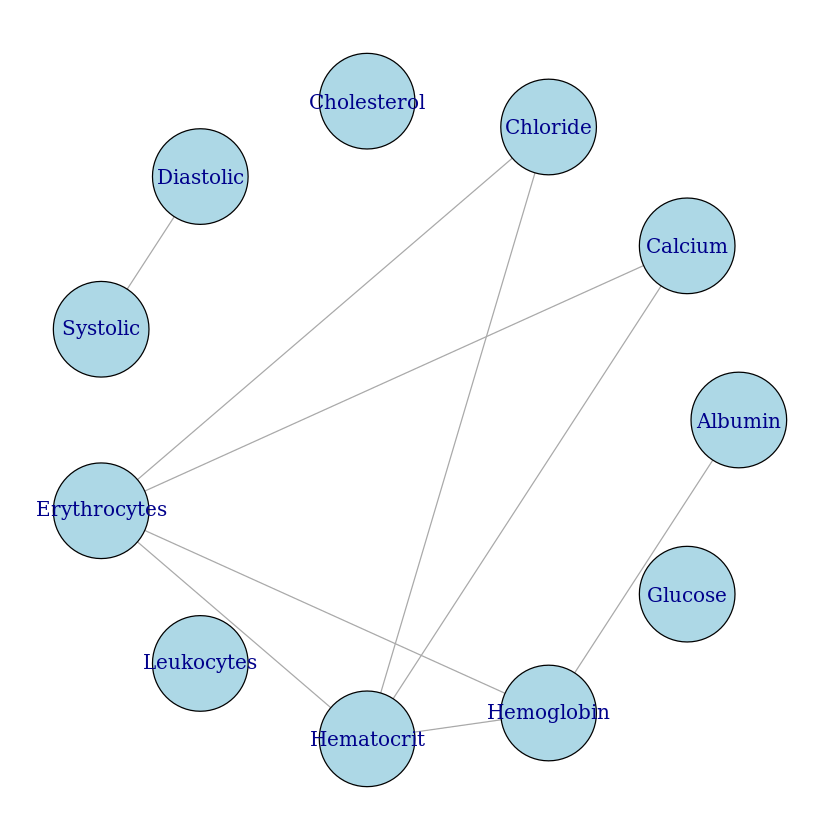}\\
    (a) $\tau = 0.1$, Control & (b) $\tau = 0.1$, Exposed \\[6pt]
    
    \includegraphics[width=0.4\textwidth]{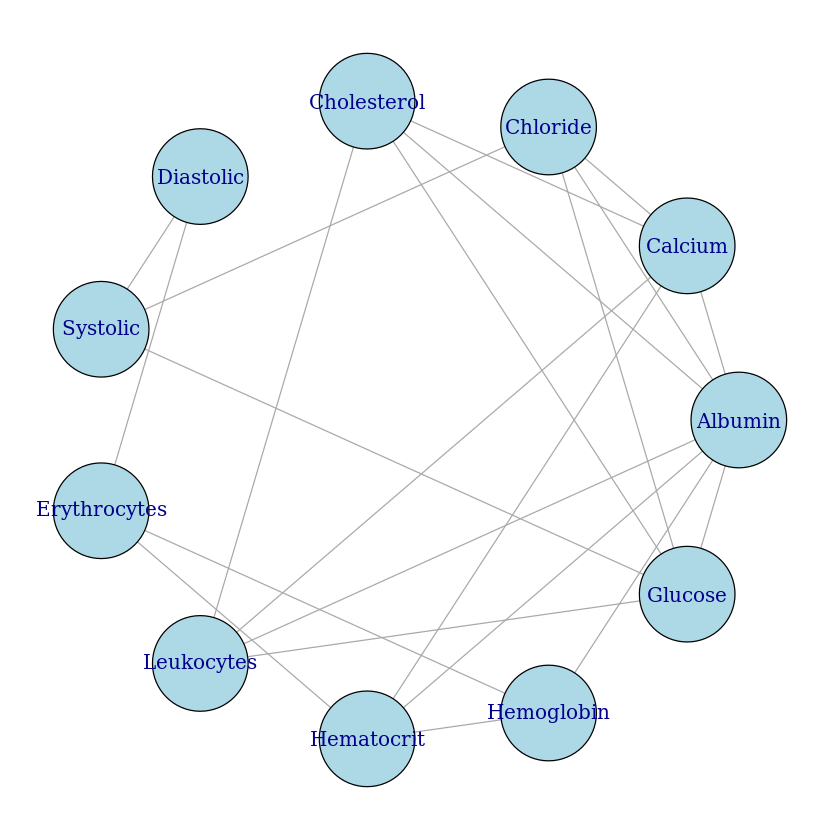} & \includegraphics[width=0.4\textwidth]{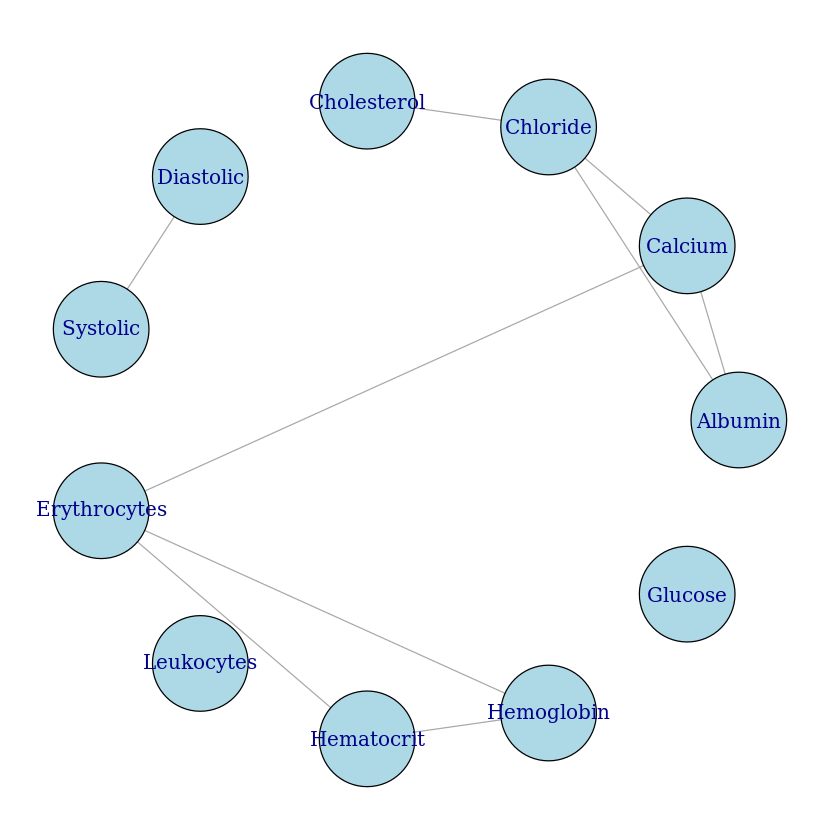}\\
    (c) $\tau = 0.5$, Control & (d) $\tau = 0.5$, Exposed  \\[6pt]
          
    \includegraphics[width=0.4\textwidth]{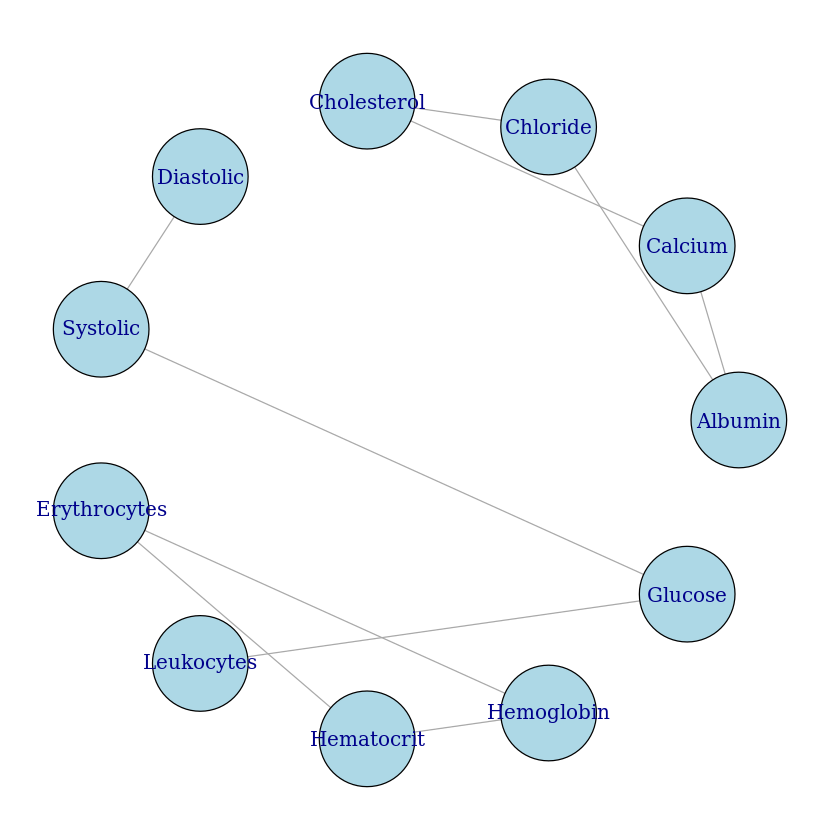} & \includegraphics[width=0.4\textwidth]{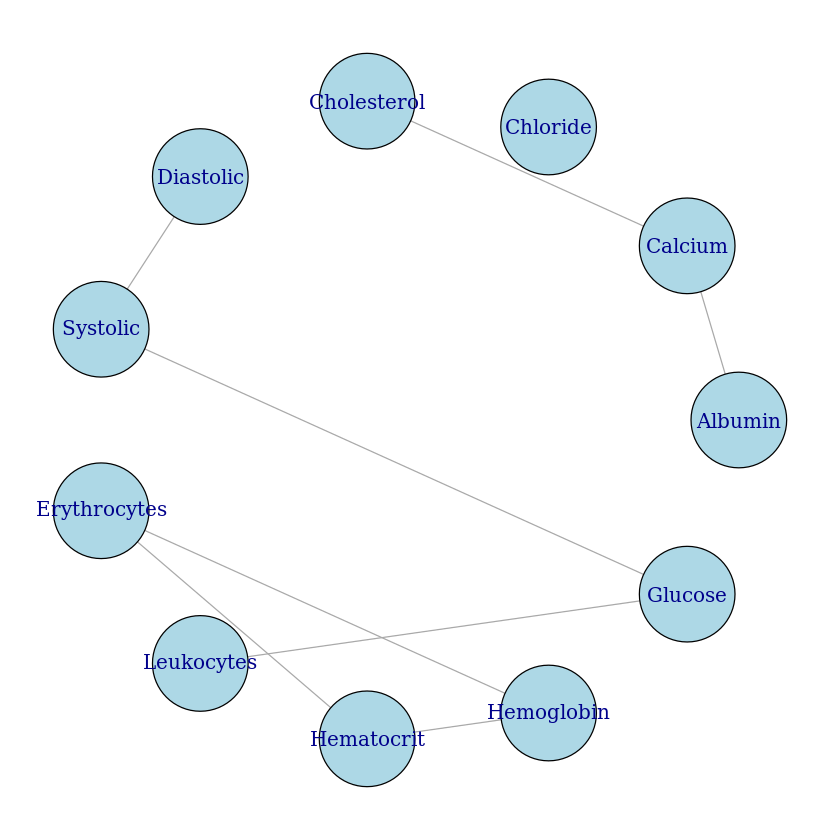} \\
    (e) $\tau = 0.9$, Control & (f) $\tau = 0.9$, Exposed \\[6pt]
    
    \\
\end{tabular}
    
\caption{The graphs on the left are the result of aggregating 15 replicates with majority voting for edges in the All of Us control population for $\tau \in \{ 0.1, 0.5, 0.9\}$. The right side features the QuACC graphical models in the All of Us MitoD population for the same $\tau$. All hypothesis testing in these graphs was performed with $\alpha = 0.1$ as the threshold for significance.  }
\label{f:allofusqgmdoubled}

\end{figure}

To illustrate uncertainty in structure learning, we provide All of Us QuACC graphs when the threshold of significance, $\alpha$, is doubled and halved below. Above we follow the standard $\alpha = 0.05$ threshold. The results for when this is restricted to $\alpha = 0.025$ can be seen in Figure \ref{f:allofusqgmchalved}. When this is instead doubled to $\alpha = 0.1$, the results can be seen in Figure \ref{f:allofusqgmdoubled}.

\section*{Appendix B: QuACC estimates using Generalized Random Forests}
Machine learning algorithms offer more flexibility in accommodating the complex associations in biological system, although it comes with a cost of prediction accuracy (i.e. large variance in predicted values). 
Depending on individual applications, different model strategies can be used. In this section, we present the asymptotic behaviors of QuACC when generalized random forest is used to estimate the conditional quantile.  We first note the additional conditions required for proper behavior of generalized random forests:

\begin{description}
    \item[C6] The conditional exceedance probabilities $p(L_{i} > l | Z_i = z)$ are Lipschitz continuous in $Z$ for all $L \in R$ for $L \in \{Y, X\}$
    \item[C7] The densities $f_{L}(l | Z = z), L \in \{Y, X\}$ are uniformly bounded from above
\end{description}

\begin{corollary}
Suppose conditions C1-C3 and C6-C7 hold for an arbitrary partition of size $n_k$, then
\begin{multline}
(n_k / s)^{1/2} \left\{ \widehat{\rho}^k_{\tau}(Y, X|Z) - \rho_{\tau}(Y,X|Z)_0 \right\} 
\overset{p}{\to} N(0, \kappa_{Y} {\sigma}^2_{{Q}_{Y}} \kappa_{Y} + \kappa_{X} {\sigma}^2_{{Q}_{X}} \kappa_{X} +  2 \kappa_{Y} \kappa_{X} V_{XY} (\tau))
\end{multline}

Where ${\sigma}^2_{{Q}_{Y}}, {\sigma}^2_{{Q}_{X}}$ are the asymptotic variances of quantile convergence for (generalized) quantile random forest estimators \citep{athey2019generalized}: 
\begin{align*}
\begin{split}
\frac{{\sigma}^2_{{Q}_{Y}}}{(n_{-k}/s)} = \frac{{H}_{n_{-k}, L}}{{f}_Y(Q_{Y}(\tau | Z))^{2}} \\
\frac{{\sigma}^2_{{Q}_{X}}}{(n_{-k}/s)} = \frac{{H}_{n_{-k}, X}}{{f}_X(Q_{X}(\tau | Z))^{2}}
\end{split}
\end{align*}

Where $H_{n_{-k}, L} = var\{\sum_{i=1}^{n_{-k}} \alpha_i(L) \psi_{Q_{L}}(L)\}$ is the variance of the adaptive weights aggregated over all trees, $\alpha_i$, multiplied by the score function $\psi_{Q_{L}} (L_i) = \tau - I\{ L_i \leq Q_{L}(\tau | Z) \}$, and $s$ is the number of samples given to each tree for $L \in \{X,Y\}$.

Here $V_{XY} (\tau)$ and $V(\tau)$ as defined as seen in Theorem \ref{theorem:conv}. 

\textbf{Proof:} Under conditions C1-C3 and C6-C7, ${\sigma}^2_{{Q}_{Y}}, {\sigma}^2_{{Q}_{X}}$ assume the above functional form. Substituting in these terms in the result for Theorem \ref{theorem:conv} completes this proof.  

\label{corollary:rfconv}
\end{corollary}

\section*{Appendix C: Normalization of QuACC to [-1, 1]}
\label{a:quacc-normalized}
The QuACC statistic can be scaled to the range of [-1, 1] to better align with correlation coefficients. In a perfect concordance, $\rho_{\tau}$ equals to $\tau^2$ when $\tau<0.5$ and equals to $(1-\tau)^2$ when $\tau\ge 0.5$; On the other hand, in a perfect discordance, $\rho_\tau = 0$ . Therefore,  we can normalized QuACC by $\rho_{\tau}^*(Y,X|Z)$ defined as follows. 

$$
\rho_{\tau}^*(Y,X|Z) = 
\begin{cases}      
\frac{\rho_{\tau}(Y,X|Z) - (1 - \tau)^2}{(1 - \tau) - (1 - \tau)^2} & \text{if } \tau \geq 0.5 \text{ and } \rho_{\tau}(Y,X|Z) > (1 - \tau)^2 \\

\frac{\rho_{\tau}(Y,X|Z) - (1 - \tau)^2}{(1 - \tau)^2} & \text{if } \tau \geq 0.5 \text{ and } \rho_{\tau}(Y,X|Z) \leq (1 - \tau)^2 \\

\frac{\rho_{\tau}(Y,X|Z) - \tau^2}{\tau - \tau^2} & \text{if } \tau < 0.5 \text{ and } \rho_{\tau}(Y,X|Z) > \tau^2 \\

\frac{\rho_{\tau}(Y,X|Z) - \tau^2}{\tau^2} & \text{if } \tau < 0.5 \text{ and } \rho_{\tau}(Y,X|Z) \leq \tau^2
\end{cases}
$$
Under independent Null, the normalized $\rho_{\tau}^*(Y,X|Z) = 0$. If they are perfectly concordant, either jointly above or below their regression planes together, then normalized  $\rho_{\tau}^*(Y,X|Z)$ is 1. If they are perfectly discordant, ie. they ever jointly above or below their regression planes, then $\rho_{\tau}^*(Y,X|Z)$ is 1. 

\section*{Appendix D: Pairwise QuACC Tables under full conditions}
\label{a:pairwise-quacc}

\begin{figure}[hh]

\centering
\begin{tabular}{cc}
    \includegraphics[width=0.4\textwidth]{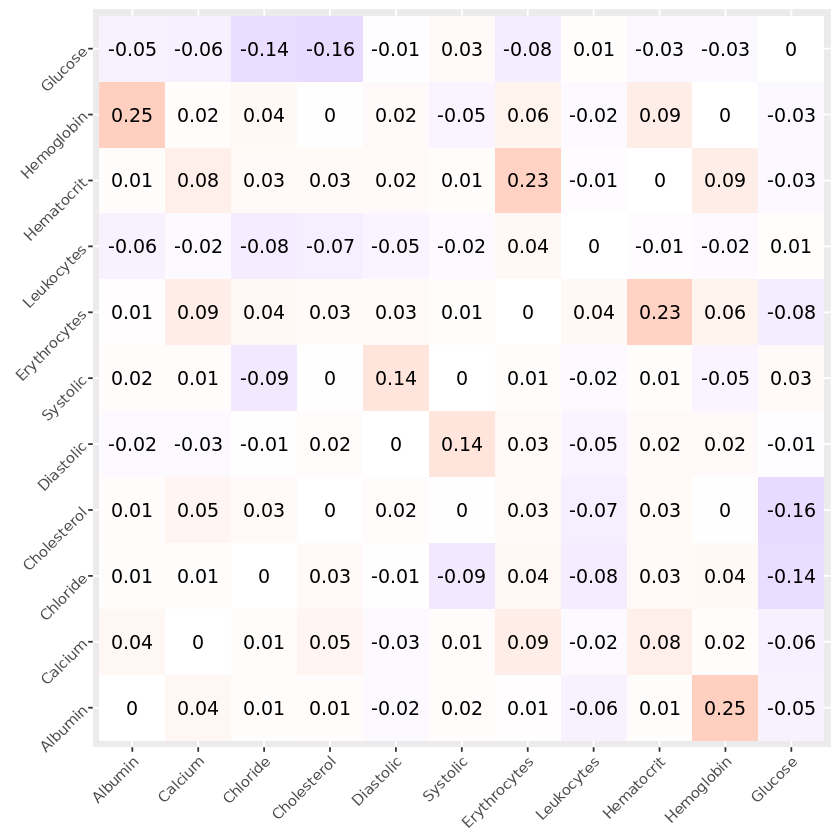} & \includegraphics[width=0.4\textwidth]{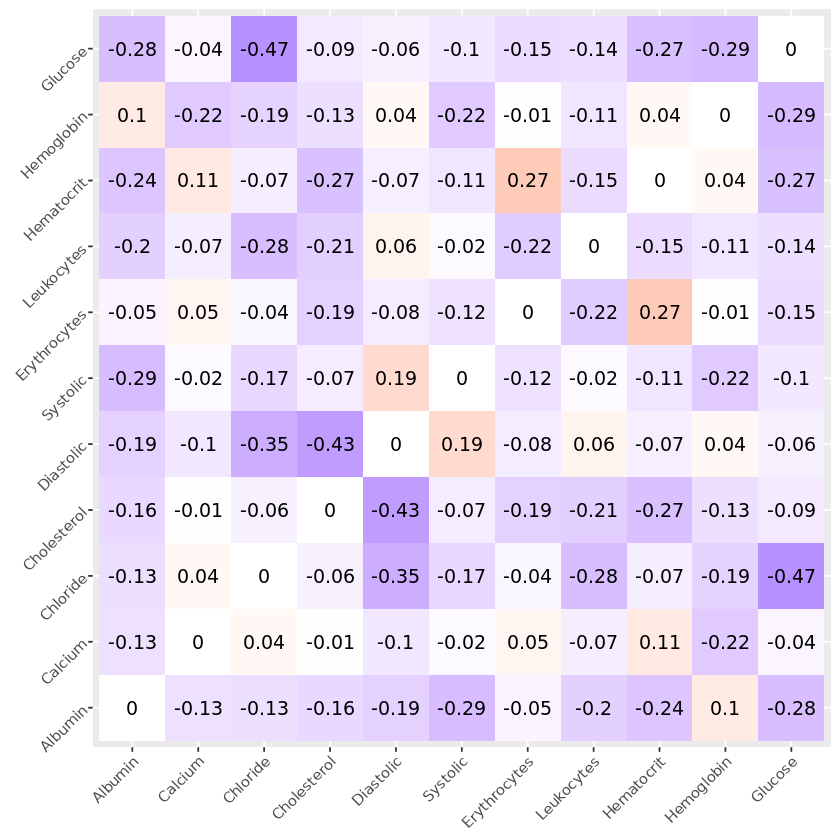}\\
    (a) $\tau = 0.1$, Control & (b) $\tau = 0.1$, Exposed \\[6pt]
    
    \includegraphics[width=0.4\textwidth]{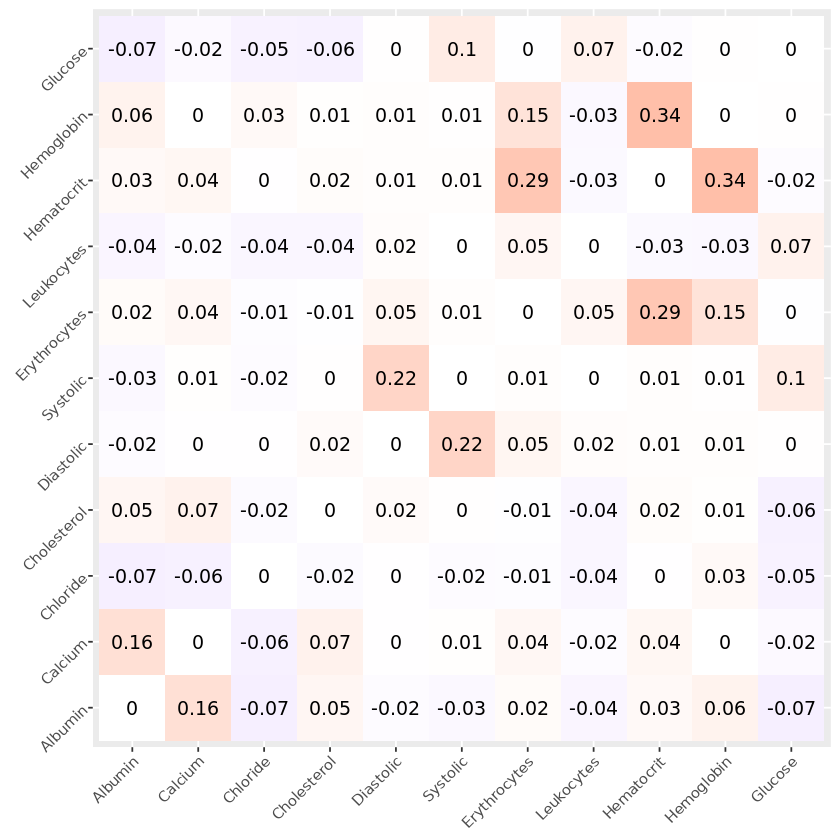} & \includegraphics[width=0.4\textwidth]{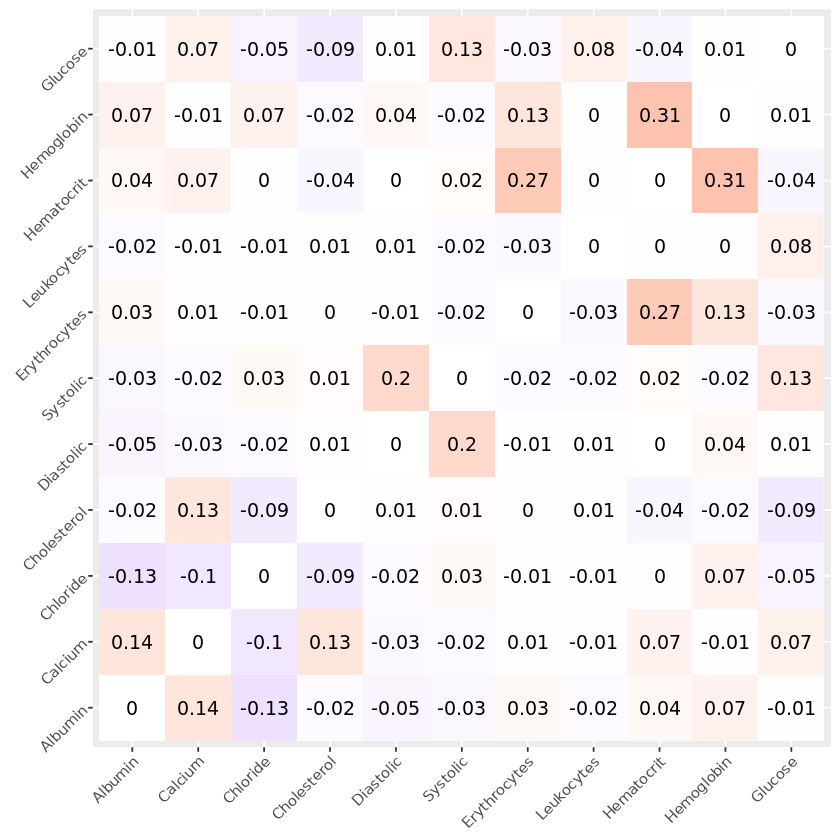}\\
    (c) $\tau = 0.5$, Control & (d) $\tau = 0.5$, Exposed  \\[6pt]
          
    \includegraphics[width=0.4\textwidth]{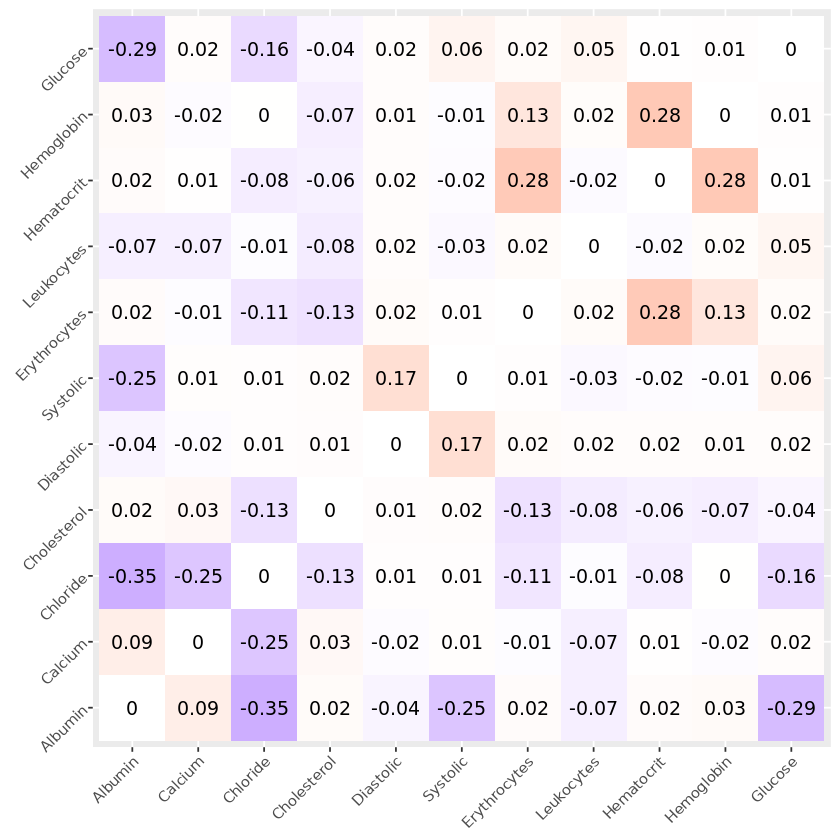} & \includegraphics[width=0.4\textwidth]{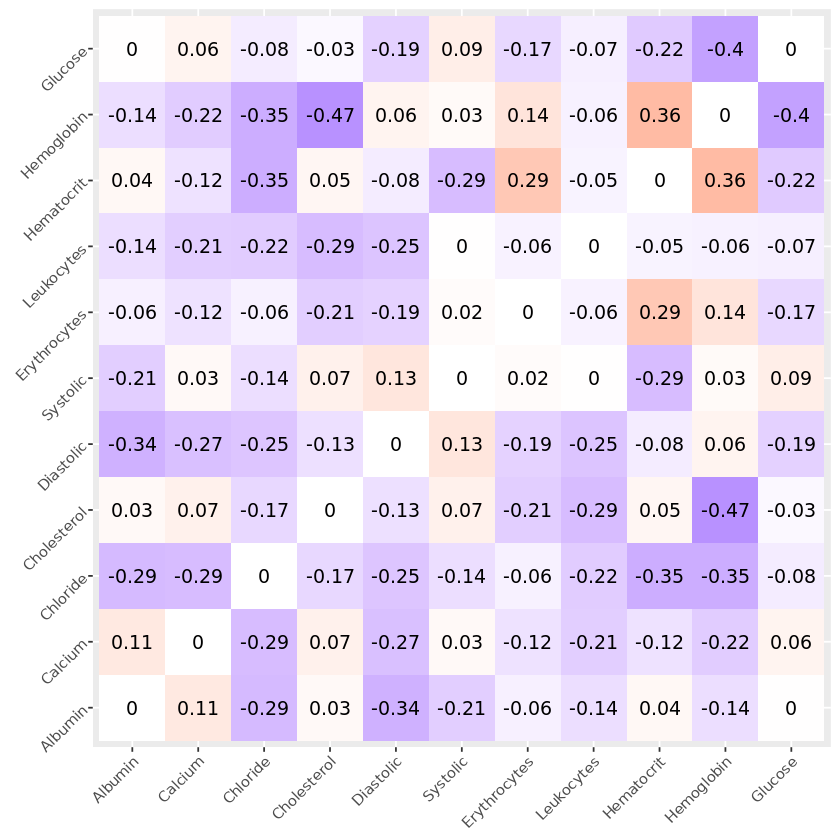} \\
    (e) $\tau = 0.9$, Control & (f) $\tau = 0.9$, Exposed \\[6pt]
    
    \\
\end{tabular}
    
\caption{The graphs on the left are the pairwise maximally conditional ($Z = V \setminus \{ X, Y\}$) QuACC statistics on the All of Us control population for $\tau \in \{ 0.1, 0.5, 0.9\}$. The right side features the pairwise maximally conditional QuACC statistics for the MitoD population for the same $\tau$. Note that testwise deletion was performed in calculating the QuACC for each pair. }

\label{f:supplement-pairwise-mrf-control-and-exposed}

\end{figure}

Figure \ref{f:supplement-pairwise-mrf-control-and-exposed} presents the pairwise QuACC that is conditioning all the other variables in the data. From the figure, we observe that  

\begin{itemize}
    \item the tails experience more (dis)concordance relative to the median 
    \item the exposed population experiences more (dis)concordance relative to the control group
\end{itemize}

These findings support the hypothesis that individuals with mitochondrial deficiencies experience a higher degree of biomarker dysregulation than their control counterparts. 

Note that because test-wise deletion was performed in constructing these pairwise tables, sample sizes differ in each QuACC.  Additionally, note that in Figure \ref{f:supplement-pairwise-mrf-control-and-exposed}, the presence of nonzero effects compared to the tables in Figure \ref{f:pairwise-marginal-control-and-exposed} may signify the presence of collider bias. This phenomenon is observed when comparing MRFs, which condition on all other variables, to conditional independence graphs as MRFs tend to be less sparse than the latter do to the presence of this bias.

\section*{Appendix E: Synthetic Experiments with Tail and Mean Effects}
Below are additional results on synthetic datasets where the data generating process is modified such that there may be a mean effect, $\alpha_i \sim Uniform(-0.4, 0.4)$. This is in contrast to the simulations presented in the main text where $\alpha_i = 0, \forall i$ 

\begin{table}[!ht]
\centering
\begin{tabular}{|| c | c c c c||} 
 \hline
 $n$ & $\tau$ & Precision & Recall & Structural Hamming Distance \\ 
 \hline\hline
 500 & 0.1 & 0.776 (0.201) & 0.311 (0.121) & 0.158 (0.033) \\ 
 \hline
 & 0.5 & 0.793 (0.176) & 0.464 (0.152) & 0.132 (0.041)  \\
 \hline
  & 0.9 & 0.812 (0.143) & 0.579 (0.152) & 0.111 (0.402) \\ 
 \hline

 1000 & 0.1 & 0.917 (0.121) & 0.501 (0.136) & 0.110 (0.033) \\ 
 \hline
 & 0.5 & 0.801 (0.114) & 0.647 (0.149) & 0.103 (0.039) \\
 \hline
  & 0.9 & 0.822 (0.115) & 0.676 (0.134) & 0.096 (0.033) \\ 
 \hline
 
 5000 & 0.1 & 0.915 (0.092) & 0.764 (0.139) & 0.062 (0.034) \\ 
 \hline
 & 0.5 & 0.706 (0.087) & 0.834 (0.141) & 0.102 (0.040) \\
 \hline
  & 0.9 & 0.728 (0.089) & 0.811 (0.129) & 0.100 (0.035) \\ 
 \hline
\end{tabular}
\caption{Mean and standard deviation of precision, recall, and Hamming distance for predicted edges obtained via QuACC versus true edges across 100 replicates of a linear quantile dependent dataset with underlying graph depicted in Figure \ref{f:simgraph}. The linear coefficients $\alpha_i \sim Uniform(-0.4,0.4)$. }
\label{t:discoverymetrics}
\end{table}

\begin{table}[h!]
\centering
\begin{tabular}{|| c | c c c||} 
 \hline
 $n$ & Precision & Recall & Structural Hamming Distance \\ 
 \hline\hline
 500 & 0.705 (0.114) & 0.717 (0.146) & 0.118 (0.043) \\ 
 \hline

 1000 & 0.684 (0.097) & 0.797 (0.137) & 0.116 (0.042) \\ 
 \hline
 
 5000 & 0.599 (0.097) & 0.877 (0.105) & 0.146 (0.51) \\ 
 \hline
\end{tabular}
\caption{Mean and standard deviation of precision, recall, and Hamming distance for predicted edges obtained via partial correlation conditional independence testing versus true edges across 100 replicates of a linear quantile dependent dataset with underlying graph depicted in Figure \ref{f:simgraph}. The linear coefficients $\alpha_i \sim Uniform(-0.4,0.4)$. }
\label{t:discoverymetricsgaussian}
\end{table}

\section*{Appendix F: Technical Proofs}
Recall that  $ \mathcal{D} =\{(X_j, Y_j, Z_j):j=1,...,n\}$ is a random sample of size $n$.
Let $\Gamma_X$ and $\Gamma_Y$ be the support of $X$ and $Y$ respectively. 
We now define two score functions $$S_n(a, b) = \frac{1}{n} \sum_{i=1}^n I\{Y_i> a, X_i >b\},$$ 
and $$S(a, b) = E_{(X,Y)} I\{Y> a, X>b\} = P(Y> a, X>b).$$ Following the law of large numbers, $|S_n(a, b) - S(a, b)|=o_p(1)$ for any following $(a,b) \in\Gamma_X\otimes \Gamma_Y$. Lemma \ref{lemma:uc} establishes the uniform convergence $S_n(a, b)$ over a compact space of $(a,b)$ with a decreasing diameter. 

\begin{lemma} \label{lemma:uc}  Let $\Gamma_{1,n} =\{a: |a - a_0| < c\ell_n\}$ and $\Gamma_{2,n} =\{b: |b - b_0| < c\ell_n\}$ for some constant $c$ and descending sequence $\ell_n$, and $\Gamma_n =\Gamma_{1,n}\bigcup\Gamma_{1,2} $ we have the following uniform convergence by
\begin{equation}
\sup_{(a,b)\in \Gamma_n} d_n^{-1}| S_n(a, b) - S_n(a_0, b_0) - S(a,b) + S(a_0,b_0)| = o_p(1),
\end{equation}
where $d_n\rightarrow 0$, $\ell_n\rightarrow 0$, $\ell_n/d_n\rightarrow 0$ as $n\rightarrow \infty$. 
\end{lemma}
\paragraph{Proof of Lemma \ref{lemma:uc}}
Let
$u(a, b, a_0, b_0) = | S_n(a, b) - S_n(a_0, b_0) - S(a,b) + S(a_0,b_0)|.$ By the triangle inequality
\begin{equation*}
u(a, b, a_0, b_0) \leq | S_n(a, b) - S_n(a_0, b_0)| + |  S(a,b) - S(a_0,b_0)|
\end{equation*}

The second term in the triangle inequality, $|  S(a,b) - S(a_0,b_0)|$, is equivalent to
\begin{align*}
\begin{split}
|  S(a,b) - S(a_0,b_0)| & = | P(Y > a, X > b) - P(Y > a_0, X > b_0) | \\
& = |  \int_{a}^{\infty} \int_{b}^{\infty} f_{YX}(y, x) dx dy - \int_{a_0}^{\infty} \int_{b_0}^{\infty} f_{YX}(y, x) dx dy |
\end{split}
\end{align*}

Under condition C2, there exist a $ \delta,$ that bounds the conditional densities of $Y$ and $X$ given $Z$, such that $\delta \geq \sup_{(x,y)}f_{Y|Z}(y) \vee f_{X|Z}(x)$. It follows that
$$
\begin{array}{ccc}
\sup_{(a,b)\in\Gamma_n} d_n^{-1}|  S(a,b) - S(a_0,b_0)| 
& \leq  & 
\sup_{(a,b)\in\Gamma_n}\delta d_n^{-1}\{|a_0 - a| + |b - b_0|\} = o_p(1).  \\
\end{array}
$$


On the other hand, the first term $| S_n(a, b) - S_n(a_0, b_0) |$ is bounded by the proportion of times $I\{Y_i > a, X_i > b \}$ and $I\{ Y_i > a_0, X_i > b_0 \}$ disagree (are not mutually (dis)concordant for a given $i$.
\begin{align*}
\begin{split}
| S_n(a, b) - S_n(a_0, b_0) | \leq | \frac{1}{n} \sum_i & I \{ Y_i > max(a_0, a) \} * I \{ |X_i - b_n | \leq |b_0 - b| \} + \\
& I \{ X_i > max(b, b_0) \} * I \{ |Y_i - a_n| \leq |a_0 - a| \} |
\end{split}
\end{align*}

A less restrictive upper bound is given by 
\begin{align*}
\begin{split}
| S_n(a, b) - S_n(a_0, b_0) | \leq | \frac{1}{n} \sum_i &  I \{ |X_i - b | \leq |b_0 - b| \} +  I \{ |Y_i - a| \leq |a_0 - a| \} |
\end{split}
\end{align*}

It follows from Lemma 2.2 of \citet{he2000parameters} that  $| S_n(a, b) - S_n(a_0, b_0) |$ has the same convergence rate as $| S(a, b) - S(a_0, b_0) |$, i.e. 
\begin{align*} 
\begin{split}
\sup_{(a,b)\in\Gamma_n} d_n^{-1} | S_n(a, b) - S_n(a_0, b_0) | = o_p(1)
\end{split}
\end{align*}

Lemma \ref{lemma:uc} follows naturally from the uniform convergence of both terms.




\begin{lemma}
Suppose conditions C1 - C4 hold for an arbitrary partition of size $n_k$, then asymptotics for the test set concordance are given by
\begin{align*}
\begin{split}
n_k^{1/2}\sum_{j=1}^{n_k} I(Y_j \lessgtr {Q}_{Y} (\tau | Z_j), X_j \lessgtr {Q}_{X} (\tau | Z_j)) - p(Y_j \lessgtr {Q}_{Y} (\tau | Z_j), X_j \lessgtr {Q}_{X} (\tau | Z_j)) \\ 
\sim AN(0, V(\tau))
\end{split}
\end{align*}

Where $V(\tau)$ is defined as 
$$
V(\tau) = \begin{cases}      
\Scale[0.75]{ (1 - \tau)^2 (1 - (1 - \tau))^2 + 
(1 - 4 ( 1 -\tau) + 4 (1 - \tau)^2) [p(Y > {Q}_{Y}(\tau | Z), X > {Q}_{X}(\tau | Z)) - (1 - \tau)^2]} & \text{if } \tau \geq 0.5 \\
\Scale[0.75]{ \tau^2 (1 - \tau)^2 + (1 - 4 \tau + 4 \tau^2) [ p(Y < {Q}_{Y}(\tau | Z), X < {Q}_{X}(\tau | Z))  - \tau^2 ]} & \text{if } \tau < 0.5.  \\
\end{cases}  
$$

When $\tau < 0.5$, any value in $[0, \tau]$ can be substituted in for $p(Y < {Q}_{Y}(\tau | Z), X < {Q}_{X}(\tau | Z))$ to test that specific level of concordance. Likewise, when $\tau \geq 0.5$, any value in $[0, 1 - \tau]$ can be substituted in for $p(Y > {Q}_{Y}(\tau | Z), X > {Q}_{X}(\tau | Z))$ 
\label{lemma:test-set}
\end{lemma}

\subsection*{Proof of Lemma \ref{lemma:test-set}}

Denote $ Q_Y (\tau | Z) $ and $ Q_X ( \tau |Z) $ as, respectively, the true conditional quantile functions of $ Y$ and $X$ given $Z$. $ \mathcal{D} =\{(X_j, Y_j, Z_j):j=1,...,n\}$ is a random sample of size $n$. 

Following the cross-fitting algorithm, we partition the data $\mathcal{D}$ into $k$ folds with equal sample sizes, and $n_k$ is the size of the $k$th partition.

Define the testing set error $S_{n_k} (Q_Y (\tau | Z), Q_X (\tau | Z) )
- S (Q_Y (\tau | Z), Q_X (\tau | Z))$ as
\begin{align*}
\begin{split}
S_{n_k} (Q_Y (\tau | Z), Q_X (\tau | Z) )
- S (Q_Y (\tau | Z), Q_X (\tau | Z)) = & \frac{1}{n_k}\sum_{j=1}^{n_k} I(Y_j < {Q}_{Y} (\tau | Z_j), X_j < {Q}_{X} (\tau | Z_j)) - \\
& p(Y_j < {Q}_{Y} (\tau | Z_j), X_j < {Q}_{X} (\tau | Z_j))  
\end{split}
\end{align*}

Under the null hypothesis of independence of concordance, we wish to evaluate 
$$
\begin{array}{ccc}
 V(\tau) =  Var( \sum_j I(Y_j < {Q}_{Y} (\tau | Z_j)) *  I(X_j < {Q}_{X} (\tau | Z_j)) ) 
\end{array}
$$

We can rewrite $V(\tau) =  Var( I(Y < {Q}_{Y} (\tau | Z)) * I(X < {Q}_{X} (\tau | Z)) )$. We can subtract the nominal means and define a centralized statistic: 
\begin{align*}
\begin{split}
 V(\tau) = & Var( I(Y < {Q}_{Y} (\tau | Z)) * I(X < {Q}_{X} (\tau | Z)) ) \\
 = & Var( (I(Y < {Q}_{Y} (\tau | Z)) - \tau)  * (I(X < {Q}_{X} (\tau | Z)) - \tau) )
\end{split}
\end{align*}

For events $W_1 = I(Y < {Q}_{Y} (\tau | Z)) - \tau, W_2 = I(X < {Q}_{X} (\tau | Z)) - \tau$, we have 
\begin{align*}
\begin{split}
Var(W_1 * W_2) = & Var(W_1) Var(W_2) + Var(W_1) (E[W_2])^2 + Var(W_2) (E[W_1])^2 \\
 & + cov(W_1^2, W_2^2) - (cov(W_1, W_2))^2 - 2 cov(W_1, W_2)E[W_1] E[W_2] - (E[W_1])^2 (E[W_2])^2 
\end{split}
\end{align*}

It is fact that $E[I(Y < {Q}_{Y} (\tau | Z))] = p(Y < {Q}_{Y} (\tau | Z)) = \tau$ and $E[I(X < {Q}_{X} (\tau | Z))] = p(X < {Q}_{X} (\tau | Z)) = \tau$ for the true conditional quantiles.
\begin{align*}
\begin{split}
Var(W_1 * W_2) = & Var(W_1) Var(W_2) + cov(W_1^2, W_2^2) - (cov(W_1, W_2))^2 
\end{split}
\end{align*}

We simplify the variance above term by term. We can model each indicator as $I(Y_j < {Q}_{Y} (\tau | Z_j)), I(X_j < {Q}_{X} (\tau | Z_j)) \sim Bernoulli(\tau)$. The variance of these Bernoulli distributed variables is $\tau (1 - \tau)$. 
\begin{align*}
\begin{split}
 V(W_1) V(W_2) =  \tau (1 - \tau) * \tau (1 - \tau) = \tau^2 (1 - \tau)^2
\end{split}
\end{align*}

The second term $cov(W_1^2, W_2^2)$ can be simplified as follows
\begin{align*}
\begin{split}
cov(W_1^2, W_2^2) = & cov(I(Y < {Q}_{Y}(\tau | Z)) - 2 \tau I(Y < {Q}_{Y}(\tau | Z)) + \tau^2, I(X < {Q}_{X}(\tau | Z)) - 2 \tau I(X < {Q}_{X}(\tau | Z)) + \tau^2) \\
= & cov(I(Y < {Q}_{Y}(\tau | Z)) - 2 \tau I(Y < {Q}_{Y}(\tau | Z)), I(X < {Q}_{X}(\tau | Z)) - 2 \tau I(X < {Q}_{X}(\tau | Z))) \\
= & E[I(Y < {Q}_{Y}(\tau | Z)) I(X < {Q}_{X}(\tau | Z)) - 4 \tau I(Y < {Q}_{Y}(\tau | Z)) I(X < {Q}_{X}(\tau | Z)) + \\
& 4 \tau^2 I(Y < {Q}_{Y}(\tau | Z)) I(X < {Q}_{X}(\tau | Z)) ]
\end{split}
\end{align*}

Where the expectation $E[I(Y < {Q}_{Y}(\tau | Z)) I(X < {Q}_{X}(\tau | Z))] = p(Y < {Q}_{Y} (\tau | Z), X < {Q}_{X} (\tau | Z))$ is the joint concordance. Thus 
\begin{align*}
\begin{split}
cov(W_1^2, W_2^2) = p(Y < {Q}_{Y}(\tau | Z), X < {Q}_{X}(\tau | Z)) (1 - 4\tau + 4 \tau^2)
\end{split}
\end{align*}

The final term $(cov(W_1, W_2))^2 = (E[W_1 W_2])^2 = \tau^2 (1 - 4 \tau + 4 \tau^2)$. 

Under conditional independence, $p(Y < {Q}_{Y}, X < {Q}_{X}) = \tau^2$ and it is clear that the covariance terms cancel. The complete variance $V(\tau)$ can be written
\begin{align*}
\begin{split}
 V(\tau) = & Var( I(Y < {Q}_{Y} (\tau | Z)) * I(X < {Q}_{X} (\tau | Z)) ) \\
 = & \tau^2 (1 - \tau)^2 + p(Y < {Q}_{Y}(\tau | Z), X < {Q}_{X}(\tau | Z)) (1 - 4\tau + 4 \tau^2) - \tau^2 (1 - 4 \tau + 4 \tau^2)
\end{split}
\end{align*}

The mean of IID indicator variables, $I(Y < {Q}_{Y} (\tau | Z), X < {Q}_{X} (\tau | Z))$, where each indicator has variance $V(\tau)$, can be normally approximated by the Lindeberg-Levy central limit theorem as 
\begin{align*}
\begin{split}
n_k^{1/2}\sum_{j=1}^{n_k} I(Y_j < {Q}_{Y} (\tau | Z_j), X_j < {Q}_{X} (\tau | Z_j)) - p(Y_j < {Q}_{Y} (\tau | Z_j), X_j < {Q}_{X} (\tau | Z_j)) \\ = \sqrt{n_k} (S_{n_k} (Q_Y (\tau | Z), Q_X (\tau | Z) )
- S (Q_Y (\tau | Z), Q_X (\tau | Z))) \sim AN(0, V(\tau))
\end{split}
\end{align*}
where $V(\tau) = \tau^2 (1 - \tau)^2 + (1 - 4 \tau + 4 \tau^2) [p(Y < {Q}_{Y}(\tau | Z), X < {Q}_{X}(\tau | Z)) - \tau^2]$.

The case for joint concordance above the true conditional quantiles follows by symmetry. This results in a complete
$$
V(\tau) = \begin{cases}      
\Scale[0.75]{ (1 - \tau)^2 (1 - (1 - \tau))^2 + 
(1 - 4 ( 1 -\tau) + 4 (1 - \tau)^2) [p(Y > {Q}_{Y}(\tau | Z), X > {Q}_{X}(\tau | Z)) - (1 - \tau)^2]} & \text{if } \tau \geq 0.5 \\
\Scale[0.75]{ \tau^2 (1 - \tau)^2 + (1 - 4 \tau + 4 \tau^2) [ p(Y < {Q}_{Y}(\tau | Z), X < {Q}_{X}(\tau | Z))  - \tau^2 ]} & \text{if } \tau < 0.5.  \\
\end{cases}  
$$.

\paragraph{Proof of Theorem 1} Recall that we denote $ Q_Y (\tau | Z) $ and $ Q_X ( \tau |Z) $ as, respectively, the true conditional quantile functions of $Y$ and $X$ given $Z$, and denote $\mathcal{D} =\{(X_j, Y_j, Z_j):j=1,...,n\}$ as a random sample of size $n$. Following the cross-fitting algorithm, we partition the data $\mathcal{D}$ into $k$ folds with equal sample sizes, and $n_k$ is the size of the $k$th partition. We denote $\widehat{Q}^{-k}_{Y}$ and $\widehat{Q}^{-k}_{X}$ are estimated quantile function from the training set excluding the $k$-th fold. Depending on the estimation algorithms employed, $\widehat{Q}^{-k}_{Y}$ and $\widehat{Q}^{-k}_{X}$ may exhibit different convergence rates and asymptotic distributions. In this work, we assume that both $\widehat{Q}^{-k}_{Y}$ and $\widehat{Q}^{-k}_{X}$ are consistent estimators, with a convergence rate of $b_n^{1/2}$, and that they are asymptotically normal. Specifically, we have
\begin{equation} \sqrt{b_n}\left\{\left(\begin{matrix} 
\widehat{Q}^{-k}_{Y}\\
\widehat{Q}^{-k}_{X} \end{matrix}\right) - \left(\begin{matrix} Q_{Y}\\
Q_{X} \end{matrix}\right)\right\} \sim AN \left(0, \left(\begin{matrix} \sigma_X(\tau)^2 & V_{XY}(\tau)\\ V_{XY}(\tau) & \sigma_Y(\tau)^2 \end{matrix}\right)\right),  \label{eq:an}
\end{equation}
where $\sigma_X(\tau)^2$ and $\sigma_Y(\tau)^2$ are the asymptotic variances of $\widehat{Q}^{-k}_{X}$ and $\widehat{Q}^{-k}_{Y}$, respectively, and $V_{XY}(\tau)$ is their asymptotic covariance. In the case of linear quantile regression, both $\widehat{Q}^{-k}_{Y}$ and $\widehat{Q}^{-k}_{X}$ achieve root-$n$ consistency, implying $b_n = n_k$. For models with increasing number of parameters, such as semiparametric partially linear quantile regression, or other nonparametric approaches, the estimators are root-$(s/n)$ consistent, where $s$ represents the dimension of the model, i.e. $b_n = n_k/s$.

Following the Lemma 1, we have 
$$b_n^{1/2}\left | S_{n_k} (\widehat{Q}^{-k}_{Y}, \widehat{Q}^{-k}_{X} ) -
S_{n_k} (Q_Y (\tau | Z), Q_X (\tau | Z) )-S (\widehat{Q}^{-k}_{Y}, \widehat{Q}^{-k}_{X} )+S (Q_Y (\tau | Z), Q_X (\tau | Z)  )\right| = o_p(1)$$
Since $\widehat{\rho}_\tau^k (Y, X|Z) =S_{n_k} (\widehat{Q}^{-k}_{Y}, \widehat{Q}^{-k}_{X} )$,   and $\rho_\tau (Y, X|Z)_0 = S (Q_Y (\tau | Z), Q_X (\tau | Z))$, the convergence above implies that
\begin{equation}    
(b_n)^{1/2}(\widehat{\rho}_\tau (Y, X|Z) - \rho_\tau (Y, X|Z)_0) = I_{n1} + I_{n2} +o_p(1) 
\end{equation}
where 
$$I_{n1} = b_n^{1/2}\left\{S_{n_k} (Q_Y (\tau | Z), Q_X (\tau | Z) )
- S (Q_Y (\tau | Z), Q_X (\tau | Z)) \right\} $$
and 
$$I_{n2} = b_n^{1/2}\left\{
S (\widehat{Q}^{-k}_{Y}, \widehat{Q}^{-k}_{X} )-S (Q_Y (\tau | Z), Q_X (\tau | Z)  ) \right\}$$

The asymptotic behaviors of QuACC depends on those of $I_{n1}$ and $I_{n2}$. The two parts are independent as they utilize independent training and testing set data. $I_{n1}$ reflects the sampling uncertainty in testing data, while $I_{n2}$ represents the estimation uncertainty. 

We start with deriving the asymptotics of $I_{n2}$. 
Following the bivariate Taylor expansion, we Taylor expand $S(\widehat{Q}^{-k}_{Y}, \widehat{Q}^{-k}_{X} )$ around the true quantiles $(Q_Y (\tau | Z), Q_X (\tau | Z)  )$, which leads the following approximation
$$
\begin{array}{ccc}
\lefteqn{b_n^{1/2}S (\widehat{Q}^{-k}_{Y}, \widehat{Q}^{-k}_{X} ) }\\ 
& \approx &b_n^{1/2} \left[ \frac{1}{n_k}\sum_{j=1}^{n_k} \frac{ \partial }{\partial {Q}_{Y} (\tau | Z_j)} p( Y_j < {Q}_{Y} (\tau | Z_j), X_j < {Q}_{X} (\tau | Z_j) ) \right] (\widehat{Q}^{-k}_{Y} (\tau | Z_j) - {Q}_{Y} (\tau | Z_j)) + \\
& & b_n^{1/2}\left[ \frac{1}{n_k}\sum_{j=1}^{n_k} \frac{ \partial }{\partial {Q}_{X} (\tau | Z_j)} p(Y_j < {Q}_{Y} (\tau | Z_j), X_j <  {Q}_{X} (\tau | Z_j) ) \right] (\widehat{Q}^{-k}_{X} (\tau | Z_j) - {Q}_{X} (\tau | Z_j)) +  \\
& & o_p(1) \\ 
& \approx & b_n^{1/2}\left[ \frac{1}{n_k}\sum_{j=1}^{n_k} f({Q}_{Y} (\tau | Z_j)| X_j <  {Q}_{X} (\tau | Z_j)) \: G_j ({Q}_{X} (\tau | Z_j)) \right] (\widehat{Q}^{-k}_{Y} (\tau | Z_j) - {Q}_{Y} (\tau | Z_j))  + \\
& & b_n^{1/2}\left[ \frac{1}{n_k}\sum_{j=1}^{n_k}  g( {Q}_{X} (\tau | Z_j) | Y_j < {Q}_{Y} (\tau | Z_j)) \: F_j ({Q}_{Y} (\tau | Z_j)) \right]  (\widehat{Q}^{-k}_{X} (\tau | Z_j) - {Q}_{X} (\tau | Z_j))  +  \\
& & o_p(1)  \\
\end{array}
$$






Where $f(.|*), g(.|*)$ are the conditional PDFs and $F, G$ are the marginal CDFs of $Y, X$, respectively. Denote $A, B, C, D$ as vectors of dimension $n_k \times 1$ whose $j$th components are given respectively by
\begin{align*}
\begin{split}
A_j = f ({Q}_{Y} (\tau | Z_j)| X_j < {Q}_{X} (\tau | Z_j)) \\
B_j = g ({Q}_{X} (\tau | Z_j) | Y_j < {Q}_{Y} (\tau | Z_j)) \\
C_j = F_j ({Q}_{Y} (\tau | Z_j)) \\
D_j = G_j ({Q}_{X} (\tau | Z_j))
\end{split}
\end{align*}
Define $\kappa_{Y} = lim_{n_k \rightarrow \infty} {n_k}^{-1} A^T D  $ and $\kappa_{X} = lim_{n_k \rightarrow \infty} {n_k}^{-1} B^T  C $. 
\begin{align*}
\begin{split}
I_{n2} = & \kappa_{Y}b_n^{1/2}(\widehat{Q}^{-k}_{Y} (\tau | Z) - {Q}_{Y} (\tau | Z)) + \\
& \kappa_{X}b_n^{1/2}(\widehat{Q}^{-k}_{X} (\tau | Z) - {Q}_{X} (\tau | Z)) + o_p(1)
\end{split}
\end{align*}

It follows from \eqref{eq:an} that 
$I_{n2}= AN(0, \kappa_{Y}{\sigma}^2_{Y} \kappa_{Y} + \kappa_{X} {\sigma}^2_{X} \kappa_{X} + \frac{2 \kappa_{Y} \kappa_{X}}{n_k} V_{XY}(\tau))$. 
On the other hand, it follows from Lemma 2 that 
$ n_k^{1/2} (\left\{S_{n_k} (Q_Y (\tau | Z), Q_X (\tau | Z) )
- S (Q_Y (\tau | Z), Q_X (\tau | Z)) \right\}) = AN(0, V(\tau))$, which indicates 
$I_{n1} = {(b_n/n_k)}^{1/2} AN(0, V(\tau))$. When $b_n = n_k$, i.e., when the conditional quantile estimates achieve root-$n$ consistency, we have
$$ (n_k)^{1/2}(\widehat{\rho}_\tau (Y, X|Z) - \rho_\tau (Y, X|Z)_0) = AN(0, \kappa_{Y}{\sigma}^2_{Y} \kappa_{Y} + \kappa_{X} {\sigma}^2_{X} \kappa_{X} + \frac{2 \kappa_{Y} \kappa_{X}}{n_k} V_{XY}(\tau)+V(\tau)).$$
However, when $b_n / n_k \rightarrow 0$, $I_{n1}$ is $o_p(1)$. Thus, the asymptotic distribution of $\widehat{\rho}_\tau (Y, X|Z)$ is given by
$$ (b_n)^{1/2}(\widehat{\rho}_\tau (Y, X|Z) - \rho_\tau (Y, X|Z)_0) = AN(0, \kappa_{Y}{\sigma}^2_{Y} \kappa_{Y} + \kappa_{X} {\sigma}^2_{X} \kappa_{X} + \frac{2 \kappa_{Y} \kappa_{X}}{n_k} V_{XY}(\tau)).$$

\end{document}